\documentclass{aa} 
\usepackage{graphicx}
\usepackage{gensymb}
\usepackage{txfonts}
\usepackage[hidelinks]{hyperref}
\hypersetup{
    colorlinks=true,
    allcolors=blue,}

\begin{document}

  \title{
  Dust and power: Unravelling the merger-active galactic nucleus connection in the second half of cosmic history\thanks{The catalogue of galaxies analysed is available in electronic form at the CDS via anonymous ftp to cdsarc.u-strasbg.fr (130.79.128.5) or via http://cdsweb.u-strasbg.fr/cgi-bin/qcat?J/A+A/.}
  }

  \subtitle{}

  \author{A. La Marca
     \inst{1,2}\thanks{\email{a.la.marca@sron.nl}}
     \and
     B. Margalef-Bentabol\inst{1}
     \and
     L. Wang
     \inst{1, 2}
     \and
     F. Gao
     \inst{2,3,4}
     \and
     A. D. Goulding
     \inst{5}
     \and
     G. Martin
     \inst{6,7,8}
     \and
     V. Rodriguez-Gomez
     \inst{9}
     \and
     S. C. Trager
     \inst{2}
     \and
     G. Yang
     \inst{1,2}
     \and
     R. Davé
     \inst{10,11}
     \and
     Y. Dubois
     \inst{12}
     }

  \institute{SRON Netherlands Institute for Space Research, Landleven 12, 9747 AD Groningen, The Netherlands
  \and
  Kapteyn Astronomical Institute, University of Groningen, Postbus 800, 9700 AV Groningen, The Netherlands
  \and
  Department of Astronomy, Nanjing University, Nanjing 210093, China
  \and
  Key Laboratory of Modern Astronomy and Astrophysics (Nanjing University), Ministry of Education, Nanjing 210093, China
  \and
  Department of Astrophysical Sciences, Princeton University, 4 Ivy Lane, Princeton, NJ 08544, USA
  \and
  Korea Astronomy and Space Science Institute, 776 Daedeok-daero, Yuseong-gu, Daejeon 34055, Korea
  \and
  Steward Observatory, University of Arizona, 933 N. Cherry Ave, Tucson, AZ 85719, USA
  \and
  School of Physics and Astronomy, University of Nottingham, University Park, Nottingham NG7 2RD, UK
  \and
  Instituto de Radioastronomía y Astrofísica, Universidad Nacional Autónoma de México, A.P. 72-3, 58089 Morelia, Mexico
  \and
  School of Physics and Astronomy, University of Edinburgh, Edinburgh EH9 3HJ, UK
  \and
  Department of Physics and Astronomy, University of the Western Cape, Bellville, 7535, South Africa
  \and
  Institut d’Astrophysique de Paris, Sorbonne Université, CNRS, UMR 7095, 98 bis bd Arago, 75014 Paris, France
       }

  \date{Received 09 October 2023; Accepted 14 August 2024}

 
 \abstract
  {}
  {Galaxy mergers represent a fundamental physical process under hierarchical structure formation, but their role in triggering active galactic nuclei (AGNs) is still unclear. 
  We aim to investigate the merger-AGN connection using state-of-the-art observations and novel methods for detecting mergers and AGNs.}
  {We selected stellar mass-limited samples at redshift $z<1$ from the Kilo-Degree Survey (KiDS), focussing on the KiDS-N-W2 field with a wide range of multi-wavelength data. 
  We analysed three AGN types, selected in the mid-infrared (MIR), X-ray, and via spectral energy distribution (SED) modelling.
  To identify mergers, we used convolutional neural networks (CNNs) trained on two cosmological simulations. 
  We created mass- and redshift-matched control samples of non-mergers and non-AGNs. 
  }
  {We first investigated the merger-AGN connection using a binary AGN/non-AGN classification. 
  We observed a clear AGN excess (of a factor of $2-3$) in mergers with respect to non-mergers for the MIR AGNs, along with a mild excess for the X-ray and SED AGNs. This result indicates that mergers could trigger all three types, but are more connected to the MIR AGNs. 
  About half of the MIR AGNs are in mergers but it is unclear whether mergers are the main trigger. 
  For the X-ray and SED AGNs, mergers are unlikely to be the dominant triggering mechanism. 
  We also explored the connection using the continuous AGN fraction $f_{AGN}$ parameter. 
  Mergers exhibit a clear excess of high $f_{AGN}$ values relative to non-mergers, for all AGN types. 
  We unveil the first merger fraction $f_{merger} - f_{AGN}$ relation with two distinct regimes. 
  When the AGN is not very dominant, the relation is only mildly increasing or even flat, with the MIR AGNs showing the highest $f_{merger}$. 
  In the regime of very dominant AGNs ($f_{AGN}\geq0.8$), $f_{merger}$ shows the same steeply rising trend with increasing $f_{AGN}$ for all AGN types. 
  These trends are also seen when plotted against AGN bolometric luminosity.
  We conclude that mergers are most closely connected to dust-obscured AGNs, generally linked to a fast-growing phase of the supermassive black hole. Such mergers therefore stand as the main (or even the sole) fuelling mechanism of the most powerful AGNs.
  }
  {}

  \keywords{Galaxies: interactions -- Galaxies: active -- Galaxies: evolution -- Techniques: image processing}

  \maketitle
%

\section{Introduction}

In the context of the widely accepted $\Lambda$-cold dark matter ($\Lambda$CDM) cosmology, structure formation proceeds in a hierarchical fashion, involving frequent mergers of smaller structures. 
During these merger events, the baryonic components of lower-mass dark matter halos collide and coalesce, eventually forming a single more massive galaxy \citep{conselice_evolution_2014,somerville_physical_2015}. 
This process is driven by gravitational forces, which pull and distort the galaxies, rearranging their stars and gas and sometimes leading to dispersion-dominated systems \citep[][and references therein]{toomre_galactic_1972,somerville_physical_2015}. 
In addition to mass assembly and morphological transformation, mergers are also expected to impact a wide range of galactic properties. 
For example, mergers have been shown to cause an increase in the star formation rate (SFR) in some cases \citep[e.g.][]{ellison_galaxy_2013,knapen_interacting_2015,martin_limited_2017,martin_role_2021,pearson_effect_2019,cibinel_early-_2019}.

Mergers may also lead to the inflow of gas onto the central supermassive black holes \citep[SMBHs,][]{barnes_fueling_1991,hopkins_unified_2006,blumenthal_go_2018}.
Various simulations predict that mergers can fuel SMBH accretion, initiating the active galactic nuclei (AGN) phase \citep[e.g.][]{di_matteo_energy_2005,hopkins_cosmological_2008,blecha_power_2018}.
However, some simulations also suggest that mergers may be responsible for only a minority of the AGN population \citep{di_matteo_black_2003,alexander_what_2012,martin_normal_2018,bhowmick_supermassive_2020,byrne-mamahit_interacting_2022,smethurst_evidence_2023}.
Observationally, the merger-AGN connection remains a topic of intense debate \citep{heckman_coevolution_2014}, with many studies either supporting \citep{koss_merging_2010,ellison_galaxy_2011,ellison_definitive_2019,hwang_activity_2012,satyapal_galaxy_2014,kocevski_are_2015,bickley_agns_2023,tanaka_galaxy_2023} or refuting \citep{grogin_agn_2005,reichard_lopsidedness_2009,cisternas_bulk_2011,kocevski_candels_2012,sabater_triggering_2015,mechtley_most_2016,smethurst_secularly_2019} this link.
In addition, there may be a dependence on AGN luminosity, with mergers playing a more important role in triggering more luminous AGNs \citep{urrutia_evidence_2008,treister_major_2012,glikman_major_2015,weigel_fraction_2018,ellison_definitive_2019,pierce_agn_2022,bickley_agns_2023}, although some counterclaims also exist \citep[][]{villforth_host_2017,hewlett_redshift_2017}.

There are many potential factors for the mixed observational results. 
One major issue is that AGNs are a complex and multi-faceted phenomenon, which releases a large amount of radiation from radio to X-rays \citep[see][for a review]{alexander_what_2012}.
However, not all AGNs emit simultaneously across the entire spectrum. 
Thus, selecting AGNs within different frequency windows may result in AGNs (and host galaxies) with distinct properties \citep[][]{yang_ceers_2023}. 
To study the merger-AGN connection, various methods have been employed to identify AGNs, such as mid-infrared (MIR) colour selection \citep[e.g.][]{satyapal_galaxy_2014,goulding_galaxy_2018,ellison_definitive_2019}, X-ray selection \citep[e.g.][]{koss_merging_2010,kocevski_are_2015,hewlett_redshift_2017}, optical emission line ratios and radio selection based on flux, and/or morphology \citep[e.g.][]{ellison_galaxy_2015,gordon_effect_2019,gao_mergers_2020, bickley_agns_2023}.


A second problem concerns the way mergers are selected, via visual classification \citep[][]{darg_galaxy_2010,tanaka_galaxy_2023}, the close-pair method \citep[][]{knapen_interacting_2015,davies_galaxy_2015}, non-parametric morphological statistics \citep[][]{lotz_new_2004,pawlik_shape_2016}, or machine learning (ML) and deep learning (DL) techniques \citep[][]{goulding_galaxy_2018,bottrell_deep_2019,pearson_effect_2019,nevin_accurate_2019, bickley_convolutional_2021}. 
Visual classification is hard to reproduce, time-consuming, and suffers from low accuracy and incompleteness \citep{huertas-company_catalog_2015}. 
The close-pair method typically requires highly complete spectroscopic data and so is observationally expensive. 
Moreover, this method cannot identify post-mergers. 
Morphological statistics are reproducible and relatively quick to compute but rely on high-quality imaging data. 
Consequently, misclassifications can rise significantly at higher redshifts \citep{huertas-company_catalog_2015}. 
Supervised ML and DL techniques are both efficient and reproducible. 
In recent years, various studies have applied convolutional neural networks (CNNs) with varying degrees of success \citep[e.g.][]{ackermann_using_2018,pearson_effect_2019,wang_towards_2020,ferreira_galaxy_2020,bickley_convolutional_2021,bickley_agns_2023,ciprijanovic_deepmerge_2020,ciprijanovic_deepmerge_2021}. 
The performance of these methods is fundamentally limited by the quality of classification labels in the training data. 
For instance, visual labels are biased towards the most conspicuous mergers. 


Lastly, merger and AGN timescales are very different, which could further complicate studies aiming to elucidate the merger-AGN connection. 
An AGN duty cycle usually lasts $\sim 10-100$ Myr \citep{marconi_local_2004}, while galaxy interaction features can survive up to a few giga-years \citep{moreno_interacting_2019}. 
Additionally, the inflowing gas needs time to fall onto the SMBH and turn on the AGN phase.
Therefore, some AGNs may be in their 'off' state when their host galaxies appear to be merging, biasing towards fewer AGNs detected in mergers \citep{villforth_morphologies_2014,shabala_delayed_2017}.
Some models predict that most of the SMBH and galaxy growth occur in an early obscured phase \citep[e.g.][]{blecha_power_2018}, followed by a 'blowout' phase in which AGN feedback limits SMBH growth and ejects gas from the galaxy \citep[e.g.][]{ishibashi_agn-starburst_2016}. 


In this study, we re-visit the relationship between mergers and the triggering of AGN, using large stellar mass limited galaxy samples at redshifts $z<1$ selected from the Kilo-Degree Survey \citep[KiDS;][]{de_jong_kilo-degree_2013}, with rich ancillary data from the X-ray to the sub-millimetre (sub-mm). 
Specifically, we address two aspects of the merger-AGN connection, namely, (1) we consider whether mergers trigger AGNs and, if so, whether this is an important triggering mechanism for different types of AGNs; (2) how the role of mergers changes with increasing AGN fraction. In other words, we explore whether galaxies hosting more dominant AGNs are more likely to be mergers.
To shed new light on these questions, we adopted innovative approaches for identifying AGNs and mergers. 
For detecting AGNs, in addition to standard binary methods (such as the MIR colour selection technique), we also employed spectral energy distribution (SED) modelling.
Thanks to the availability of high-quality multi-wavelength data, we are able to characterise AGN contribution relative to the host galaxy. 
To identify mergers, we trained CNNs and applied the trained model to the Hyper-Suprime-Cam Subaru Strategic Program \citep[HSC-SSP;][]{aihara_hyper_2018} survey images of the KiDS galaxies. 
To mitigate issues with visual classification, we trained CNNs on mock observations generated from cosmological hydrodynamical simulations, where galaxy merger histories are available. 

This paper is organised as follows. 
In Sect. \ref{sect:data}, we first present a summary of the multi-wavelength observations used in this work.
Then, we introduce the two simulations, Illustris TNG \citep{pillepich_simulating_2018} and Horizon-AGN \citep{dubois_dancing_2014}, used to train CNNs to detect mergers.  
In Sect. \ref{sect:Methods}, we describe the SED fitting process to detect AGNs and the CNN-based merger classifier. 
In Sect. \ref{sect:Results}, we present our results on how mergers are related to the binary AGN and non-AGN classification, as well as to the continuous AGN fraction parameter. We also present a set of detailed comparisons among our findings and previous works.
Finally, we summarise our main conclusions and future outlook in Sect. \ref{sect:Conclusions}. 
Throughout the paper, we assume a flat $\Lambda$CDM universe with $\Omega_M=0.2865$, $\Omega_{\Lambda}=0.7135$, and $H_0=69.32$ km s$^{-1}$ Mpc$^{-1}$ \citep{hinshaw_nine-year_2013}.
Unless otherwise stated, all magnitudes are in the AB system.

\section{Data}\label{sect:data}

\begin{figure*}
  \includegraphics[width=12cm]{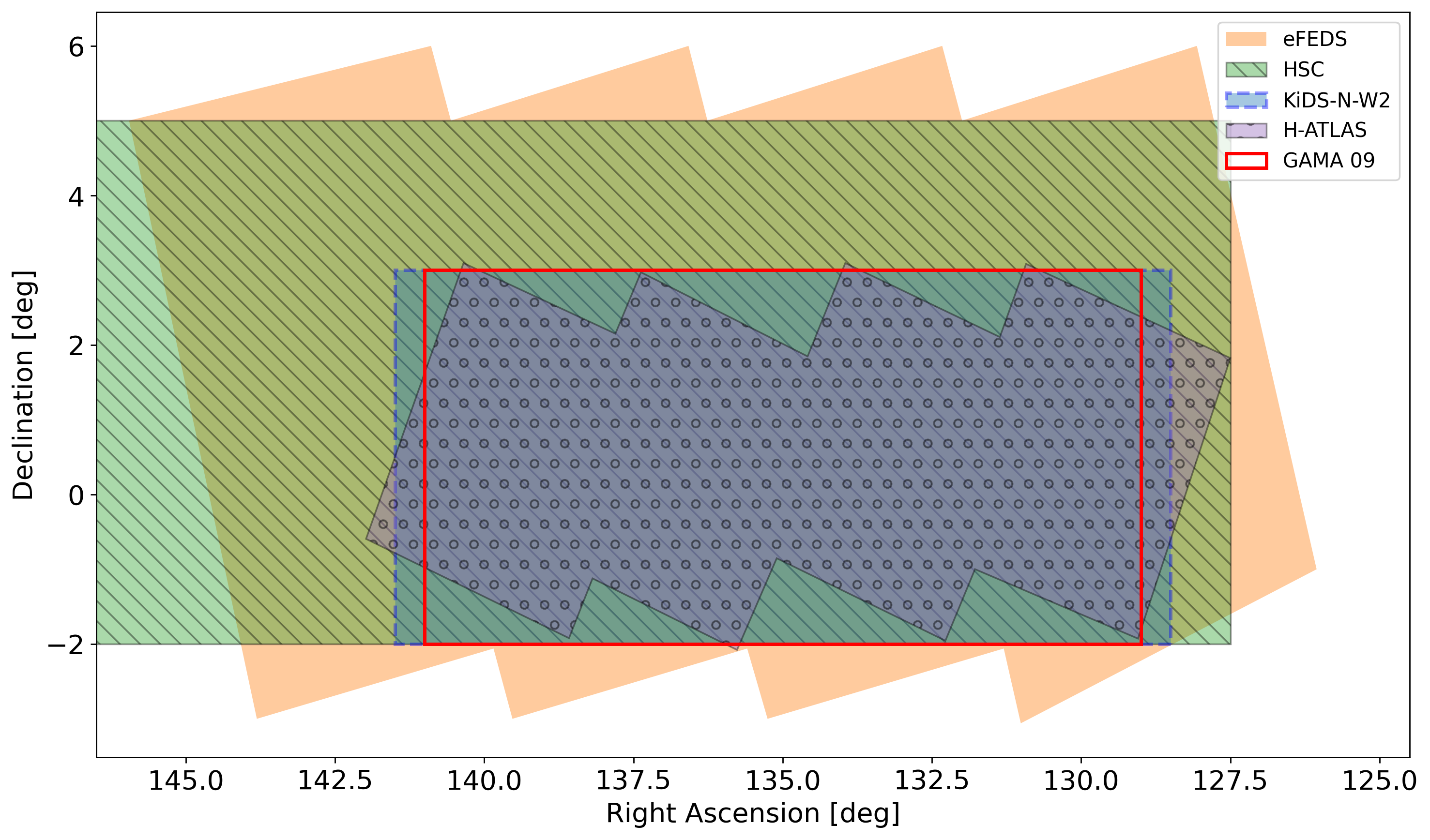}
  \\
  \sidecaption
  \includegraphics[width=12cm]{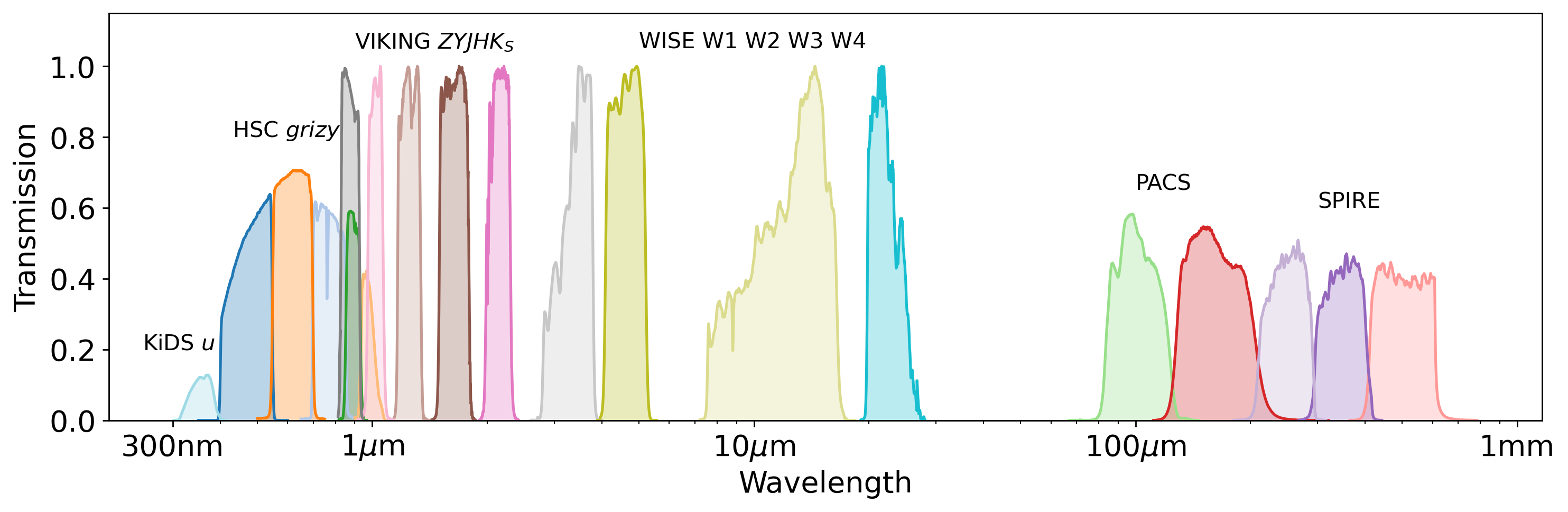}
  \caption{Multi-wavelength data used in this study. \textit{Top:} Footprints of the surveys. The eFEDS coverage is not corrected for vignetting. We do not show the WISE survey area as it covers the entire sky.
  \textit{Bottom:} Transmission curves of the filters, from the optical to the sub-mm. For clarity, we do not show the eROSITA soft X-ray band. Similarly, we only display the KiDS $u-$band as the other KiDS bands overlap with the HSC filters. 
  }
  \label{fig:filters}
\end{figure*}



In this section, we first introduce the various datasets from the multi-wavelength galaxy surveys used to construct our sample. 
Then we briefly describe the cosmological hydrodynamic simulations used to train our CNN-based merger classifier.

\subsection{Observations}\label{sect:observ}

We focussed our study on one of the KiDS equatorial fields, the KiDS-N-W2 field, which spans $128.5^{\circ}\leq R.A.\leq141.5^{\circ}$ and $-2^{\circ}\leq Dec. \leq 3^{\circ}$ \citep{kuijken_fourth_2019}.
KiDS-N-W2 benefits from extensive multi-wavelength coverage.
Moreover, this field contains one of the Galaxy and Mass Assembly survey \citep[GAMA;][]{driver_galaxy_2011} equatorial fields, the GAMA-09 field. 
GAMA is a spectroscopic survey of $r<19.8$ mag galaxies selected from the Sloan Digital Sky Survey \citep[SDSS;][]{york_sloan_2000}, which facilitates calibration of photometric redshifts. 
Thanks to its large area (65 deg$^2$) and panchromatic coverage, KiDS-N-W2 is an ideal field for this study.
Figure~\ref{fig:filters} shows the footprints and filters of the various surveys. Below we describe the individual surveys.

\subsubsection{X-ray data}\label{sect:xray}

The extended ROentgen Survey with an Imaging Telescope Array \citep[eROSITA;][]{predehl_erosita_2021} X-ray instrument observed $\sim 140$ deg$^2$ in this area, as part of the eROSITA Final Equatorial Depth Survey \citep[eFEDS;][]{brunner_erosita_2022}.
\citet{brunner_erosita_2022} performed source detection in the 0.2$-$2.3 keV range to create the eFEDS source catalogue, referred to as the main sample. 
\citet{salvato_erosita_2022} identified the multi-wavelength counterparts of these sources to derive crucial information such as photometric redshifts photo-$z$ or $z_{phot}$\footnote{The eROSITA/eFEDS main point source counterparts catalogue is available at the CDS via anonymous ftp to cdsarc.u-strasbg.fr (130.79.128.5) or via \url{http://cdsarc.u-strasbg.fr/viz-bin/cat/J/A+A/661/A3}.}.
Point sources are separated into Galactic and extragalactic samples. 
We selected extragalactic point sources (\texttt{CTP\_CLASS} 'likely' or 'secure') with a reliable counterpart (\texttt{CTP\_quality}$\geq2$) from the eFEDS main sample in the KiDS-N-W2 field.

\subsubsection{Optical and Near-IR (NIR) data}\label{sect:hsc}

We selected optical and NIR data from the combined VISTA Kilo-degree INfrared Galaxy survey \citep[KiDS-VIKING][]{kuijken_fourth_2019, edge_vista_2013} and the HSC-SSP survey. 
KiDS imaged a large part of the sky in $ugri$ \citep[][]{de_jong_kilo-degree_2013}.
The limiting magnitudes for the fourth data release (DR4) are $r\sim 25$ and $i\sim 23.6$ mag at 5$\sigma$ \citep{de_jong_third_2017}. 
The mean seeing is $0.70\arcsec$ in the $r$ band and $0.80\arcsec$ in $i$. 
The VIKING survey observed the same region in $ZYJHK_S$, reaching $K_S\sim21$ mag at 5$\sigma$ \citep[][]{de_jong_third_2017}.
A detection catalogue was used for list-driven photometry on the KiDS and VIKING images \citep{kuijken_fourth_2019}. 
A final nine-band KiDS-VIKING (hereafter KV) catalogue was produced\footnote{The KV catalogue is available at \url{http://archive.eso.org/.}}, containing PSF- and aperture-matched photometry using the {\sc GAaP} technique \citep[Gaussian Aperture and PSF photometry,][]{kuijken_gravitational_2015}. Photo-$z$ were also included, estimated using the Bayesian Photometric Redshift code \citep[BPZ;][]{benitez_bayesian_2000}. The normalised median absolute deviation $\sigma_m$ of  $(z_{phot} - z_{spec})/(1 + z_{spec})$ is 0.061. At $z_{phot}<0.9$, $11.8\%$ of the sources have $|\Delta z/(1+z_{spec})|>3\sigma_m$  \citep{wright_kidsviking-450_2019}.
We selected sources in KiDS-N-W2 and adopted the {\sc GAaP} photometry. 
However, these aperture fluxes are not optimised for total fluxes \citep{kuijken_gravitational_2015}. 
Following \citet{wright_kidsviking-450_2019}, we applied a scale factor, $F$, which is the ratio of the $r$-band AUTO flux to the {\sc GAaP} $r$-band flux $F=f_{AUTO}/f_{\sc GAaP}$. 

We used the HSC-SSP images to detect mergers, as they are deeper and sharper than KiDS.
The HSC-SSP is a ground-based imaging survey that, to date, has covered $\sim 1200$ deg$^2$ in \emph{grizy}, with a pixel resolution of 0.168$\arcsec/$pixel \citep[][]{aihara_hyper_2018}. 
The KiDS-N-W2 field was observed as part of its 'Wide' survey, down to $i\sim26$ mag at 5$\sigma$ for point sources, with an average seeing of 0.61$\arcsec$ in $i$ \citep{aihara_third_2022}\footnote{ 
All HSC-SSP data products are accessible at \url{https://hsc-release.mtk.nao.ac.jp/}.}.
We made use of the DR2 forced photometry catalogue \citep[][]{aihara_second_2019}, which includes spectroscopic $z$ (spec-$z$) when available from other surveys. 
The HSC pipeline provided model-fitted photometry obtained with the {\sc cModel} algorithm \citep{huang_characterization_2018}.
As we used the KV photo$-z$, we limited the HSC-SSP catalogue to $i<24$ mag, as this is roughly the KiDS depth. 
The HSC-SSP DR2 has significant background subtraction issues that require an aggressive star mask to be used \citep{aihara_second_2019}. 
Therefore we rejected masked objects using the flags \texttt{mask pdr2 bright objectcenter} and \texttt{pixelflags bright objectcenter}. 
Moreover, we excluded sources with the following pixel flags, \texttt{pixelflags edge, pixelflags interpolatedcenter, pixelflags saturatedcenter, pixelflags crcenter, pixelflags bad} \citep{aihara_first_2018}, and with failed {\sc cModel} fits.
We downloaded the $i$-band coadded images of each galaxy in our sample from the HSC-SSP DR3 \citep{aihara_third_2022}.
Details on the cutouts are provided in Sect. \ref{sect:img_prep}.

{\renewcommand{\arraystretch}{1.1}
\begin{table*}[t]
  \caption{Number of galaxies in our sample.}
  \begin{center}
    \begin{tabular}{lll}
    \hline 
    Survey & Nr. galaxies & Summary \\
    \hline
    HSC-SSP \& KV & 1\,075\,815 & HSC $i\leq24$ mag; matching radius $1\arcsec$\\
    WISE & 72\,875 & W1 or W2 S/N$\geq3$; W1 and W2 S/N$\geq1$; matching radius $1.4\arcsec$\\
    H-ATLAS (250, 350 and 500 $\mu$m) & 5\,492 & Optical counterparts reliability $R\geq0.80$; matching radius $1.5\arcsec$\\
    H-ATLAS (100 and 160 $\mu$m) & 4\,990 & - \\
    eFEDS soft band & 3\,139 & Only extragalactic point sources in the main sample; matching radius $1.5\arcsec$\\
    \hline
    \end{tabular} 
  \end{center}
  \label{tab:counts}
  \tablefoot{
The columns correspond to the survey name, number of galaxies detected, matching radius used and other important selections applied to a specific survey. KV and HSC-SSP are listed together as we required galaxies to be detected by both surveys.
}
\end{table*}
}

\subsubsection{Mid-IR and sub-mm data}\label{sect:ir_data}

The NASA Wide-field Infrared Survey Explorer \citep[WISE;][]{wright_wide-field_2010} mapped the entire sky at 3.4, 4.6, 12, and 22 $\mu$m (i.e. bands W1, W2, W3, and W4). 
The latest Data Release, AllWISE \citep{cutri_explanatory_2013,cutri_vizier_2014}, combined data from the cryogenic and NEOWISE \citep{mainzer_neowise_2011} post-cryogenic phases\footnote{The AllWISE catalogue is available at \url{https://irsa.ipac.caltech.edu/frontpage/}.}, reaching a $5\sigma$ depth of 16.96, 15.95, 11.46, and 8.04 mag Vega (54, 71, 730, and 5000 $\mu$Jy) in W1, W2, W3, and W4, respectively \citep{cutri_explanatory_2013}. 
We selected objects with a signal-to-noise ratio (S/N) $>1$ in W1 and W2, and S/N$\geq$3 in W1 or W2.
Furthermore, we only included sources below the saturation limits, which are W1 $>8$ mag and W2 $>7$ mag (Vega).
We converted the WISE magnitudes from Vega to AB using the conversions provided by the WISE team\footnote{\url{https://wise2.ipac.caltech.edu/docs/release/allsky/expsup/sec4_4h.html}.}.

The {\it Herschel} Astrophysical Terahertz Large Area Survey \citep[H-ATLAS;][]{valiante_herschel-atlas_2016} observed $\sim$600 deg$^2$ at 100, 160, 250, 350, 500 $\mu$m.
Applying the source detection software \textsc{MADX} \citep[][]{maddox_madx_2020} on the 250 $\mu$m background-subtracted map resulted in a catalogue containing point source fluxes and aperture photometry for resolved sources. 
SDSS optical counterparts were identified using a likelihood ratio technique \citep{bourne_herschel-atlas_2016}.
We selected sources with reliable optical counterparts \citep[reliability $R\geq0.80$;][]{bourne_herschel-atlas_2016} from the DR1 catalogue\footnote{All H-ATLAS data are available at \url{http://www.h-atlas.org/}.}, which includes all sources $>2.5\sigma$ at 250 $\mu$m and  $\geq4\sigma$ at 250, 350, or 500 $\mu$m. 
Additionally, detections $\geq3\sigma$ at 100 and 160 $\mu$m are included. 
The 1$\sigma$ limits are 14.7, 16.3, 7.4, 9.4, and 10.2 mJy at 100, 160, 250, 350, and 500 $\mu$m, respectively \citep{valiante_herschel-atlas_2016}. 
Finally, we multiplied the 250, 350, and 500 $\mu$m fluxes by the $K_{ColE}$ parameter given in Table 5.7 in the Spectral and Photometric Imaging Receiver (SPIRE) handbook \citep[see also][Appendix A1]{valiante_herschel-atlas_2016}.

\subsubsection{Catalogue cross-matching and summary}\label{sect:catalogue}

\begin{figure}
    \centering
    \includegraphics[width=.3\textwidth]{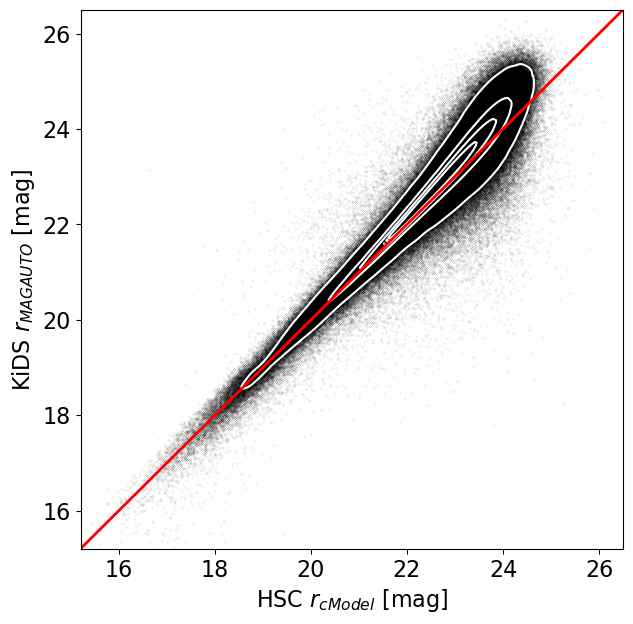}
    \includegraphics[width=.3\textwidth]{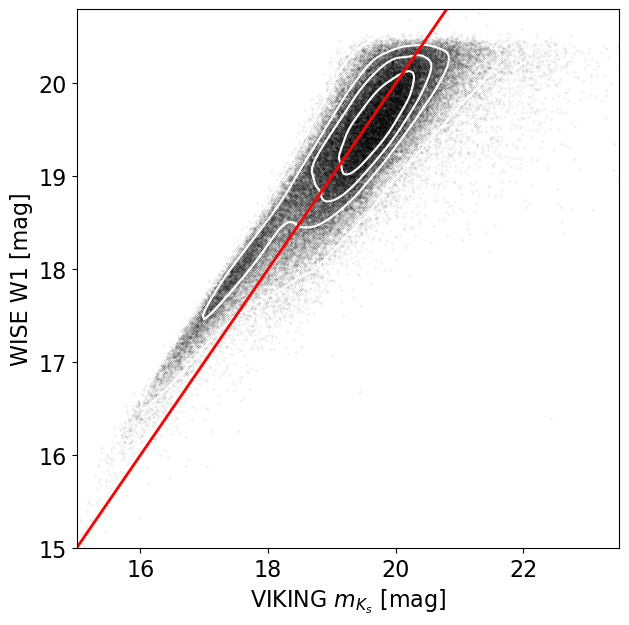}
    \caption{Flux comparison. \textit{Top:} KiDS $r_{MAGAUTO}$ vs HSC $r_{cModel}$. 
    \textit{}{Bottom}: WISE W1 vs VIKING $K_S$ magnitude. The red line indicates the one-to-one line. The white lines show the density contours.}
    \label{fig:match_comp}
\end{figure}

First, we cross-matched the KV catalogue and the HSC-SSP DR2 catalogue, adopting a $1\arcsec$ search radius. 
Requiring sources to be detected in both surveys ensured that both HSC images and KV photo$-z$ were available.
For matching to other surveys, we used HSC coordinates, given its superior resolution.
We matched the HSC-KV sample with AllWISE, using a $1.4\arcsec$ radius \citep{goulding_galaxy_2018}. 
We retained the sources with no WISE counterparts but removed WISE detections with multiple optical counterparts \citep[see also][]{toba_hyper-luminous_2015,jarrett_galaxy_2017}, which account for $\sim$0.4\% of the HSC-KV sample.
We compare KiDS and HSC $r$-band magnitudes, and VIKING $K_S$ magnitudes with WISE W1 magnitudes in Fig. \ref{fig:match_comp}. 
Most of the KiDS-HSC matches lie close to the 1:1, with increased scatter at fainter magnitudes. 
Likewise, W1 magnitudes correlate with $m_{K_S}$, with a tail toward the fainter end.
Finally, H-ATLAS and eFEDs measurements are included by cross-matching to their optical counterparts, using a search radius of 1.5$\arcsec$. 
We prioritised spec$-z$ when available, followed by photo$-z$ from eFEDS, and then photo$-z$ from the KV catalogue.
The last one constituted the largest redshift source, $\sim$98\% of the total sample.

To perform the data cleaning, we first selected objects at $0.1\leq z \leq 1.0$, as HSC imaging does not allow merger identification at higher $z$  \citep{goulding_galaxy_2018}.
We then removed stars identified by the HSC pipeline by excluding detections with \texttt{prob\_star} $=1$ for sources without spec-$z$.
Finally, we applied the recommended KV quality flags cuts, \texttt{IMAFLAGS\_ISO} $=0$ and {\sc SExtractor} $<4$, to exclude objects with bad photometry and bad detections \citep[][]{kuijken_fourth_2019}.
After data cleaning, our sample contained just over one million galaxies. 
Final galaxy counts and band coverage details are presented in Table \ref{tab:counts}.

\subsection{The IllustrisTNG and Horizon-AGN simulations}\label{sect:simul}

The IllustrisTNG project is a cosmological hydro-dynamical simulation that consists of three volumes varying in size and resolution \citep{pillepich_simulating_2018,nelson_IllustrisTNG_2019}.
We used the TNG$-$100 and TNG$-$300 boxes. 
The latter was chosen because of its larger volume and larger number of more massive galaxies. 
On the other hand, TNG$-$100 has a factor of 8 better baryonic matter resolution ($1.4\times10^6$ M$_{\odot}$) and is used to extend the sample to lower masses.
We selected galaxies between $z=0.1$ and $1$, corresponding to simulation snapshot numbers $50-91$. 
The TNG$-$300 and TNG$-$100 samples were restricted to stellar mass $M_*>5\times10^9$ M$_{\odot}$ and $>10^9$ M$_{\odot}$ respectively, to ensure that all galaxies have a reasonable number of particles. 
A complete merger history is available for each galaxy \citep{rodriguez-gomez_merger_2015}, identified using the \textsc{Subfind} algorithm \citep{springel_populating_2001}. 
To construct the merger sample, we tracked the merger trees and selected galaxies that had a merger event in the last 300 Myr or will merge in the next 800 Myr. 
We only considered major mergers with stellar mass ratios up to 4. Galaxies within the defined temporal window, mass and $z$ ranges, were labelled as mergers. 
In total, we found 280\,753 mergers.
To create a balanced dataset of non-merging galaxies, we selected galaxies that did not meet those criteria. 
Since this population is much larger, we randomly chose an equal number of non-mergers. 


The Horizon-AGN simulation is a cosmological hydro-dynamical simulation of (100 Mpc $h^{-1})^3$ comoving volume, with a stellar mass particle resolution of $2\times10^6$ M$_{\odot}$ \citep{dubois_dancing_2014,dubois_horizon-agn_2016}.
Horizon-AGN includes various sub-grid models such as star formation and feedback from stars and AGNs. 
Galaxies are identified with a recent version of the \textsc{AdaptaHOP} algorithm, updated for building merger trees \citep{tweed_building_2009}. 
We selected galaxies with $0.1<z<1$ and $M_*>10^9$ M$_{\odot}$ and adopted the same merger definition to build a sample of mergers and non-mergers. 
Additionally, to increase sample size, we produced two images for each galaxy from two different projections and ended up with 232\,296 mergers and 161\,374 non-mergers. 
Finally, we divided both simulation samples into four redshift intervals, $z$-bin 1 $=[0.1, 0.31)$, $z$-bin 2 $=[0.31, 0.52)$, $z$-bin 3 $=[0.52, 0.76)$, and $z$-bin 4 $=[0.76, 1.00]$, with each bin containing a similar number of galaxies. 

\subsection{Mock images preparation}\label{sect:img_prep} 

Observational effects, such as the presence of background galaxies and noise are crucial for improving the performance of CNN-based merger classifiers \citep{bottrell_deep_2019,huertas-company_hubble_2019,rodriguez-gomez_optical_2019,snyder_automated_2019}. 
To emulate HSC observations, first, we generated synthetic $i$-band images from the simulations, with a physical size of 160 $\times$ 160 kpc and the same pixel resolution as the HSC images. 
This size choice is motivated by the expected maximum radial separation between merging galaxies in the chosen time window \citep[][]{moreno_interacting_2019}. 
Then, the synthetic images were convolved with the real $i$-band PSF, retrieved from the HSC-SSP database. 
Lastly, we added Poisson noise and injected each image into cutouts of real HSC observations, which are described below. 

To make cutouts of HSC observations without bright sources and/or artefacts, first, we generated a catalogue of low-$z$ and relatively bright sources that we want to avoid. 
We selected these objects from the HSC-SSP source catalogue using the following criteria, $129^{\circ}\leq R.A.\leq 180^{\circ}$ and  $-2^{\circ}\leq dec. \leq 2^{\circ}$, photo-$z\leq1$ (with reduced $\chi^2\leq3$), $g_{cModel}\leq26.0$ mag, $r_{cModel}\leq25.6$ mag, $i_{cModel}\leq25.4$ mag, $z_{cModel}\leq24.2$ mag, and $y_{cModel}\leq23.4$ mag. 
We still allowed our sky cutouts to contain possible faint sources and higher-$z$ background galaxies, as would happen in real observations. 
Then, we generated random sky coordinates while ensuring that there is no catalogued bright or low-$z$ source within 21$\arcsec$ (based on the density of the sources to be avoided). 
These coordinates were used as the centres of our sky cutouts. 
After that, we examined each cutout to remove possible pixel defects. 
In particular, we removed post-stamps with bad pixels, saturated pixels, unmasked NaN, and possible missed bright objects. 
A complete list of mask flags is given by \citet{bosch_hyper_2018}.

Our network architecture requires images of a single size. 
Thus we resized all images to a common size for each $z$-bin.
At the midpoint of each redshift bin, a physical size of 160 kpc would correspond to 320, 192, 160, and 128 pixels, respectively. 
While resizing the images, we kept a constant physical size of 160 kpc. 
After that, we cropped the central part of the images using a different size according to the $z$-bin.
We selected the central $120\times120$, $96\times96$, $80\times80$, and $64\times64$ pixels for the galaxies in $z$-bins 1, 2, 3, and 4, respectively. 
Normalisation was applied following \citet{bottrell_deep_2019}, so that all images were in hyperbolic arcsin-scale in the range $0-1$ and were scaled to maximise the contrast for the central target. 
Below, we summarise the normalisation steps: 
\begin{enumerate}
  \renewcommand{\labelenumi}{\it \roman{enumi})}
  \item We took the hyperbolic arcsin of the sky-subtracted images. Values $<-7$ were converted to NaNs. 
  \item We computed the median of each image, $a_{min}$, and the 99th percentile, $a_{max}$, for central boxes of side 40, 32, 26, and 22 pixels, in $z$-bins 1, 2, 3, and 4, respectively. 
  \item Values $<a_{min}$ were set to $a_{min}$, including the NaNs. Values $>a_{max}$ were set to $a_{max}$. The resulting clipped images were normalised by subtracting $a_{min}$ and dividing by $a_{max}-a_{min}$.
\end{enumerate}

\section{Methods}\label{sect:Methods}

\subsection{CIGALE SED fitting}\label{sect:CIGALE}

\begin{table*}[ht]
\small
\caption{Fitting parameters in the CIGALE runs.}
  \centering
  \begin{tabular}{lll}
  \hline
  \multicolumn{3}{c}{\textbf{\large Star-Formation History}} \\
  \hline
  delayed $\tau$+starburst & e-folding time of the main population & 1000, 3000, 5000, 8000 Myrs \\
  & age the main population & 4000-13000 (step 1000) Myrs \\
  & e-folding time of the late starburst population & 9000, 13000 Myrs \\
  & age of the late starburst population & 1, 50, 150 Myrs \\
  & mass fraction of the late starburst population & 0.01, 0.1, 0.25 \\
  \hline
  \multicolumn{3}{c}{\textbf{\large Single Stellar Population}}\\
  \hline
  \citet{bruzual_stellar_2003} & IMF & Chabrier (1) \\
  & metallicity & solar (0.02) \\
  \hline
  \multicolumn{3}{c}{\textbf{\large Dust attenuation}}\\
  \hline
  modified starburst & colour excess & 0.0, 0.1, 0.25, 0.4, 0.55, 0.7, 0.85, 1.0\\
  & reduction factor & 0.25, 0.5, 0.75 \\
  & Slope delta of the power law modifying the attenuation curve & -0.45, -0.3, -0.2, -0.1, 0 \\
  & Extinction law & Milky Way (1) \\
  \hline
  \multicolumn{3}{c}{\textbf{\large Dust emission}}\\
  \hline
  Draine 2014 & Mass fraction of PAH & 0.47 \\
  & minimum radiation field & 5, 15, 25\\
  & power-law slope $\alpha$ in dM/dU $\propto U^\alpha$ & 2 \\
  & Fraction illuminated from minimum to maximum radiation field & 0.02 \\
  \hline
  \multicolumn{3}{c}{\textbf{\large X-ray}}\\
  \hline
  X-CIGALE X-ray & AGN photon index $\Gamma$ & 1.8 \\
  \citep{yang_x-cigale_2020,yang_fitting_2022} & power slope $\alpha_{ox}$ & -2.0, -1.8, -1.6, -1.4, -1.2\\
  & Max deviation from the $\alpha_{ox}-L_{\nu,2500\AA}$ relation & 0.4\\
  & AGN X-ray angle coefficients ($a_1$, $a_2$) & (0.5, 0)\\
  \hline
  \multicolumn{3}{c}{\textbf{\large AGN template}}\\
  \hline
  SKIRTOR & Average edge-on optical depth at 9.7 $\mu$m & 3, 7, 11 \\
  \citep{stalevski_3d_2012} & torus density radial parameter $p$ & 1 \\
  & torus density angular parameter $q$ & 1\\
  & Angle between the equatorial plane and edge of the torus & 40$\degree$ \\
  & viewing angle & 10, 30, 50, 70$\degree$ \\
  & AGN fraction, $f_{AGN}$ & 0, 0.1, 0.2, 0.3, 0.45, 0.6, 0.75, 0.9, 0.99\\
  & Restframe wavelength range where $f_{AGN}$ is computed & 3-30 $\mu$m \\
  & extinction law of polar dust & SMC (0) \\
  & E(B-V) of polar dust & 0, 0.2, 0.4, 0.8\\
  & temperature of polar dust & 100\\
  & emissivity of polar dust & 1.6\\
  \hline
  \end{tabular}  
  \label{tab:cigale_parameter}
\end{table*}

We used the SED fitting and modelling tool Code Investigating GALaxy Emission \citep[CIGALE;][]{burgarella_star_2005,noll_analysis_2009,boquien_cigale_2019} to derive properties such as stellar mass.
The latest version includes AGN models and can fit data from the X-ray to the radio \citep[version 2022.1\footnote{Every CIGALE version is accessible at \url{https://cigale.lam.fr/}};][]{yang_x-cigale_2020,yang_fitting_2022}.
CIGALE assumes energy balance so that the energy absorbed by dust is re-emitted in the IR/sub-mm.
We used a delayed$-\tau$ star-formation history (SFH), which is able to model both early- and late-type galaxies, using small and large $\tau$ respectively \citep{boquien_cigale_2019}.
We also included an optional exponential burst component for possible recent star formation. 
We chose the \citet{bruzual_stellar_2003} single stellar population (SSP) model, Chabrier initial mass function (IMF), solar metallicity, \citet{calzetti_dust_2000} dust attenuation law and \citet{draine_andromedas_2014} dust emission models. 
We included the X-ray module for modelling X-ray emission from both AGNs and galaxies. 
CIGALE offers two AGN template libraries. 
The first is the \citet[][]{fritz_revisiting_2006} smooth AGN model that assumes a flared disk geometry for the dust distribution and includes a distribution function of graphite and silicate grains.
The second is the SKIRTOR model \citep{stalevski_3d_2012}, which also assumes a flared disk geometry but models the dusty torus as a two-phase medium consisting of high-density clumps and a low-density medium between the clumps.
We used the SKIRTOR model as it is more physical \citep{yang_x-cigale_2020}.
To accurately determine the AGN component, we only applied the described configuration to galaxies with at least one MIR or X-ray measurement. 
For the remaining galaxies, we only used CIGALE to estimate their stellar masses $M_*$ without including the X-ray and AGN modules, significantly reducing computing time. 
A summary of the fitted CIGALE parameters can be found in Table \ref{tab:cigale_parameter}.

\begin{figure}
  \centering
  \includegraphics[width=.45\textwidth]{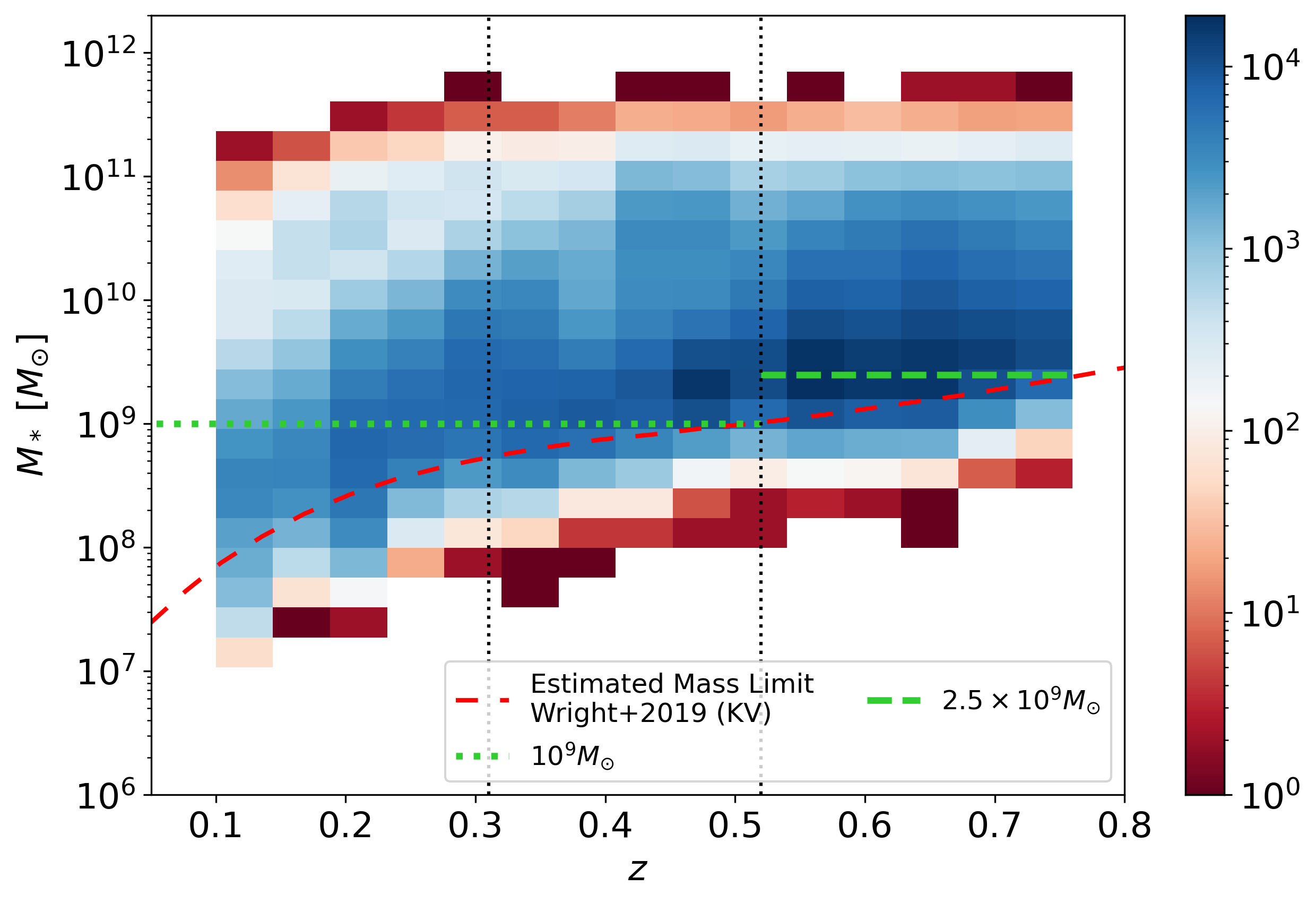}
  \caption{Stellar mass distribution as a function of redshift shown as a 2D histogram with logarithmic scaling as indicated by the colour bar. The red line is the mass limit of the KV survey. The vertical lines indicate the $z$-bins and the corresponding mass limits are shown in the legend.}
  \label{fig:m_star}
\end{figure}

We selected a reliable sample by discarding SED fits with reduced $\chi^2>5$. 
We used the Bayesian results in the CIGALE output instead of the best-fit values. 
This approach includes the effect of intrinsic degeneracies between parameters, which allows us to obtain more reliable estimates and uncertainties \citep{boquien_cigale_2019}.
In Fig. \ref{fig:m_star}, we show the stellar mass distribution of the entire sample as a function of $z$.
Hereafter, we focus on the first three redshift bins, given the difficulty in identifying mergers in the last $z$-bin (see Sect. \ref{sect:CNN-perf}).
To construct a stellar mass limited sample within each $z$-bin, we used the mass limits estimated by \citet{wright_kidsviking-450_2019} for the KV survey in combination with the simulation limits.
This is a reasonable choice, as we required all galaxies to be detected by the KV survey and our models were trained on simulated galaxies with $M_*\geq 10^9$ M$_{\odot}$ (see Sect. \ref{sect:CNN}).
We defined the limits as M$_*\geq 10^9$ M$_{\odot}$ in the first two redshift bins ($0.1\leq z < 0.52$), and M$_*\geq 2.5\times 10^9$ M$_{\odot}$ in $z$-bin 3 ($0.52\leq z < 0.76$). 
After applying the limits, we were left with a mass-complete sample of 69\,140, 183\,554 and 266\,709 galaxies, in $z$-bins 1, 2, and 3, respectively. 
Of these,  9\,159, 10\,614, and 23\,108 galaxies have a fitted AGN component. We show examples of best-fit SEDs in Appendix~\ref{app:CIGALE_examples}.

\subsection{AGN selection}\label{sect:agn_sel}

\begin{figure}
    \centering
    \includegraphics[width=.45\textwidth]{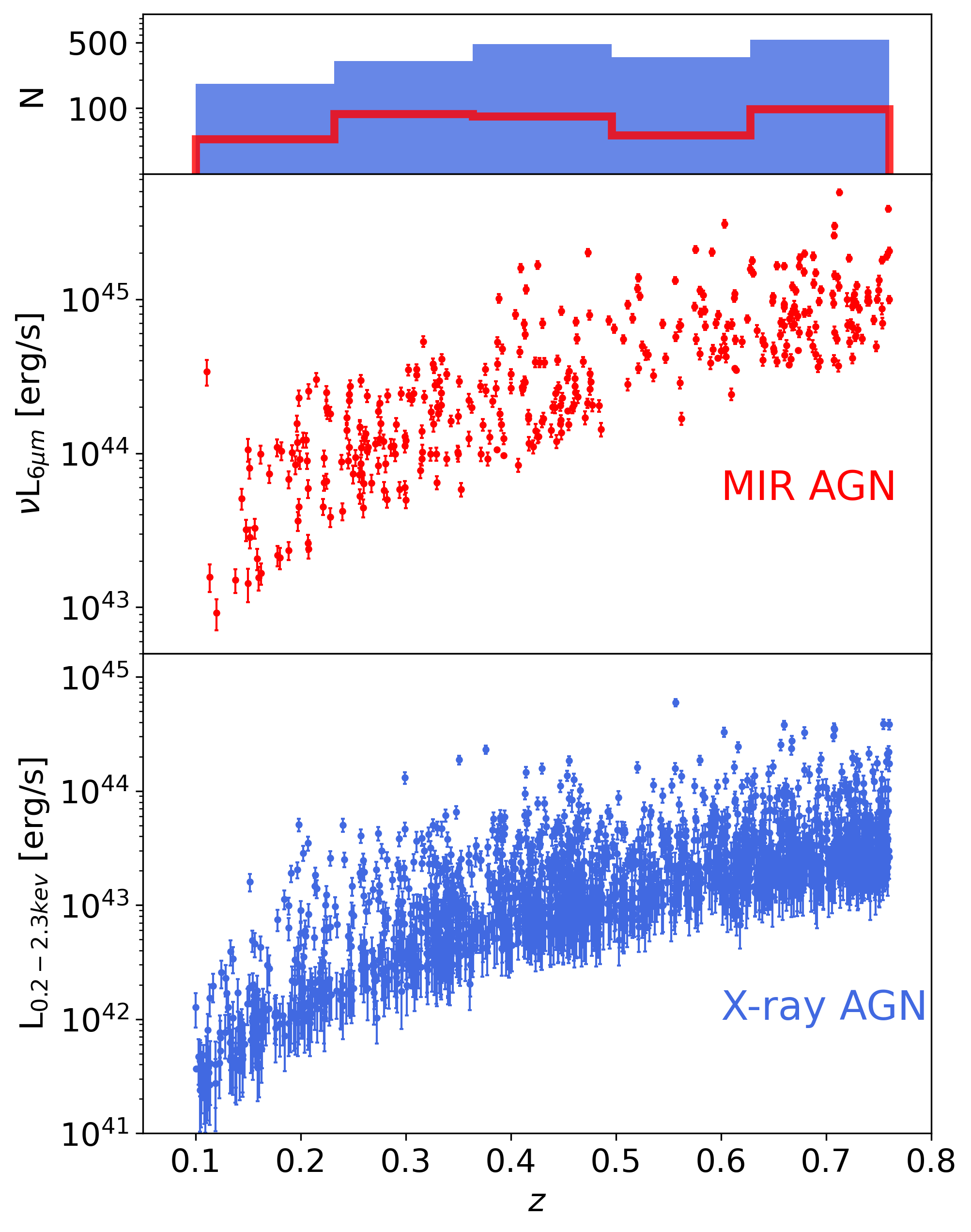}
    \caption{MIR and X-ray AGNs. \textit{Mid:} Rest-frame 6$\mu$m luminosity $L_{6\mu m}$ vs $z$ for the MIR AGNs. 
    \textit{Bottom:} Soft X-ray luminosity $L_{0.2-2.3 keV}$ vs $z$ for the X-ray AGNs. 
    \textit{Top:} Redshift distributions of the MIR and X-ray AGNs.}
    \label{fig:L_agn}
\end{figure}

\begin{figure}
    \centering
    \includegraphics[width=.4\textwidth]{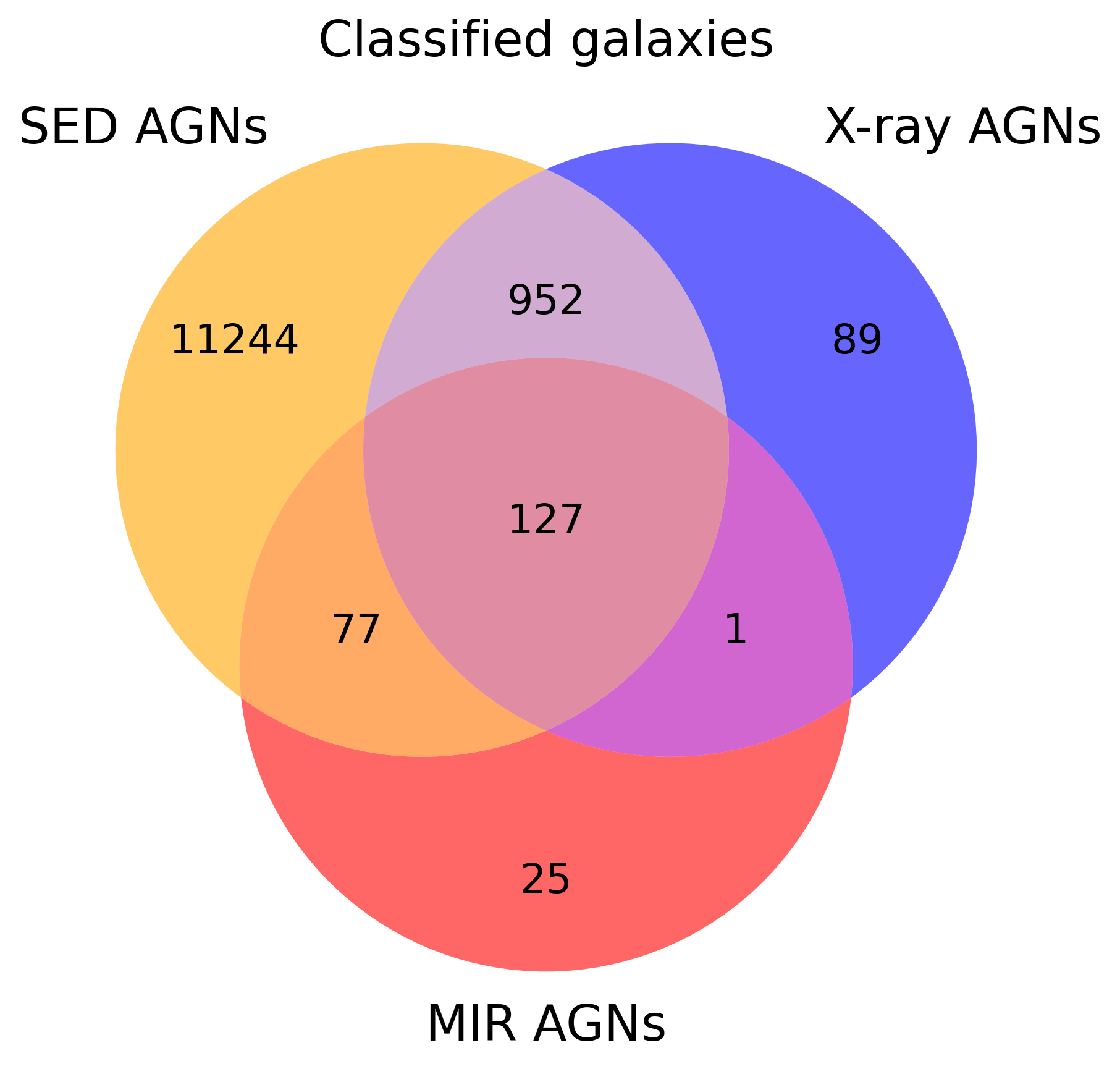}
    \caption{Venn diagram for the three AGN types for the classified galaxies (Sect. \ref{sect:CNN}). While different types of AGNs do have some overlap, each selection also identifies unique AGNs missed by other methods.}
    \label{fig:Venn-diag}
\end{figure}

To select MIR AGN, we used a single colour cut $W1-W2 > 0.8$ mag Vega \citep{stern_mid-infrared_2012} for all galaxies with $W2\leq15$ mag Vega, requiring $S/N\geq 5$ in both bands, which yielded 96, 118, and 145 AGNs in $z$-bins 1, 2, and 3, respectively. 
We used the rest-frame 6 $\mu$m luminosity to trace the AGN accretion power. 
The AGN dusty torus structure is believed to absorb the AGN bolometric luminosity produced in the X-ray/UV/optical and re-emit mainly at $\sim5-30\,\mu$m \citep{lutz_relation_2004,mateos_revisiting_2015}.
The rest-frame 6 $\mu$m flux is derived by linearly interpolating the WISE W1, W2, and W3 band fluxes.
The X-ray AGNs were selected from the eFEDS main catalogue \citep{salvato_erosita_2022}, resulting in 289, 699, and 830 AGNs in $z$-bins 1, 2, and 3, respectively. 
These X-ray AGNs are dominated by X-ray unobscured sources \citep{liu_erosita_2022}.
We used the soft band luminosity, $L_{0.2-2.3\;kev}$, to trace the AGN emission power. 
In Fig. \ref{fig:L_agn}, we show the MIR and X-ray AGNs in the luminosity vs $z$ space.
The MIR AGNs lie in the range $L_{6\mu m}\simeq10^{43}-10^{46}$ erg/s, while the X-ray AGNs span over $L_{0.2-2.3\;kev} \simeq 3\times 10^{41} - 10^{45}$ erg/s. 
The X-ray luminosity range we observe is in agreement with the AGN catalogue presented by \citet{liu_erosita_2022}, which performed an X-ray spectral analysis. 
They found that about $80\%$ of the eFEDS point sources are in the eFEDS AGN catalogue. 
It is common to require $L_{0.2-2.3\;kev}>10^{42}$ erg/s to select X-ray AGNs \citep[e.g.][]{riccio_x-ray_2023}, even though some AGNs may be fainter \citep{aird_evolution_2015,liu_erosita_2022}. 
In our sample, only $\lesssim4\%$ of the X-ray AGNs show $L_{0.2-2.3\;kev} \lesssim 10^{42}$ erg/s. 
We conducted our analysis including and excluding them and found no significant difference. 
Therefore, we opted to keep these sources. 

\begin{figure}
    \centering
    \includegraphics[width=.49\textwidth]{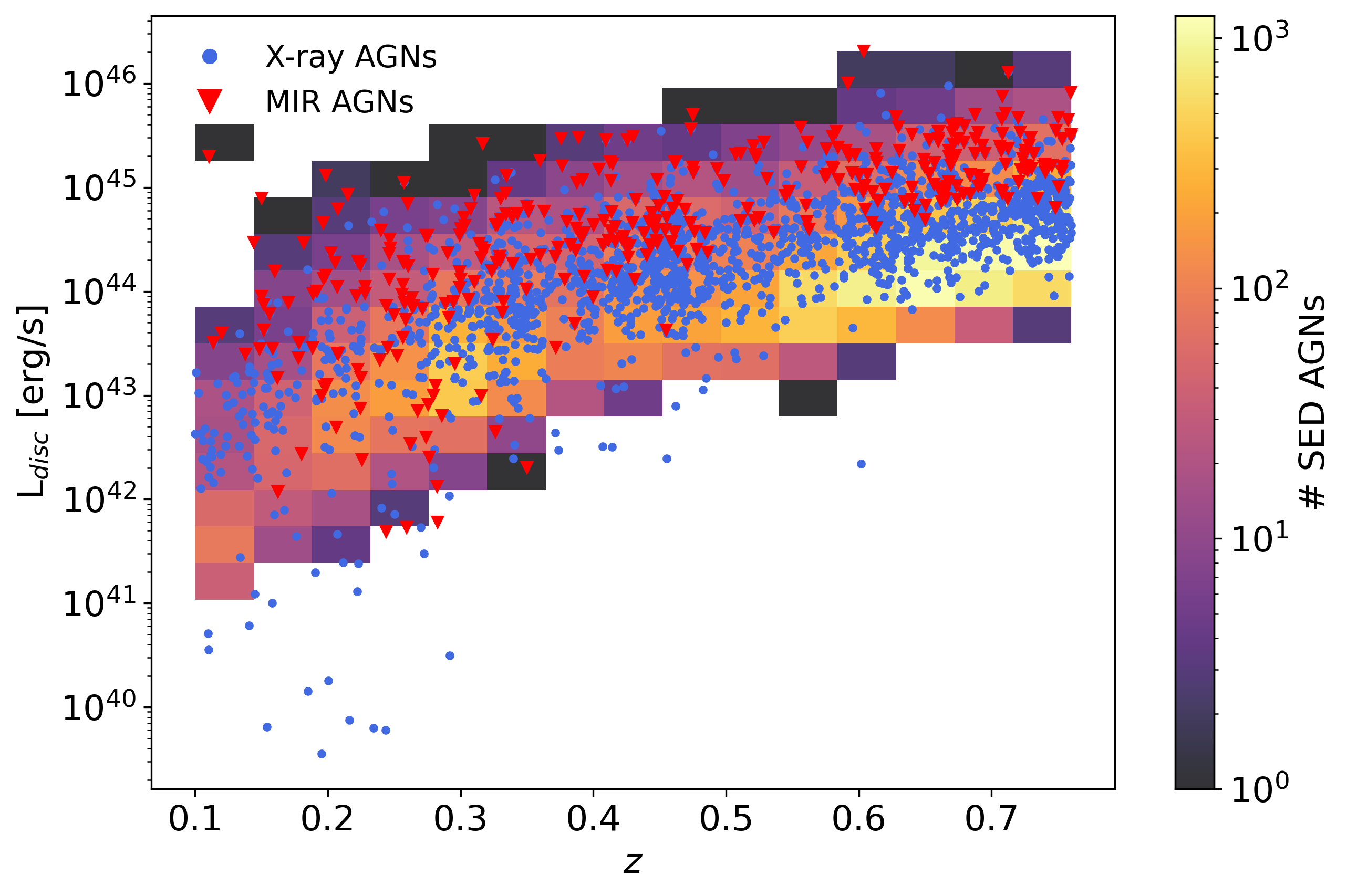}
    \caption{AGN bolometric luminosity $L_{disc}$ vs redshift for MIR AGNs (triangles) and X-ray AGNs (circles).
    We also plot the 2D distribution of the SED AGNs (with colour-coding based on the number of sources).}
    \label{fig:L_disc}
\end{figure}

\begin{figure}
  \centering
  \includegraphics[width=.49\textwidth]{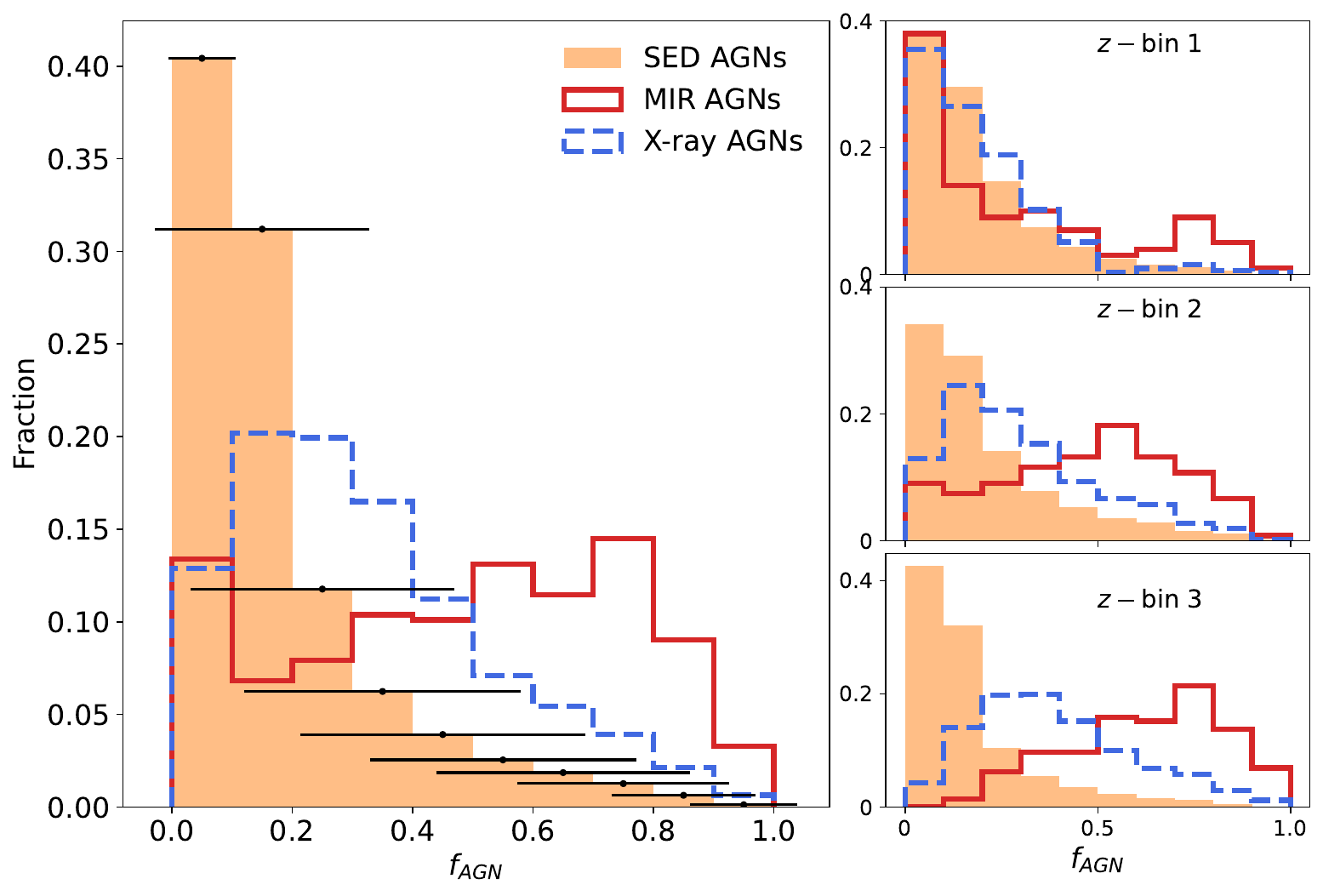}
  \caption{
  AGN fraction distributions for the SED AGNs (orange bins), MIR AGNs (solid red line), and X-ray AGNs (dashed blue line). Horizontal bars indicate the median uncertainty in $f_{AGN}$ in each bin.
  The left panel displays the distributions over the entire redshift range and the three panels on the right in individual $z$-bins. The sample shown here contains only galaxies detected in the MIR and/or in the X-ray.
  }
  \label{fig:f_agn}
\end{figure}

In addition to the MIR and X-ray AGNs, we also selected SED AGNs if the SED fitting yielded $f_{AGN}\geq 5\%$, resulting in 2\,260, 3\,125, and 11\,214 SED AGNs in $z$-bins 1, 2, and 3, respectively.
Figure \ref{fig:Venn-diag} provides the AGN counts for each AGN type for the classified galaxies (see Sect. \ref{sect:CNN}).
From the SED fits, we also computed the viewing angle-averaged intrinsic accretion-disk luminosity $L_{disc}$ \citep[][]{yang_linking_2018}  and AGN fraction ($f_{AGN}$).
The $L_{disc}$ distributions for the three AGN types are displayed in Fig. \ref{fig:L_disc}. $L_{disc}$ is numerically equivalent to the angle averaged AGN bolometric luminosity due to energy conservation. The black hole accretion rate (BHAR) is related to the AGN bolometric luminosity as follows, 
\begin{equation}
    BHAR = \frac{L_{disc}(1-\epsilon)}{\epsilon c^2} = \frac{1.59 L_{disc}}{10^{46}\mathrm{erg\,s^{-1}}} M_{\odot},
\end{equation}
where $c$ is the speed of light and $\epsilon$ is the radiative efficiency, for which we adopted a conventional value of $\epsilon = 0.1$ \citep[e.g.][]{brandt_cosmic_2015}. AGN fraction ($f_{AGN}$) is defined as the AGN strength relative to the host galaxy ($L_{AGN}/L_{tot}$). In this work, we calculated $f_{AGN}$ in the rest-frame $3-30\,\mu$m. 
In Fig. \ref{fig:f_agn}, we plot the $f_{AGN}$ distributions of the three AGN types, for the entire $z$ range and individual $z$-bins. 
The SED AGNs are dominated by galaxies with low $f_{AGN}$ (mostly $<0.3$). 
The MIR AGNs have typically high $f_{AGN}$, with the X-ray AGNs in between the other two AGN types. 

\subsection{Merger identification with CNNs}\label{sect:CNN}

\begin{table}[hb]
\caption{CNN architectures.}
\small
\begin{tabular}{lccc}
\hline
Layer type & \# Param. & Output shape & Properties \\ \hline
Input & 0 & (1,N,N) & \\
\hline
\begin{tabular}[l]{@{}l@{}}Convolutional\\ 16 filters (11,11)\end{tabular} & 1952 & (16,N/2,N/2) & \begin{tabular}[c]{@{}c@{}}2 pixels stride, \\ same padding, \\ Leaky ReLU act.\end{tabular} \\
Batch Norm. & 2N & (16,N/2,N/2) & \\
Dropout & 0 & (16,N/2,N/2) & 50\% \\
\hline
\begin{tabular}[l]{@{}l@{}}Convolutional\\ 32 filters (5,5)\end{tabular} & 12832 & (32,N/4,N/4) & \begin{tabular}[c]{@{}c@{}}2 pixels stride, \\ same padding, \\ Leaky ReLU act.\end{tabular} \\
Batch Norm. & N & (32,N/4,N/4) & \\
Dropout & 0 & (32,N/4,N/4) & 50\% \\
\hline
\begin{tabular}[l]{@{}l@{}}Convolutional\\ 64 filters (3,3)\end{tabular} & 18496 & (64,N/8,N/8) & \begin{tabular}[c]{@{}c@{}}2 pixels stride, \\ same padding,\\ Leaky ReLU act.\end{tabular} \\
Batch Norm. & N/2 & (64,N/8,N/8) & \\
Dropout & 0 & (64,N/8,N/8) & 50\% \\
\hline
Flatten & 0 & (N$^2$) & \\
Dense & (N$^2$+1)$\cdot$64 & (64) & \begin{tabular}[c]{@{}c@{}}64 units, \\ ReLU act.\end{tabular} \\
Dropout & 0 & (64) & 30\% \\
Dense & 2080 & (32) & \begin{tabular}[c]{@{}c@{}}32 units, \\ ReLU act.\end{tabular} \\
Dropout & 0 & (32) & 30\% \\
Dense & 33 & (1) & \begin{tabular}[c]{@{}c@{}}1 unit, \\ sigmoid act.\end{tabular} \\
\hline
\end{tabular}
\label{tab:CNN}
\tablefoot{
The columns are the name of the Keras layer (and the filters for the convolutional layers), the number of trainable parameters, output, and hyper-parameters for each layer. N represents the size of the input images which is 120, 96, 80, and 64 for $z$-bins 1, 2, 3, and 4, respectively.
}
\end{table}

To classify mergers and non-mergers, we used convolutional neural networks \citep[CNN;][]{lecun_gradient-based_1998}, which have multiple layers to effectively extract features from the input images. 
The lower layers consist of convolutional layers which convolve the data from the previous layer with a pre-defined number of filters, composed of sets of neurons with trainable weights and biases. 
The output is a feature map, which is passed to the next layer.
The higher layers are typically 1D fully-connected layers, where all the neurons are connected to all the neurons in the previous layer.
The CNN gives a score for each input image as the final output, which is then used for classification. 
To train the CNN, a large number of labelled images are needed to adjust the weights to match the given classification.
To develop, train, and test the CNNs we utilised the latest Keras framework for the TensorFlow platform \citep{chollet_keras_2023,abadi_tensorflow_2016}. 
For each redshift bin, we built a three-layer CNN with a common architecture to adapt it to input images with different sizes. 
Table \ref{tab:CNN} shows the adopted architectures and the associated number of parameters and properties. 
In each convolution layer, we used a stride of 2 and a Leaky Rectified Linear Unit \citep[Leaky ReLU;][]{maas_rectifier_2013} as an activation function. 
Similarly, in the fully connected layer, we made use of a ReLu activation function \citep{nair_rectified_2010}. 
The last CNN unit was provided with a sigmoid activation function to predict a score between 0 and 1. 
To prevent any over-fitting we introduced a dropout after each processing layer.
At each step during training, the dropout layer randomly sets input units to 0 with a rate specified by the user. 
To further prevent over-fitting, early stopping in the training phase was used as well. 
The hyper-parameters are reported in Table \ref{tab:CNN}, including filter numbers and sizes, dropout rates, and strides chosen based on a grid search.

\subsubsection{CNN training and performance}\label{sect:CNN-perf}

\begin{table*}[ht]
\caption{Performance of the CNNs on the simulated test sets.}
\centering
\small
\begin{tabular}{l|cccc|cccc}
          & \multicolumn{4}{c|}{IllustrisTNG}                              & \multicolumn{4}{c}{Horizon-AGN}               \\
\hline
Metric    & $z$-bin 1 & $z$-bin 2 & $z$-bin 3 & $z$-bin 4                 & $z$-bin 1 & $z$-bin 2 & $z$-bin 3 & $z$-bin 4 \\
 & $[0.1;0.31)$ & $[0.31;0.52)$ & $[0.52;0.76)$ & $[0.76;1.0]$ & $[0.1;0.31)$ & $[0.31;0.52)$ & $[0.52;0.76)$ & $[0.76;1.0]$ \\
\hline
Precision & 0.79      & 0.76      & 0.74      & \multicolumn{1}{c|}{0.73} & 0.84      & 0.72      & 0.69      & 0.62      \\
Recall    & 0.70      & 0.72      & 0.71      & \multicolumn{1}{c|}{0.63} & 0.62      & 0.66      & 0.67      & 0.69      \\
F1-score  & 0.74      & 0.74      & 0.72      & \multicolumn{1}{c|}{0.68} & 0.71      & 0.69      & 0.68      & 0.65      \\
Accuracy  & 0.76      & 0.75      & 0.73      & \multicolumn{1}{c|}{0.70} & 0.75      & 0.70      & 0.68      & 0.64     \\
\hline
\end{tabular}
\label{tab:CNN_performance_sim}
\tablefoot{Precision, recall and F1-score refer only to the merger class.
}
\end{table*}

For each $z$-bin, we constructed two CNNs with the above-described architecture, trained on mock images from IllustrisTNG and Horizon-AGN respectively. 
Both simulation samples were split into a 90\% training set and a 10\% test set. 
Galaxies from the same merger tree were put in only one of the two sets, which prevented the networks from learning any of the galaxies in the test set. 
Henceforth, we define the CNN trained on the IllustrisTNG sample as the TNG-CNN and the CNN trained on the Horizon-AGN sample as the Horizon-CNN. 
We trained each network using the corresponding simulation and $z$-bin training sample until early stopping was reached. 
Each training sample was balanced to have an equal number of mergers and non-mergers. 
To evaluate the performance of the CNNs, we used precision, recall, and $F_1$-score.
Precision is the ratio of correctly identified mergers over the total number of classified mergers.
Recall is the fraction of correctly identified mergers out of the total number of true mergers. 
$F_1$-score is the harmonic mean of precision and recall. 
We also computed the overall accuracy, which is the ratio of correctly classified objects, of both classes.
Below we detail the performance of the CNNs on the simulation test sets and a third test set based on visual inspection of real observations.


The IllustrisTNG test sets consisted of 13\,382, 13\,302, 14\,082, and 15\,400 images, balanced between mergers and non-mergers, in $z$-bins 1, 2, 3, and 4, respectively. 
The Horizon-AGN test sets contained 8\,764, 6\,984, 8\,196, and 8\,328 images in $z$-bins 1, 2, 3, and 4, respectively, equally split into mergers and non-mergers. 
We classified the IllustrisTNG and Horizon-AGN test set images using the CNNs trained on the respective training sets.
Images with predicted scores $\leq0.5$ ($>0.5$) are classified as mergers (non-mergers). 
The results are presented in Table \ref{tab:CNN_performance_sim}.
Generally, both CNNs perform worse towards higher $z$. For example, the precision of the TNG-CNN (Horizon-CNN) decreases from 79\% to 73\% (from 84\% to 62\%). 
However, the recall is roughly constant at $\sim$60-70\% for both CNNs.

\begin{figure*}[ht]
    \centering
    \includegraphics[width=.95\textwidth]{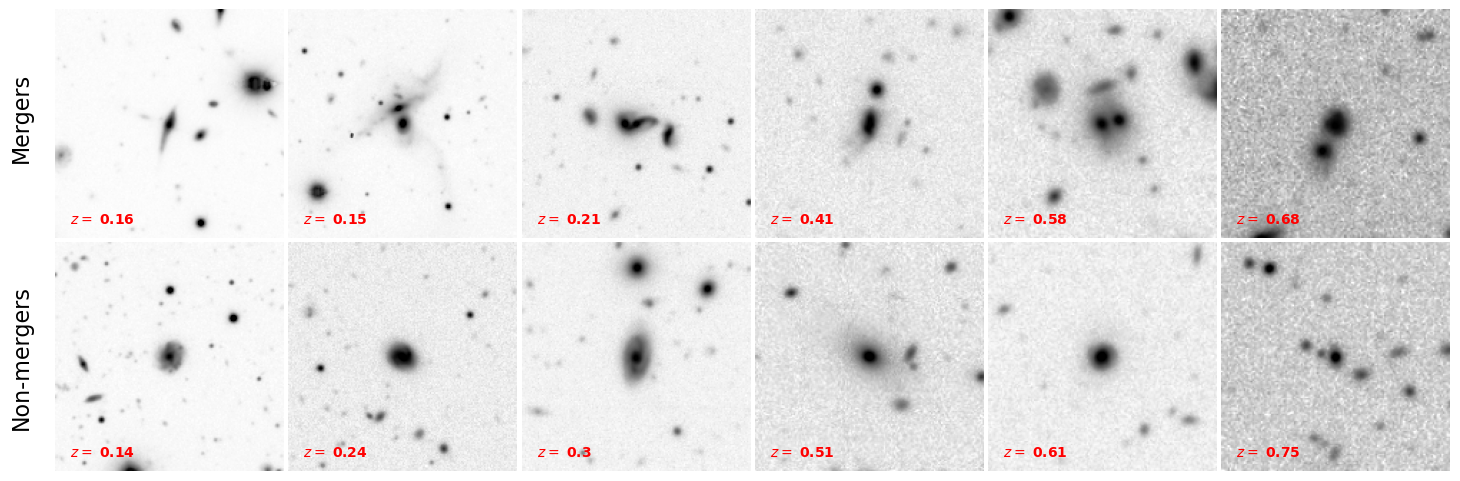}
    \caption{Examples of visually classified mergers \citep{goulding_galaxy_2018} and non-mergers (this work) at different redshifts. The HSC-SSP DR3 $i$-band cutouts have an approximate physical size of 160 kpc and have been resized to $160\times160$ pixels. An arcsinh inverted grey scale is used.}
    \label{fig:vis_examples}
\end{figure*}

\begin{figure}[ht]
  \centering
  \includegraphics[width=0.4\textwidth]{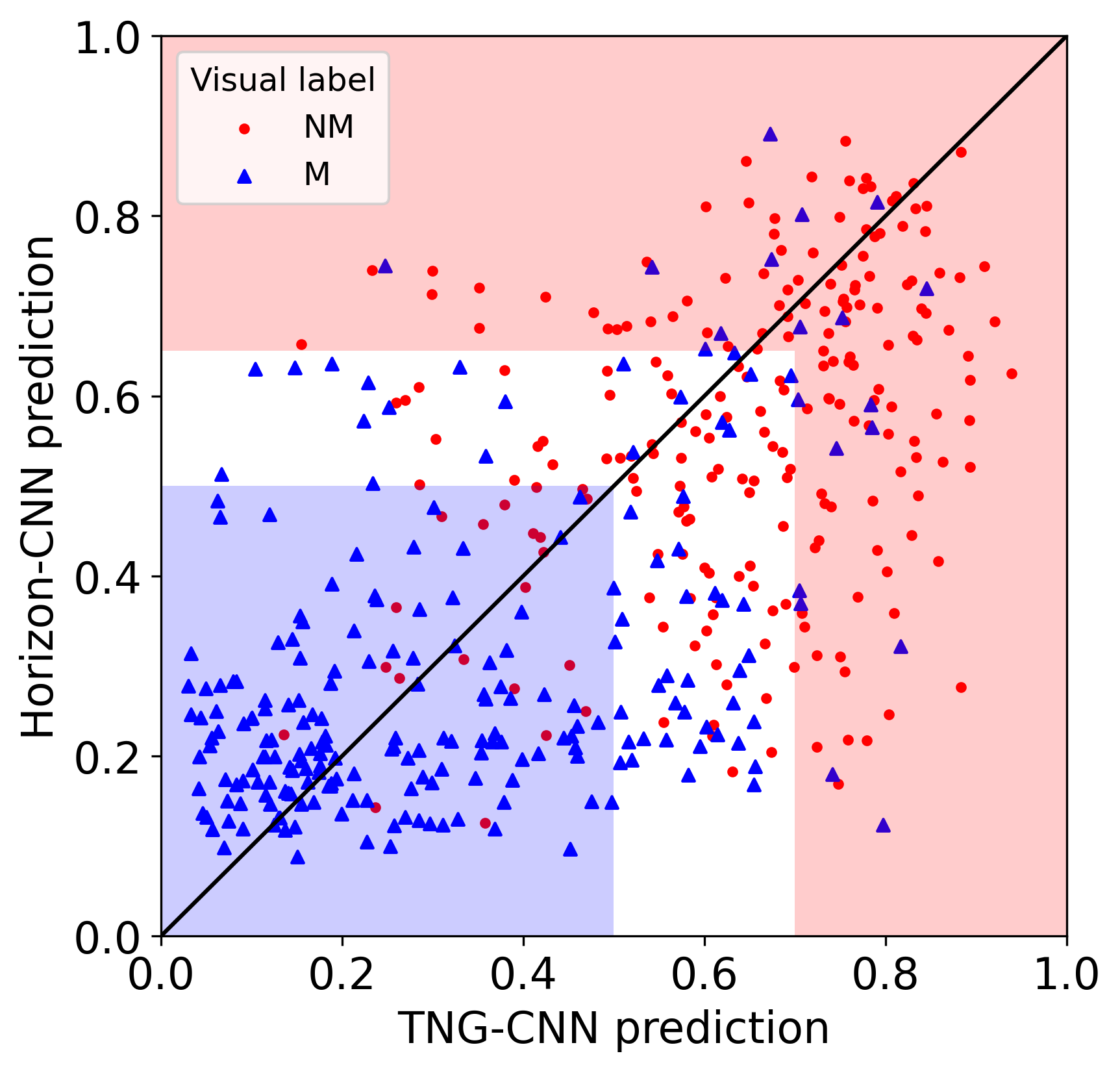}
  \caption{Comparison of the predicted scores from the TNG-CNN and Horizon-CNN for the HSC test set in $z$-bin 1. Visual labels are given in the legend. NM denotes a non-merger (red dots) and M denotes a merger (blue triangles). The red and blue shaded areas represent the combined model's non-merger and merger class definitions with two thresholds. }
  \label{fig:model_comp}
\end{figure}

\begin{figure*}[ht]
  \centering
  \includegraphics[width=0.25\textwidth]{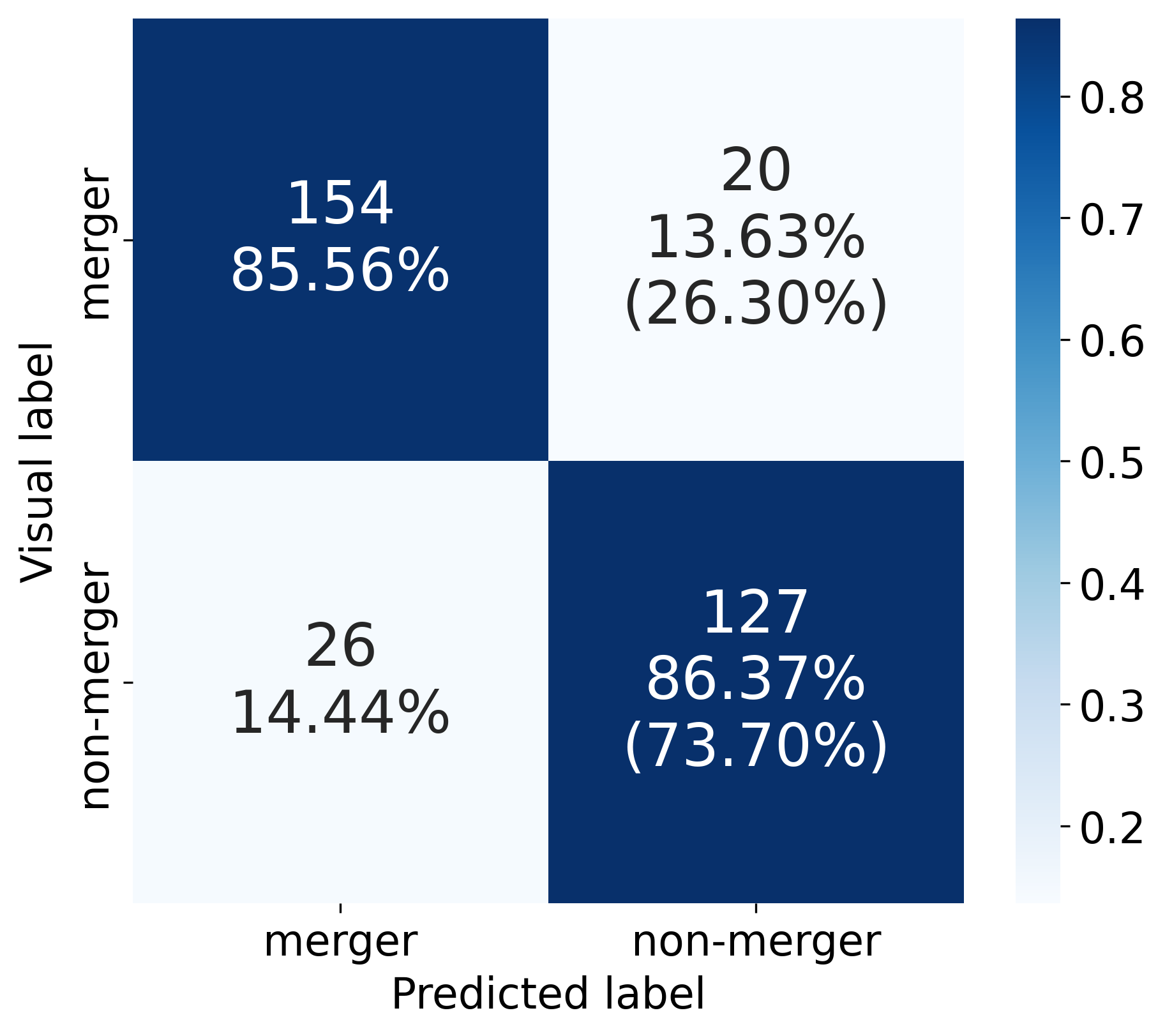}
  \includegraphics[width=0.25\textwidth]{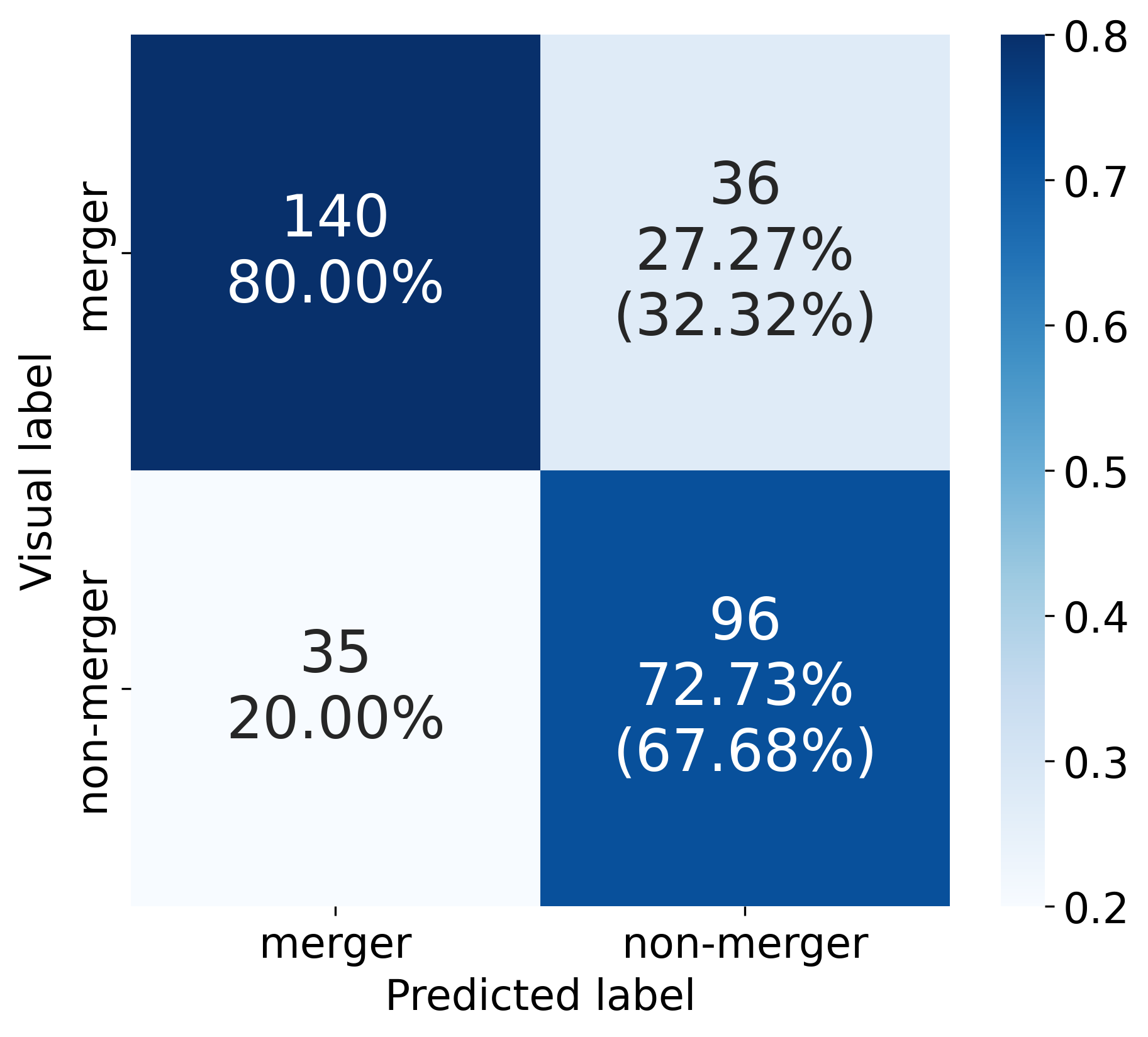}
  \includegraphics[width=0.25\textwidth]{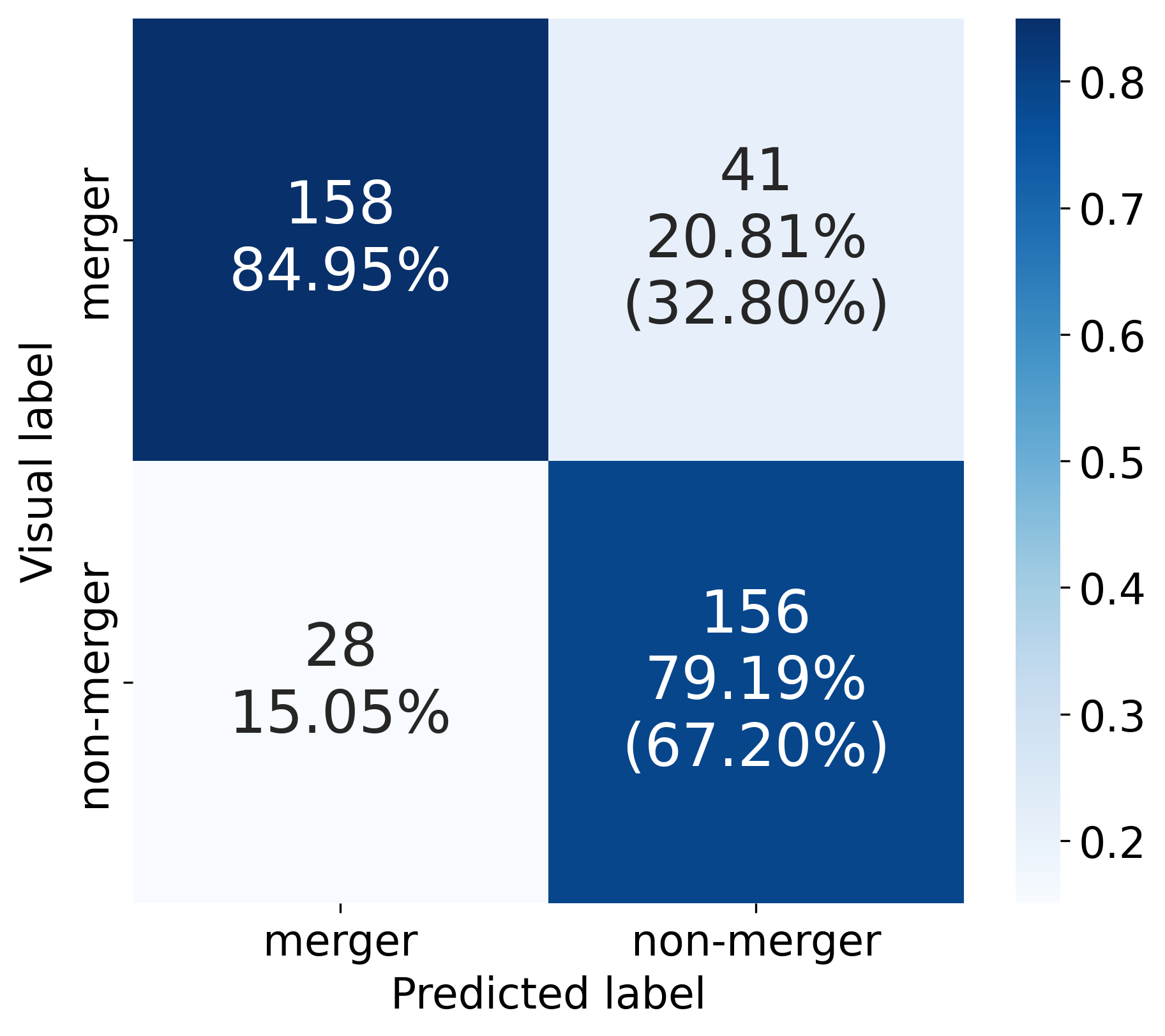}
  \caption{Confusion matrices for the combined model classification scheme using two thresholds, averaged over the five HSC test sets (left: $z$-bin 1; middle: $z$-bin 2; right: $z$-bin 3). The $x$ axis corresponds to the predicted class and the $y$ axis represents the visual label. In each cell, we also report the number of galaxies with a given visual label divided by the number of galaxies with a given predicted label. 
  For the predicted non-mergers, we report between brackets the fractions obtained using only one threshold to show the improvement obtained by adding a second threshold.}
  \label{fig:CM_comb_model}
\end{figure*}

\begin{table*}[ht]
\caption{Performance of the CNNs averaged over the five test sets containing visually classified galaxies from the HSC-SSP survey.}
  \centering
  \begin{tabular}{l|ccc|ccc}
  & \multicolumn{3}{c|}{Precision} & \multicolumn{3}{c}{Recall} \\
  \hline
  CNN model & z-bin 1 & z-bin 2 & z-bin 3 & z-bin 1 & z-bin 2 & z-bin 3 \\
 & $[0.1;0.31)$ & $[0.31;0.52)$ & $[0.52;0.76)$ & $[0.1;0.31)$ & $[0.31;0.52)$ & $[0.52;0.76)$ \\
 \hline
  TNG-CNN & $0.78\pm0.02$ & $0.67\pm0.02$ & \multicolumn{1}{c|}{$0.61\pm0.02$} & $0.74\pm 0.01$ & $0.67\pm0.02$& $0.81\pm0.01$ \\
  Horizon-CNN & $0.71\pm0.01$ & $0.76\pm0.02$ & \multicolumn{1}{c|}{$0.82\pm0.01$} & $0.84\pm0.01$ & $0.76\pm0.01$ & $0.63\pm0.01$ \\
  Comb-CNN & $0.86\pm0.02$ & $0.80\pm0.03$ & \multicolumn{1}{c|}{$0.85\pm0.02$} & $0.68\pm0.01$ & $0.58\pm0.02$ & $0.56\pm0.01$ \\
  \textit{Comb-CNN-2thr.} & $\mathit{0.86\pm0.01}$ & $\mathit{0.80\pm0.03}$ & \multicolumn{1}{c|}{$\mathit{0.85\pm0.02}$} & $\mathit{0.88\pm 0.01}$ & $\mathit{0.81\pm0.03}$ & $\mathit{0.80\pm0.03}$ \\
  \hline
  & \multicolumn{3}{c|}{F1-score} & \multicolumn{3}{c}{Accuracy} \\
  \hline
  TNG-CNN & $0.76\pm0.01$ & $0.67\pm0.02$ & \multicolumn{1}{c|}{$0.69\pm0.01$} & $0.76\pm 0.01$ & $0.67\pm0.02$ & $0.64\pm0.02$ \\
  Horizon-CNN & $0.77\pm0.01$ & $0.76\pm0.01$ & \multicolumn{1}{c|}{$0.71\pm0.01$} & $0.75\pm0.01$ & $0.76\pm0.02$ & $0.75\pm 0.01$ \\
  Comb-CNN & $0.76\pm0.01$ & $0.67\pm0.02$ & \multicolumn{1}{c|}{$0.68\pm0.01$} & $0.79\pm0.01$ & $0.72\pm0.02$ & $0.73\pm0.01$ \\
  \textit{Comb-CNN-2thr.} & $\mathit{0.87\pm0.01}$ & $\mathit{0.80\pm0.02}$ & \multicolumn{1}{c|}{$\mathit{0.82\pm0.03}$} & $\mathit{0.86\pm 0.01}$ & $\mathit{0.77\pm0.02}$ & $\mathit{0.82\pm 0.03}$ \\
  \hline
  \end{tabular}
  \label{tab:CNN_performance_1}

   \tablefoot{
The reported errors correspond to the standard deviation. Precision, recall, and F1-score refer only to the merger class.
   }
\end{table*}

We also assessed the performance of the CNNs on real HSC images.
For the mergers, we used galaxies visually classified as clear mergers by \citet{goulding_galaxy_2018}, which include 545 galaxies with $0.1\leq spec-z\leq0.9$ and $M_*>10^9$ M$_{\odot}$.
For the non-mergers, we randomly selected 2\,000 galaxies from our sample with spec$-z$ in the same redshift and mass ranges. 
We visually inspected them and found 1\,339 clear non-mergers. 
Examples of the visually classified galaxies are shown in Fig. \ref{fig:vis_examples}. 
To create a balanced test set, we randomly picked 545 non-mergers and joined them with the mergers from \citet{goulding_galaxy_2018}. 
To reduce selection bias, we repeated this step five times and created five different test sets. 
We show the performance of the CNNs, averaged over the five test sets, in Table \ref{tab:CNN_performance_1}.
Given the poor performance and the low number statistics in the last $z$-bin, we removed this bin from Table \ref{tab:CNN_performance_1} and subsequent analyses. 
Rather than applying the same threshold as for the simulation test sets, we chose a threshold $p$ to obtain the highest $F_1$-score. 
Consequently, we set  $p=0.5$ for both CNNs in $z$-bin 1, $p=0.4$ (0.5) for the TNG-CNN (Horizon-CNN) in $z$-bin 2, and $p=0.35$ (0.55) for the TNG-CNN (Horizon-CNN) in $z$-bin 3. 
In $z$-bin 1, the CNNs have comparable accuracy and $F_1-$score. The TNG-CNN is more precise and the Horizon-CNN has better recall. 
In $z$-bin 2, the TNG-CNN is much worse in every metric than the Horizon-CNN. 
In $z$-bin 3, the TNG-CNN achieves better recall, but worse precision and accuracy. 
We also made activation maps to help interpret the CNN classifications. Examples of both classes, with their corresponding activation maps, are presented in Appendix~\ref{sect:act_maps}.

\subsubsection{Combined CNN classifier}\label{sect:CNN-comb}

Studying the merger-AGN connection requires high-purity samples of mergers and non-mergers. 
To increase precision, we combined predictions from the two CNNs. 
In this combined model (Comb-CNN), galaxies for which both CNNs predicted a score $<p$ were labelled as mergers and galaxies for which at least one CNN predicted a score $>p$ were labelled as non-mergers.
The Comb-CNN shows a higher precision compared to the best single model. 
However, the recall drops significantly, which is expected as some of the galaxies classified as mergers by one CNN are now non-mergers.
To mitigate this, we then selected galaxies for which at least one CNN predicted a score $>$ another threshold $p'$ as non-mergers. 
This new threshold is set to $p'=0.7$ for the TNG-CNN in all $z$-bin, $p'=0.65$ for the Horizon-CNN in $z$-bin 1, and $p'=0.7$ for $z$-bins 2 and 3. 
Galaxies outside the two defined regions are unclassified, as illustrated in  
Fig. \ref{fig:model_comp} for $z$-bin 1.
The performances of the Comb-CNN using one and two thresholds are summarised in Table \ref{tab:CNN_performance_1} and the confusion matrices are compared with each other in Fig. \ref{fig:CM_comb_model}. 
When using the Comb-CNN with two thresholds, we were able to improve both precision and recall as well as reduce contamination in the non-merger class.

We applied this combined model with two thresholds to our stellar mass-limited samples. 
We prepared the HSC images in the same way as the mock images. 
A galaxy is labelled as a merger if both CNNs predict a score $<p$, and as a non-merger if one of the two predictions is $>p'$, according to the $z$-bin and the CNN. 
Unless otherwise stated, unclassified galaxies are excluded from our analysis. 
In $z$-bin 1, we classified 1\,272 mergers, 5\,808 non-mergers, and rejected 2\,079 galaxies. 
In $z$-bin 2, we classified 1\,073 mergers, 6\,514 non-mergers, and removed 3\,027 galaxies. 
In $z$-bin 3, there are 2\,060 mergers, 15\,841 non-mergers, and 5\,207 galaxies removed. 
Examples of visually inspected galaxies and our classifications are shown in Appendix \ref{sect:example_img}.

\section{Results}\label{sect:Results}

In this section, we first study the merger-AGN connection using a binary AGN and non-AGN classification for the three AGN types. 
Then, we analyse the merger-AGN relation using the continuous $f_{AGN}$ parameter and AGN bolometric luminosity. 

\subsection{Merger-AGN connection using a binary AGN classification}\label{sect:binary_agn_res}

In the first half of the analysis, we performed two experiments: (1) a comparison of AGN frequency in merging and non-merging galaxies, where a higher AGN frequency in mergers would suggest that mergers can indeed trigger AGN activity; and (2) a comparison of merger fractions in AGN and non-AGN host galaxies. 
If the majority of AGN host galaxies show merging signs, it could indicate that mergers are dominant in activating AGNs. 
It is crucial to construct proper control samples as AGN occurrence can depend on properties such as stellar mass and redshift. 
Following \cite{gao_mergers_2020}, for each merger in our sample, we identified a non-merger counterpart that met the following conditions:
\begin{equation}\label{z_eq}
  |z_{control} - z_{sample}|\leq 0.05
\end{equation}
and
\begin{equation}\label{mass_eq}
  |log \; M_{*, control} - log \; M_{*, sample}| \leq 0.1 \; dex.
\end{equation}
We only included mergers that have $\geq10$ non-merger counterparts and randomly picked ten of them to add to the control sample. 
For each AGN, we looked for possible non-AGN galaxies (defined as galaxies not classified as X-ray or SED AGNs, and with $W1-W2<0.5$) satisfying Eqs. \ref{z_eq} and \ref{mass_eq}.
We only kept AGNs that have $\geq10$ non-AGN counterparts and randomly chose 10 of them to create the non-AGN control sample. 
Some sources in the control groups may appear more than once. 

When calculating fractions of mergers, we considered only classified galaxies, namely $f_{merger} = N_{merger}/(N_{merger}+N_{non-merger})$. 
We refer to this as our main classification. 
In some cases, we also show the results for the worst-case scenario, in which all unclassified galaxies are considered as non-mergers.

\subsubsection{AGN frequency in mergers versus non-mergers}\label{sect:agn_freq}

\begin{figure}
  \centering
  \includegraphics[width=.49\textwidth]{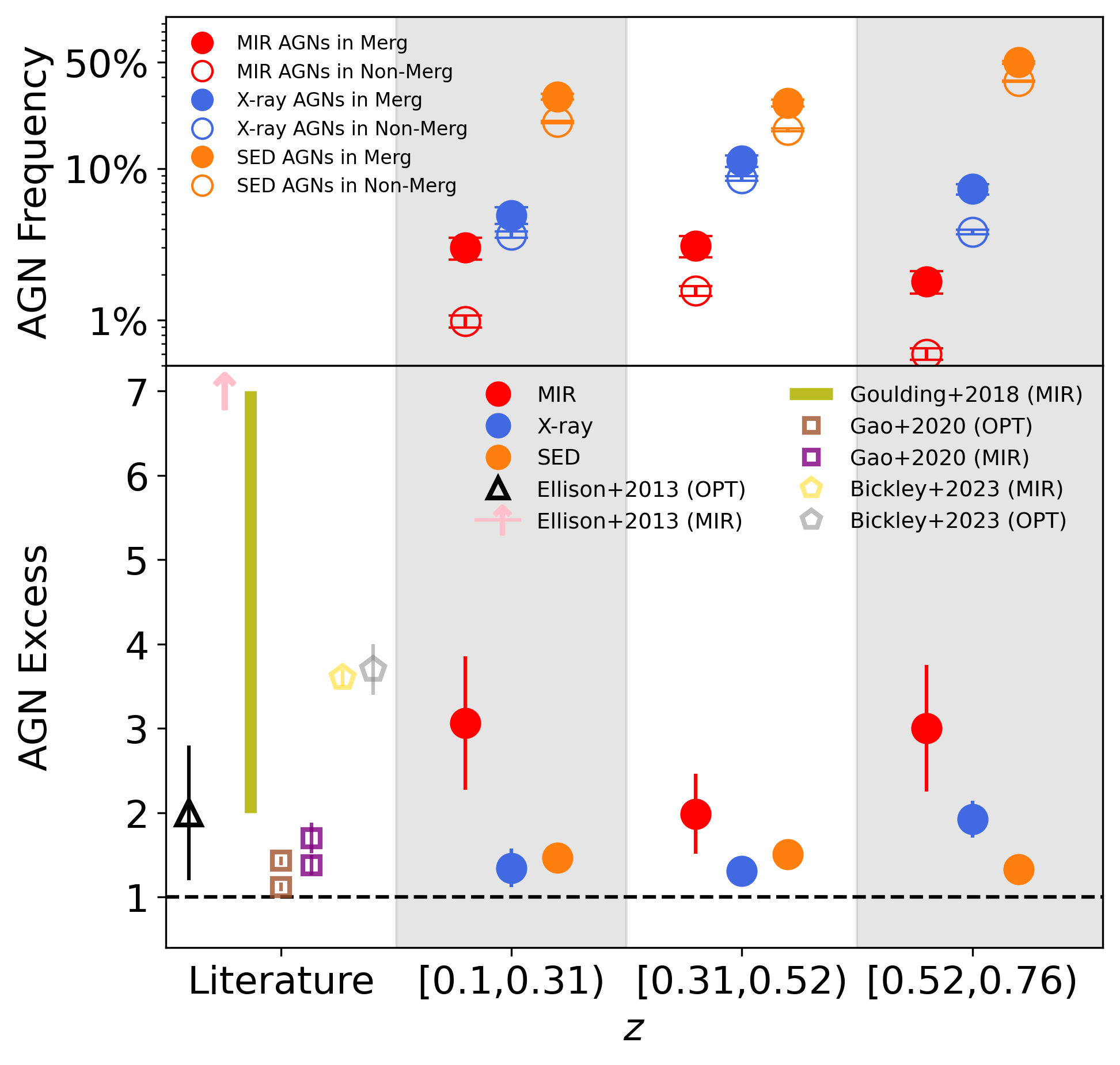}
  \caption{AGN frequency in mergers vs non-mergers. \textit{Top:} Frequency of AGNs (MIR AGN: red; X-ray AGN: blue; SED AGN: orange) in mergers (filled symbols) and non-mergers (empty symbols) per $z$-bin. 
  Errors are calculated using binomial statistics. 
  \textit{Bottom:} Ratio of AGN frequency in mergers relative to that in non-mergers (i.e. AGN excess). 
  The dashed line indicates no excess. Previous measurements are shown on the left. \citet{ellison_galaxy_2013} found a MIR AGN excess of 13.3, shown as an arrow for readability.
  }
  \label{fig:AGN_freq}
\end{figure}

\begin{table*}[h]
\caption{Frequencies of MIR-, X-ray-, and SED-selected AGNs in mergers and non-mergers, in different redshift bins. }
\small
  \centering
  \begin{tabular}{lcccccc}
  \hline
  & \multicolumn{2}{c}{MIR AGNs} & \multicolumn{2}{c}{X-ray AGNs} & \multicolumn{2}{c}{SED AGNs} \\
  $z$ bin & merger & non-merger (control) & merger & non-merger (control) & merger & non-merger (control) \\
  \hline
  $[0.1,0.31)$ & $3.0\pm 0.5 \%$ & $0.98 \pm 0.09 \%$ & $4.9\pm 0.6 \%$ & $3.7 \pm 0.2 \%$ & $29.8\pm 1.3\%$ & $20.3 \pm 0.4 \%$ \\
  & (38/1260) & (124/12600) & (62/1260) & (461/12600) & (375/1260) & (2555/12600) \\
  \hline
  $[0.31,0.52)$ & $3.1\pm0.5 \%$ & $1.56\pm0.12 \%$ & $11 \pm 1 \%$ & $8.6 \pm 0.3 \%$ & $27.0 \pm 1.4 \%$ & $17.9 \pm 0.4 \%$ \\
  & (33/1059) & (165/10590) & (119/1059) & (911/10590) & (286/1059) & (1901/10590) \\
  \hline
  $[0.52,0.76)$ & $1.8\pm0.3 \%$ & $0.60\pm0.05 \%$ & $7.3\pm0.6 \%$ & $3.81\pm0.13 \%$ & $50.1\pm 1.1 \%$ & $37.8\pm 0.3 \%$ \\
  & (37/2036) & (122/20360) & (149/2036) & (776/20360) & (1021/2036) & (7685/20360) \\
  \hline
  \end{tabular}
  \label{tab:agn_1}
  \tablefoot{
Errors are calculated through binomial statistics. In brackets, we provide the numbers of AGNs, for each type, over the total number of mergers and non-merger controls, in each $z$-bin. 
}
\end{table*}

The top panel in Fig. \ref{fig:AGN_freq} shows the frequencies of MIR, X-ray, and SED AGNs in mergers and non-merger controls, which are also reported in Table \ref{tab:agn_1}. 
The bottom panel of Fig. \ref{fig:AGN_freq} displays the excess, which is the ratio of the AGN frequency in mergers relative to non-mergers. 
Typically $\sim$3\% of the mergers and $\lesssim$1\% of the non-mergers contain MIR AGNs. 
This leads to a factor of 2-3 excess of MIR AGNs in mergers, which is the highest among the three AGN types. 
Around $5-10\%$ of the mergers host X-ray AGNs and a similar frequency is found in the non-mergers, except in $z$-bin 3 where mergers show an excess of $\sim2$. 
Much higher fractions of mergers and non-mergers contain SED AGNs, 
which is expected as they are weaker AGNs identified via SED fitting. 
However, we find a consistently low SED AGN excess at $\sim1.4$ across the three $z$-bins.

Our findings are reasonably consistent with most previous studies. 
As shown in Fig. \ref{fig:AGN_freq}, \citet{bickley_agns_2023} found a MIR AGN excess of $3-4$ in post-mergers at $z<0.3$. 
\citet{gao_mergers_2020} studied merging galaxies at $z<0.1$ from SDSS and at $z<0.6$ from GAMA. They determined a slightly lower MIR AGN excess at $\sim1.5-2$. 
\citet{goulding_galaxy_2018} found that mergers at $z<0.9$ are a factor of $\sim 2-7$ more likely to contain MIR AGNs than non-interacting galaxies. 
\citet{ellison_galaxy_2013} found a significantly higher MIR AGN excess ($\sim$13) in visually confirmed post-mergers at $z<0.3$. 
This could be because their sample consisted of visually conspicuous, highly disturbed post-mergers, which may host more dust-obscured AGNs \citep{bickley_agns_2023}.
In comparison, our sample also includes less disturbed systems and pre-mergers. 
On the other hand, the excess for the X-ray and SED AGNs we measure are comparable to each other and to previous measurements for optical AGNs \citep{ellison_galaxy_2013,gao_mergers_2020}. 
\citet{bickley_agns_2023} found a considerably higher optical AGN excess ($\sim 4$) in their post-mergers, similar to their MIR AGNs. 
There is also a small difference between our results for the X-ray AGNs and \citet{hewlett_redshift_2017}. 
The latter found no difference in mergers and undisturbed controls, while we see a slight excess in mergers. 
To conclude, we find a clear excess of AGN occurrence in mergers, for all three types of AGNs, which demonstrates mergers can trigger AGN activity, regardless of AGN type. 
Our findings also reveal a higher MIR AGN excess relative to the X-ray and SED AGNs. 
These results could imply the following: 
\begin{enumerate}
\renewcommand{\labelenumi}{\it \roman{enumi})}
\item Mergers are more likely to be connected to the triggering of MIR AGNs, possibly due to a physical connection (for instance, if the SMBH mass assembly in mergers preferentially occurs in a dust-obscured phase). 

\item The MIR AGNs generally have higher AGN fractions. Therefore, mergers could be more connected with triggering more luminous and/or dominant AGNs. 

\item The SED AGNs, with the lowest AGN fractions, could be triggered by secular processes. 
The same may apply to the X-ray AGNs to a lesser extent. 

\item Merger could redistribute gas and dust and increase dust obscuration. Thus detecting AGNs in the (soft) X-ray or optical would be more challenging.

\end{enumerate}

\subsubsection{Merger fraction in AGN versus non-AGN}

\begin{figure}
  \centering
  \includegraphics[width=.49\textwidth]{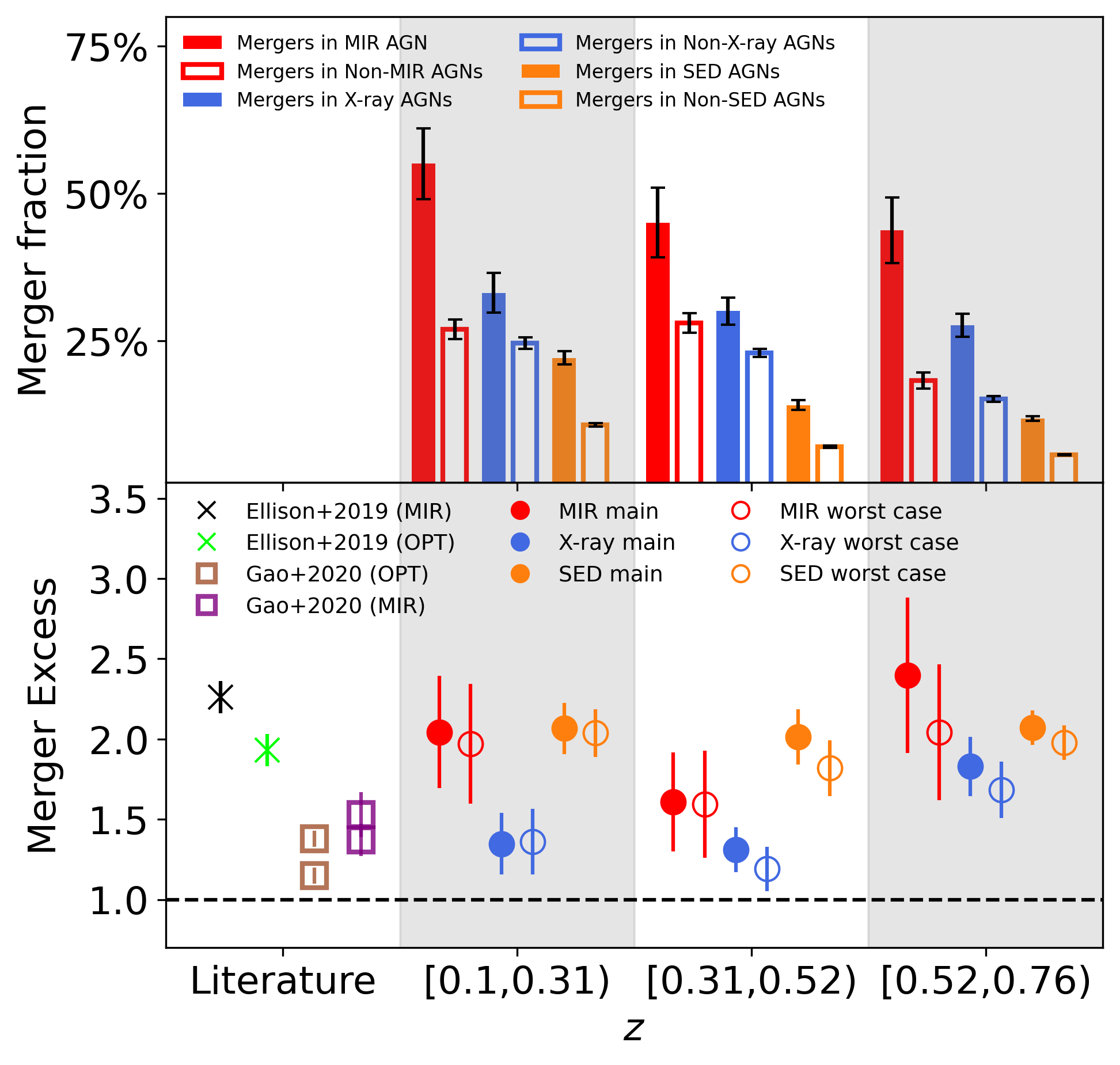}
  \caption{Merger fraction in AGNs vs non-AGNs. \textit{Top:} Merger fraction in MIR, X-ray, and SED AGNs and the respective non-AGN control samples per $z$-bin. 
  Filled bars indicate the merger fractions for MIR (red), X-ray AGNs (blue), and SED AGNs (orange), respectively.
  Empty bars correspond to the merger fractions in the non-AGN controls. 
  Errors are calculated through binomial statistics. 
  \textit{Bottom}: Ratio of the merger fraction in AGNs relative to non-AGN controls (i.e. the merger excess) for the main classification (filled symbols) and the worst-case scenario (empty symbols). 
  Dashed line indicates no merger excess.
  First column displays previous literature results.}
  
  \label{fig:merg_frac}
\end{figure}

\begin{table*}[ht]
\caption{Merger fractions in the MIR, X-ray, and SED AGNs and their respective non-AGN control sample, in different redshift bins. }
\small
  \centering
  \begin{tabular}{lcccccc}
  \hline
  & $f_{merger}$ in & $f_{merger}$ in & $f_{merger}$ in & $f_{merger}$ in & $f_{merger}$ in & $f_{merger}$ in \\
  $z$ & MIR AGNs & non-MIR AGNs & X-ray AGNs & non-X-ray AGNs & SED AGNs & non-SED AGNs \\
  & & (control) & & (control) & & (control) \\
  \hline
  \multicolumn{7}{c}{\bf Main} \\
  \hline
  $[0.1,0.31)$ & $55.1\pm 6.0 \%$ & $27.0 \pm 1.7 \%$ & $33.2\pm 3.4 \%$ & $24.6 \pm 1.0 \%$ & $22.1\pm 1.1 \%$ & $10.7 \pm 0.3 \%$ \\
  & (38/69) & (186/690) & (64/193) & (475/1930) & (324/1467) & (1571/14670) \\
  \hline
  $[0.31,0.52)$ & $45.1\pm 5.9 \%$ & $28.0\pm 1.7 \%$ & $30.1\pm 2.3 \%$ & $23.0\pm 0.7 \%$ & $14.1 \pm 0.8 \%$ & $7.0\pm 0.2 \%$ \\
  & (32/71) & (199/710) & (120/399) & (916/3990) & (245/1741) & (1223/17410) \\
  \hline
  $[0.52,0.76)$ & $43.7\pm 5.6\%$ & $18.2\pm 1.4 \%$ & $27.6\pm 1.9 \%$ & $15.1\pm 0.5\%$ & $11.8\pm 0.4 \%$ & $5.7\pm 0.1\%$ \\
  & (35/80) & (146/800) & (152/550) & (831/5500) & (965/8151) & (4677/81510) \\
  \hline
  \multicolumn{7}{c}{\bf Worst-case} \\
  \hline
  $[0.1,0.31)$ & $40.4\pm 5.1 \%$ & $20.5 \pm 1.3 \%$ & $22.9\pm 2.5 \%$ & $16.8 \pm 0.7 \%$ & $16.5\pm 0.8\%$ & $8.1\pm 0.2 \%$ \\
  & (38/94) & (193/940) & (65/284) & (478/2840) & (325/1968) & (1596/19680) \\
  \hline
  $[0.31,0.52)$ & $28.7\pm 4.2 \%$ & $18.0\pm 1.1 \%$ & $17.4\pm 1.4 \%$ & $14.6\pm 0.4 \%$ & $10.0\pm 0.6\%$ & $5.5\pm 0.2\%$ \\
  & (33/115) & (207/1150) & (120/690) & (1008/6900) & (245/2450) & (1350/24500) \\
  \hline
  $[0.52,0.76)$ & $26.4\pm3.7 \%$ & $12.9\pm0.9 \%$ & $18.6\pm1.4 \%$ & $11.0\pm0.3 \%$ & $9.1\pm 0.3\%$ & $4.6 \pm 0.1 \%$ \\
  & (38/144) & (186/1440) & (153/825) & (909/8250) & (971/10609) & (4925/106090) \\
  \hline
  \end{tabular}
  \label{tab:agn_2_worst}
  \tablefoot{
  The top half of the table corresponds to galaxies classified using the main classification scheme. The bottom half corresponds to the worst-case scenario, where all unclassified galaxies are labelled as non-mergers. 
  Errors are calculated through binomial statistics.
  In brackets are the number of mergers out of the total number of AGNs and non-AGN controls, per AGN type and $z$-bin.
  }
\end{table*}

Here we perform the reverse experiment by measuring merger fractions in AGN host galaxies and non-AGN control samples.
In Fig. \ref{fig:merg_frac}, we plot the merger fractions obtained using the main classification for identifying mergers and non-mergers in the AGN and non-AGN controls, which are also listed in the top half of Table \ref{tab:agn_2_worst}. Of the MIR AGNs, around half reside in mergers. A much lower fraction of mergers ($f_{merger}$), by $\sim20-25$\%, is observed in the MIR non-AGN controls.
The X-ray AGNs have considerably lower $f_{merger}$ ($\sim30$\%) and even lower $f_{merger}$ ($\sim15-25$\%) in the corresponding non-X-ray AGN controls.
The SED AGNs have the lowest $f_{merger}$ ($\sim10$-20\%) and the corresponding non-SED AGN control samples show roughly a factor of two lower $f_{merger}$.
We define merger excess as the merger fraction in AGNs divided by that in non-AGN controls, which is plotted in the bottom panel of Fig. \ref{fig:merg_frac}. 
The merger excess in the MIR AGNs varies from $\sim 1.6$ to $\sim2.4$ across the three $z$-bins. 
The merger excesses in the X-ray AGNs ($\sim$1.2 at $z<0.5$, and $\sim1.8$ at $z>0.5$) are lower than that in the MIR AGNs. 
The merger excess in the SED AGNs ($\sim2$) is similar to that in the MIR AGNs.

We now repeat the same experiment, using the worst-case classification scheme in which unclassified galaxies are regarded as non-mergers. 
The merger number counts and fractions are presented in the bottom half of Table \ref{tab:agn_2_worst}. 
As expected, the merger fraction decreases. 
Nevertheless, we derive similar merger excess values to those obtained with the main classification scheme, as shown in the bottom panel of Fig. \ref{fig:merg_frac}.
We also compare with previous measurements.
In $z$-bin 1, our results for the MIR AGNs agree well with \citet{ellison_definitive_2019} who analysed $z<0.25$ galaxies from the Canada France Imaging Survey. \citet{gao_mergers_2020} found a lower excess, similar to what we see in $z$-bin 2 for the MIR AGNs. 
At $z>0.5$, our merger excess in the X-ray AGNs is similar to that in optical AGNs by \citet{ellison_definitive_2019}. 
However, at $z<0.5$, our results are significantly lower but comparable to the values in \citet{gao_mergers_2020} for optical AGNs. 
Our findings for the SED AGNs are very close to excess values measured by \citet{ellison_definitive_2019} for optical AGNs. 

At face value, we might conclude that merger is the dominant mechanism for triggering the MIR AGNs, given half of them are in mergers. 
However, this interpretation may be oversimplified. 
We also detect mergers in the non-AGN controls, which could mean that mergers do not necessarily cause AGN activity. 
On the other hand, due to the large mismatch in the merger and AGN timescales, it is also possible that mergers do trigger AGNs but we happen to observe the AGN in an off-state in some systems. 
For instance, in $z$-bin 1, 55\% of the MIR AGNs reside in mergers, which could imply that mergers play a dominant role. 
However, 27\% of the non-MIR AGN controls also reside in mergers. 
Therefore, we could have two possible scenarios.
In the first case, only $55\%-27\%=28\%$ of the MIR AGNs are really triggered by mergers.
In the second scenario, $55\%+27\%=82\%$ of the MIR AGNs are triggered by mergers (but some of them happen to be switched off).
The real merger fraction could be somewhere between 28\% and 82\% if both scenarios are acting in the observed systems. 
Similar arguments could be applied to the X-ray and SED AGNs. 
For the X-ray AGNs, mergers may or may not be the dominant triggering process. 
But for the SED AGNs, it seems clear that merger is not the main mechanism even if we consider all the non-AGNs in mergers are actually AGNs. 
In summary, we cannot conclude on the extent of the role of mergers in triggering AGNs, but it is likely that mergers are more important (or even dominant) for the MIR AGNs.
On the other hand, for the X-ray and SED AGNs mergers are unlikely to be the primary trigger.

\subsection{merger-AGN connection using the continuous $f_{AGN}$ parameter}

In the second half of the analysis, we investigate the merger-AGN connection, using the continuous AGN fraction $f_{AGN}$ parameter derived from SED fitting in the rest-frame $3-30\mu$m.  
We also examine the connection using the BHARs and the equivalent bolometric luminosities.

\subsubsection{The $f_{AGN}$ parameter in mergers versus non-mergers}

\begin{figure*}[ht]
  \sidecaption
  \includegraphics[width=12cm]{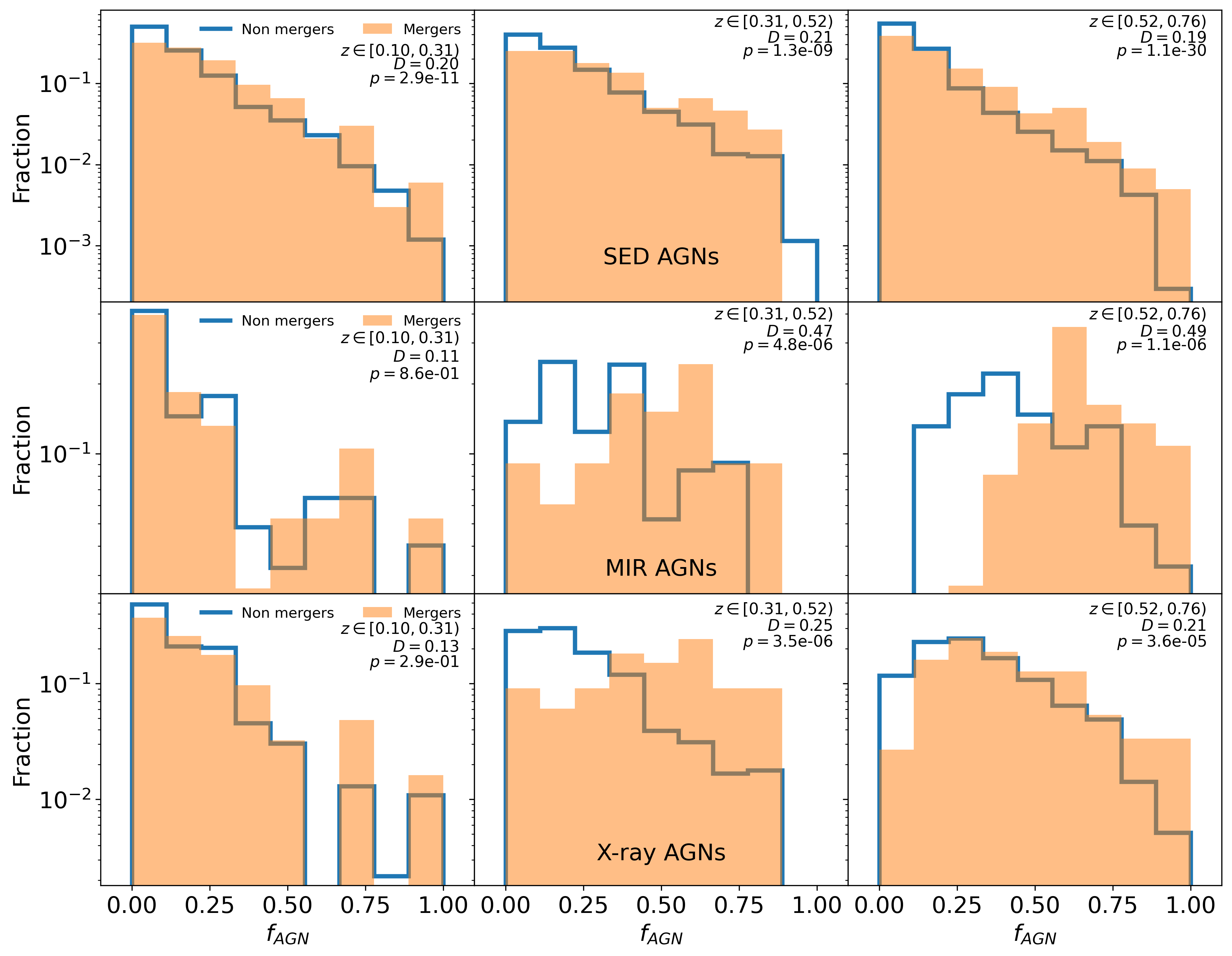}
  \caption{Normalised distributions of AGN fraction in mergers  (orange) and non-mergers (blue), as a function of redshift ($z$-bin 1: left; $z$-bin 2: middle; $z$-bin 3: right) and AGN type (SED AGN: top; MIR AGN: bottom). In each panel, we show the results of a KS test, where $D$ is the test result and $p$ is the corresponding $p-$value. For SED AGNs, the $f_{AGN}$ distribution in mergers is considerably different from that in non-mergers in all $z$-bins.
  MIR AGNs have different merger and non-merger distributions at $z\geq0.31$. 
  }
  \label{fig:fAGN_dist}
\end{figure*}

\begin{figure}[h]
    \centering
    \includegraphics[width=.45\textwidth]{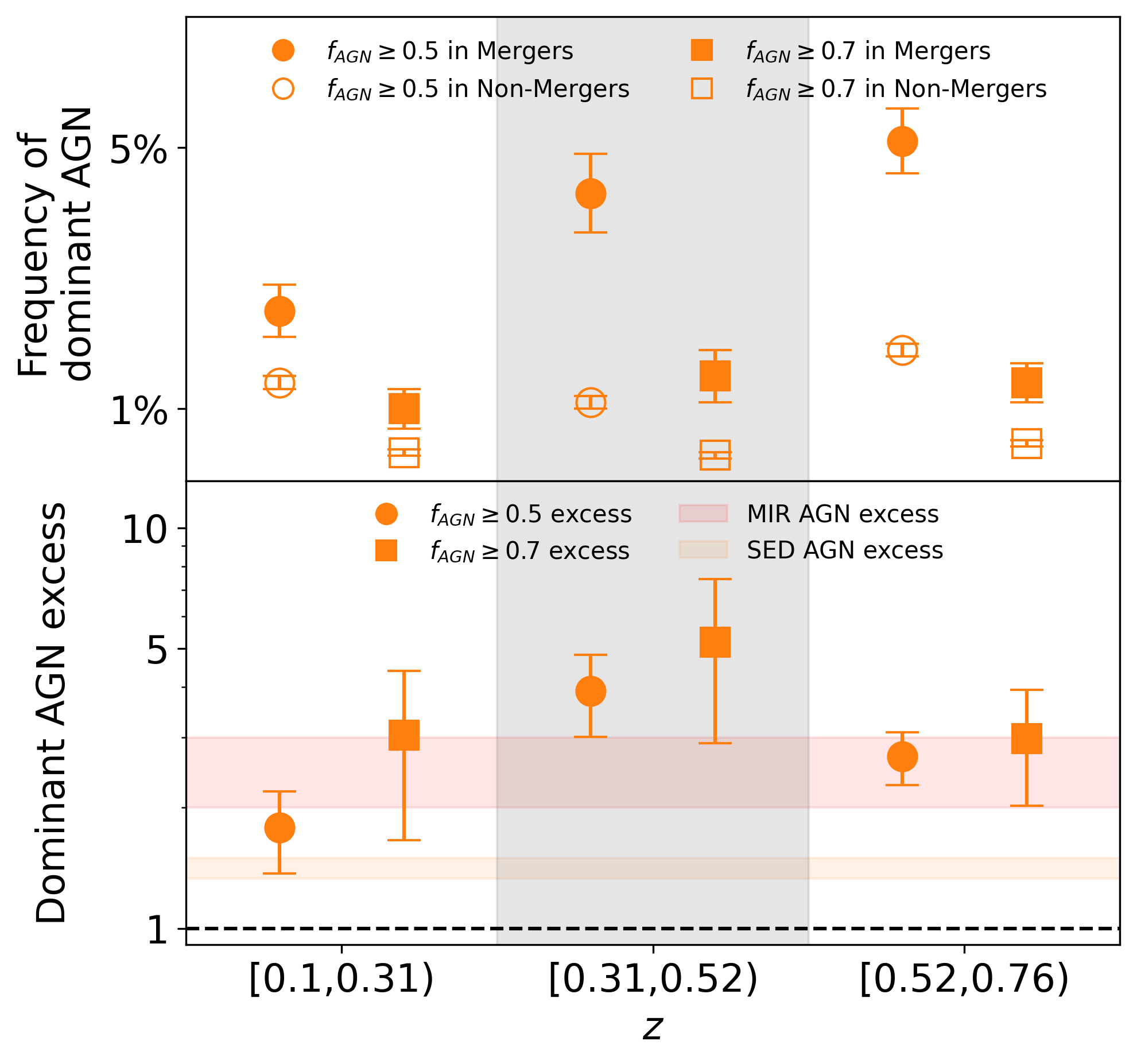}
    \caption{Dominant subset of the SED AGNs. \textit{Top:} Frequency of the SED AGNs with $f_{AGN}\geq0.5$ (circles) or $f_{AGN}\geq0.7$  (squares) in mergers (filled symbols) and non-mergers (empty symbols) as a function of redshift. Uncertainties are estimated using binomial statistics. 
    \textit{Bottom:} Excess of the dominant SED AGNs in mergers. 
    The red and orange bands display the AGN excess for the MIR and SED AGNs as a whole, respectively (Fig.~\ref{fig:AGN_freq}).
    }
    \label{fig:agn_dom}
\end{figure}

First, we compare the distribution of $f_{AGN}$ between mergers and non-mergers. 
Fig.~\ref{fig:fAGN_dist} displays the normalised distributions of $f_{AGN}$ in the merger and non-merger control samples. 
To assess the statistical significance of the differences between the two populations, we employ a two-sample Kolmogorov–Smirnov test \citep[KS test;][]{hodges_significance_1958} and report the results in each panel of Fig.~\ref{fig:fAGN_dist}.
It is clear that in all $z$-bins and for all AGN types, the $f_{AGN}$  distributions in mergers and non-merger controls are different, with a clear excess of high $f_{AGN}$ in mergers (particularly at the highest $f_{AGN}$ values). 
For the MIR AGNs, the excess of high $f_{AGN}$ values in mergers appears to be the strongest, except in $z$-bin 1 which is limited by low-number statistics.
These results again indicate that mergers are more relevant for more dominant AGNs and dust-obscured AGNs. 
We also compare the distributions of BHAR (or bolometric luminosity) between mergers and non-mergers. A similar picture emerges in the sense that mergers have a larger fraction of AGNs with high BHAR ($>10^{-2}$ M$_{\odot}/$yr) or equivalently more luminous AGNs ($L_{disc}\gtrsim10^{44}$erg/s) with respect to non-mergers. In addition, the excess of more powerful AGNs with higher BHARs is the strongest in the MIR AGNs.

To investigate further the link between mergers and the more dominant AGNs we perform an additional experiment by comparing the number of AGNs with $f_{AGN}\geq0.5$ or $\geq0.7$ in the merger sample and non-merger controls.
Here, we only consider the SED AGNs. The MIR and X-ray AGNs are excluded due to low-number statistics. 
Fig. \ref{fig:agn_dom} shows that $\gtrsim2\%$  ($\sim1\%$) of the mergers host an SED AGN with $f_{AGN}\geq0.5$ ($\geq 0.7$).
In comparison, only $\sim 1 \%$ ($\sim 0.4\%$) of the non-mergers host such a dominant AGN.
As a result, the SED AGN excess in mergers for $f_{AGN}\geq0.5$ is a factor of $\sim$2, 4, and 3 in $z$-bins 1, 2, and 3, respectively. When considering $f_{AGN}\geq0.7$, the excess factor in mergers increases to $\sim$3, 5, and 3 in $z$-bins 1, 2, and 3, respectively. 
These excess factors are comparable to the values measured for the MIR AGNs and significantly higher than those for the entire SED AGN sample (Section \ref{sect:agn_freq}).
As shown in Fig. \ref{fig:f_agn}, the SED AGNs are largely weak AGNs, while a considerable fraction of the MIR AGNs have $f_{AGN}\geq0.5$. 
Yet, when concentrating to the most powerful AGNs, the SED AGNs exhibit higher AGN excesses, comparable to the MIR AGNs.
These results hint at a scenario where mergers play a more important role in triggering more powerful AGNs, independent of selection.

\subsubsection{Merger fraction as a function of $f_{AGN}$ and BHAR}

\begin{figure}[ht]
  \centering
  \includegraphics[width=0.45\textwidth]{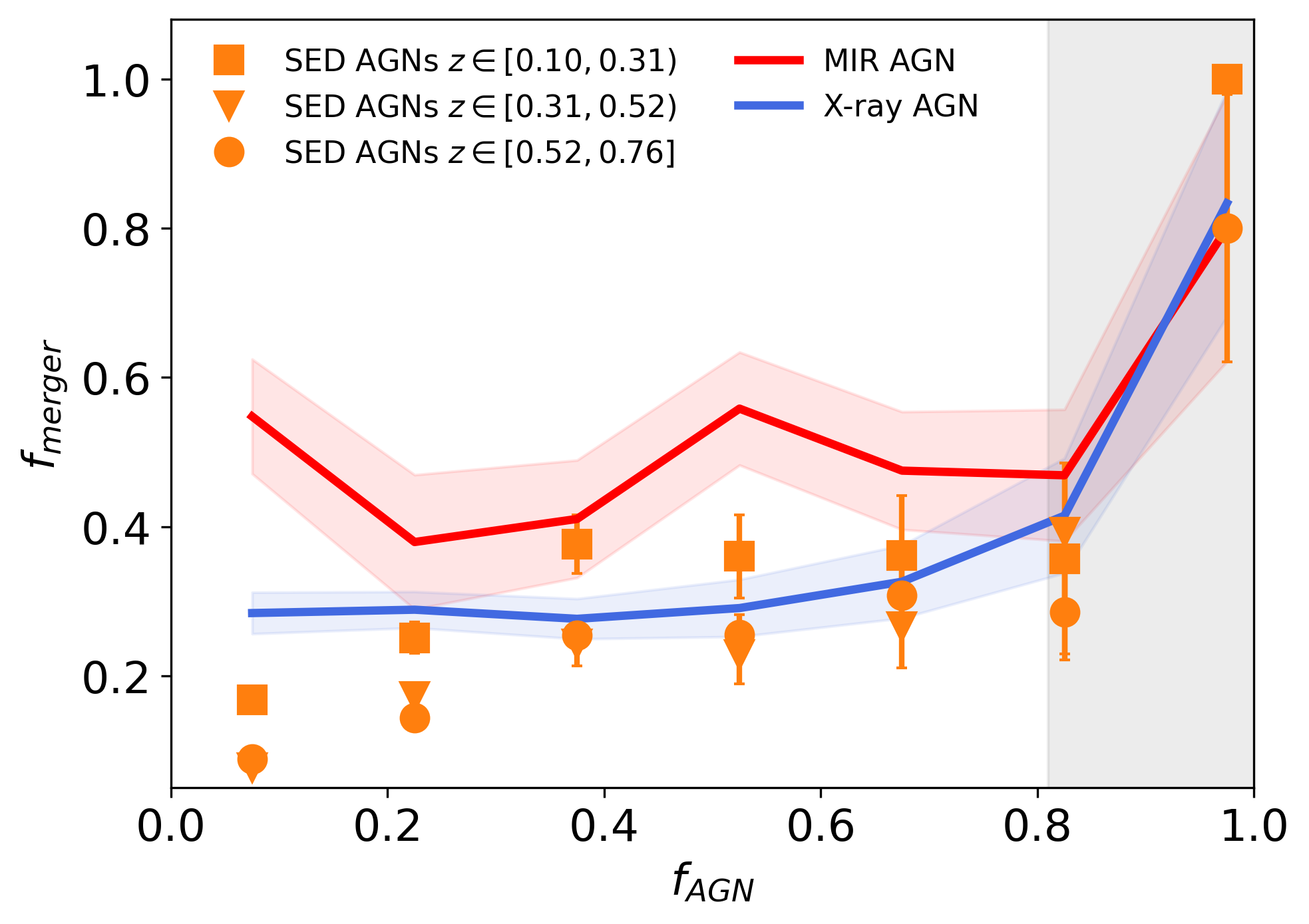}
      \includegraphics[width=.45\textwidth]{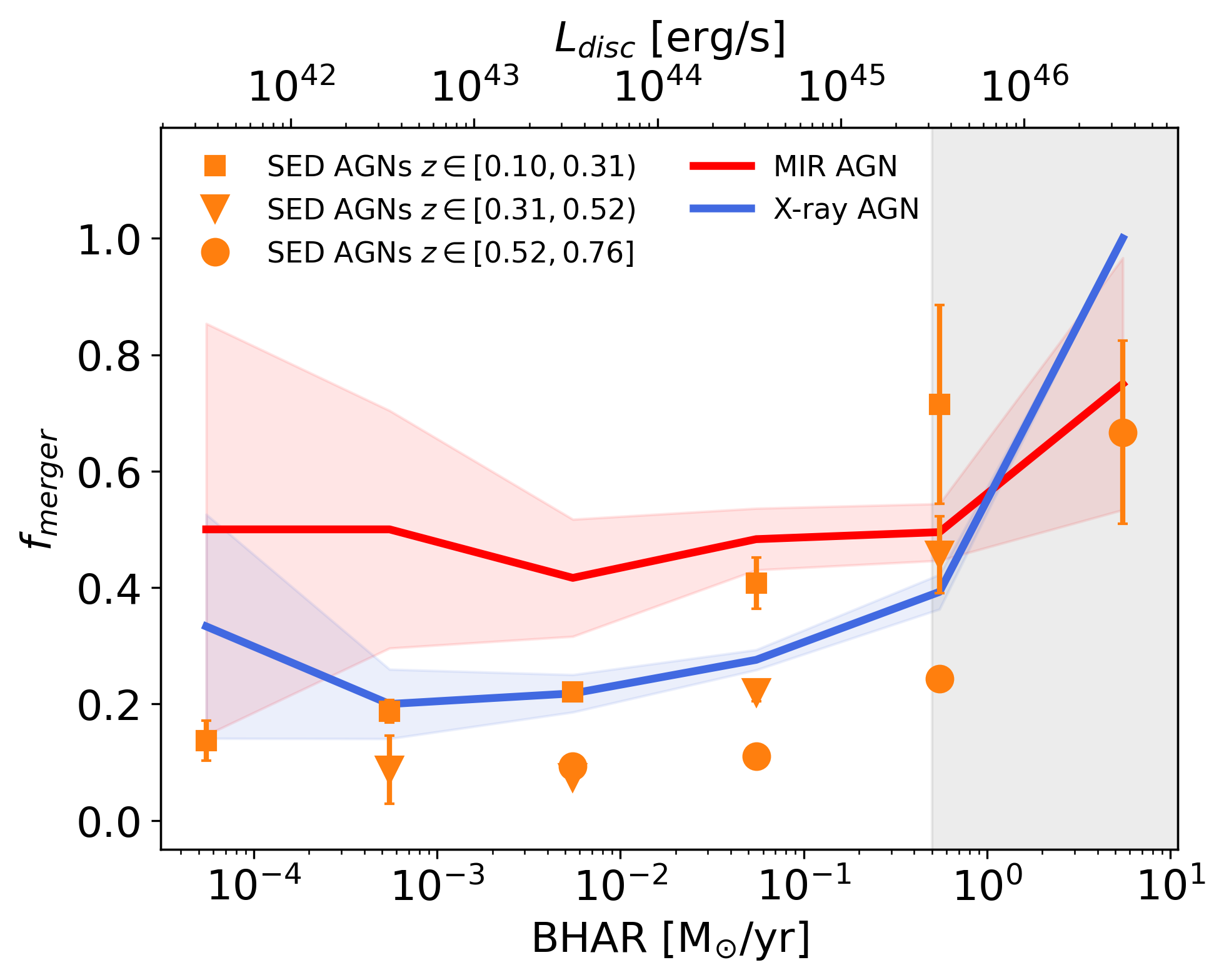}
  \caption{Merger fraction as a function of AGN fraction and BHAR. \textit{Top:} $f_{merger}$ as a function of $f_{AGN}$. The $f_{AGN}$ bin width is 0.15 and $f_{merger}$ is plotted at the centre of each bin. Errors are calculated through binomial statistics. 
  Qualitatively similar trends in $f_{merger}$ vs $f_{AGN}$ are seen for all three AGN types, namely a mildly increasing $f_{merger}$ at $f_{AGN} < 0.8$, followed by a rapid increase to $f_{merger}$ $\sim$80-100\% when the AGN is very dominant. 
  \textit{Bottom:} $f_{merger}$ as a function of BHAR and the equivalent AGN bolometric luminosity $L_{disc}$.}
    \label{fig:fmerg_bhar}
\end{figure}

Finally, we investigate how the merger fraction, $f_{merger}$, varies with AGN fraction, $f_{AGN}$.
We plot $f_{merger}$ as a function of $f_{AGN}$ in the top panel in Fig.~\ref{fig:fmerg_bhar}, which shows the SED AGNs in three $z$-bins and the entire samples of the MIR and X-ray AGNs (due to low number statistics). 
For the SED AGNs, $f_{merger}$ generally increases with $f_{AGN}$ in all $z$-bins, with a plateau around $f_{merger}\sim0.3$ at $f_{AGN}\sim0.4-0.8$. 
At $f_{AGN}>0.8$, a sharp increase in $f_{merger}$ can be seen in $z$-bin 1 and 3 (missing in $z$-bin 2 due to a lack of data). 
The MIR AGNs exhibit similar behaviour, with an almost flat relation around $f_{merger}\sim50\%$ for AGN fractions up to $f_{AGN}\simeq0.8$, followed by a sharp increase in $f_{merger}$ at $f_{AGN}>0.8$.
The X-ray AGNs also follow a similar pattern, with a mildly increasing (almost flat) $f_{merger}\sim30\%$ at $f_{AGN}<0.8$ and then a steep increase to $\sim80\%$ at the highest $f_{AGN}$ values. 

From this first attempt to characterise the relation between $f_{merger}$ and $f_{AGN}$, we find clear evidence for a connection between AGNs and mergers with two distinct regimes. 
For all three types of AGNs, the qualitative trend in $f_{merger}$ vs $f_{AGN}$ is the same. 
At $f_{AGN} > 80\%$, $f_{merger}$ rises quickly to $\sim80-100$\%. 
In addition, this steep rising trend in the three AGN types seems to overlap with each other. 
This could indicate that mergers are the dominant or even the only triggering mechanism for galaxies with a very dominant AGN, regardless of AGN type.
At $f_{AGN} < 0.8$, there appears to be a mildly increasing or even flat trend in $f_{merger}$ vs $f_{AGN}$.
This could be due to other physical processes, such as disk instability, galactic bars, and stellar winds \citep{bournaud_black_2011,garland_most_2023,ciotti_radiative_2007}, playing a more important role in fuelling SMBHs. It is also possible that the triggering of AGNs by mergers is somewhat stochastic when the AGN is relatively weak compared to the host galaxy. 
Moreover, in the regime of less dominant AGNs, $f_{merger}$ does depend on AGN type, with $f_{merger}\sim50$\% in the MIR AGN, $\sim30$\% in the X-ray AGNs, and slightly lower in the SED AGNs at around $20-30\%$. 
This could imply that when the AGN is not very dominant or relatively weak, mergers preferentially trigger dust-obscured AGNs, consistent with our findings in Section~\ref{sect:binary_agn_res}. 
We also investigate the correlation between $f_{merger}$ and AGN bolometric luminosity or BHAR. 
We split the AGNs into bins of bolometric power and calculate $f_{merger}$ in each bin according to our main classification scheme. 
We present the results in the bottom panel in Fig. \ref{fig:fmerg_bhar}.
Overall, we see a similar picture with $f_{merger}$ vs $f_{AGN}$. 
When the AGN bolometric luminosity or BHAR is very high ($L_{disc}\gtrsim 3\times 10^{45}$ erg/s, BHAR $\gtrsim$ 0.5 $M_{\odot}$/yr), $f_{merger}$ can increase very rapidly, approaching $70-100\%$.
When the AGN bolometric luminosity or BHAR is not that extreme, $f_{merger}$ can vary between $\sim$20 - 50\%, depending on the AGN type.

Some previous studies also found evidence that the most luminous AGNs preferentially reside in mergers. 
For example, \citet{gao_mergers_2020} found an increase in $f_{merger}$ as AGN luminosity increases, with $f_{merger}\geq 50\%$ for the most luminous MIR AGNs ($L_{6\mu m}>10^{44}$ erg/s or $L_{AGN}\gtrsim10^{44}$ erg/s). 
For the optical AGNs, \citet{ellison_definitive_2019} and \citet{gao_mergers_2020} observed an enhanced $f_{merger}$ ($\sim40-60\%$) at $L_{[OIII]}\gtrsim 10^{42.5}$ erg/s (equivalent to $L_{AGN}\gtrsim 10^{44}$ erg/s). 
Likewise, \citet{pierce_agn_2022} showed that the proportion of disturbed galaxies is $>60\%$ at $L_{[OIII]}>10^{42}$ erg/s.
\citet{goulding_galaxy_2018} found that AGNs with $L_{AGN}\gtrsim10^{45}$ erg/s mostly reside in mergers. 
Similarly, \citet{urrutia_evidence_2008} and \citet{glikman_major_2015} found $f_{merger}>80\%$ in luminous dust-reddened quasars ($L_{AGN}>10^{46}$ erg/s), consistent with what we observe for the most luminous AGNs.
\citet{treister_major_2012} observed a lower $f_{merger}$ for less powerful AGNs ($f_{merger}\lesssim20\%$ at $L_{AGN}<10^{44}$ erg/s), and a rapid increase to $f_{merger}\sim90\%$ at $L_{AGN}\simeq10^{46}$ erg/s. 
These findings reinforce the scenario we propose in which mergers are strongly linked to the triggering of dust-obscured AGNs and the most dominant/powerful AGNs (or even the only viable mechanism in the latter case), while less rapidly accreting AGNs may be mainly fuelled by secular processes or stochastically by mergers. 

Other studies found less convincing evidence for a dependence of $f_{merger}$ on AGN luminosity.
\citet{villforth_host_2017} analysed 20 luminous ($L_{AGN}\geq10^{45}$ erg/s) X-ray AGNs at $z=0.5-0.7$ and found no signs of enhanced $f_{merger}$. 
However, after taking into account mild disturbances, they estimated an upper limit for $f_{merger}$ of 38\%, comparable to what we find for the X-ray AGNs with $10^{45}\leq L_{disc} \leq 10^{46}$ erg/s. 
\citet{hewlett_redshift_2017} studied $\sim 100$ X-ray AGNs, spanning a wider redshift range ($0.5\leq z \leq 2.2$) and luminosity range ($10^{43}\lesssim L_{X} \lesssim 10^{45}$ erg/s), finding no evidence that mergers play a more dominant role at higher luminosities. 
Nevertheless, the fraction of galaxies classified as clear mergers and disturbed by \citet{hewlett_redshift_2017} is $\sim 20-30\%$,  in agreement with what we find. 
These apparently contradictory results could be at least partly explained by the two-regime behaviour shown in Fig. \ref{fig:fmerg_bhar}, namely mergers could be a secondary trigger for the X-ray AGNs up to $L_{disc}\simeq 5\times10^{45}$ erg/s, and only become the principal triggering mechanism for the most dominant/luminous AGNs. 


\subsubsection{Caveats on the $f_{merger}-f_{AGN}$ relation}

\begin{figure}[h]
    \centering
    \includegraphics[width=0.4\textwidth]{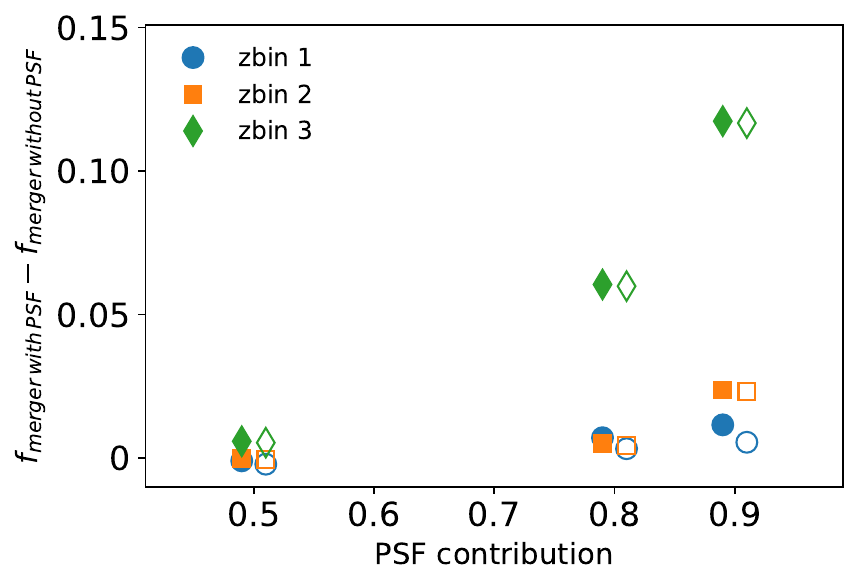}
    \caption{
    Difference between the merger fraction measured in the test sets with and without injected PSF ($f_{merger\,with\,PSF} - f_{merger\,without\,PSF}$) as a function of the contribution of the PSF. 
    The filled (empty) symbols refer to the test sets with an intrinsic merger fraction of 5\% (10\%). 
    Symbols are shifted slightly along the x-axis to prevent overlapping with each other. 
    The main difference is observed in $z-$bin 3 at PSF contribution levels of 80\% and 90\%.}
    \label{fig:fmerg_diff}
\end{figure}

Bright point sources may affect our CNN classifiers, thus influencing the measured merger fraction. 
For example, optically unobscured luminous quasars can outshine the host galaxy, potentially hampering merger detection. 
In Appendix~\ref{sect:quasars}, we investigate the impact of the SDSS quasars as they represent the brightest optical AGNs and so are most likely to affect our ability to detect merging features in the host galaxy. We find that our results on the $f_{merger}$ - $f_{AGN}$ relation are not significantly affected.

We also conducted a second experiment to assess the potential impact by injecting point sources modelled on the HSC PSF into our simulation test sets. 
The luminosity of the point sources is set to a certain fraction of the total light within a circular aperture with a radius equal to half of the PSF full-width half maximum (FWHM, $\sim0.6\arcsec$).
Hereafter, we use the terminology 'PSF contribution' to indicate this fraction, $f_{PSF} = L_{PSF,\, aper}/(L_{PSF,\, aper}+L_{galaxy,\, aper})$.
We set the injected $f_{PSF}$ to 50\%, 80\%, and 90\%.
Examples of simulated galaxies with injected PSFs are shown in Appendix~\ref{appendix:psf_images}.
We applied our merger identification to these test sets to examine changes in $f_{merger}$ as a function of the injected $f_{PSF}$, as shown in Fig.~\ref{fig:fmerg_diff}. 
The intrinsic merger fractions in the test sets are set to 5\% and 10\%, covering the range generally reported in the literature in the redshift and stellar mass range studied in this work \citep[e.g.][]{duncan_observational_2019,ferreira_galaxy_2020,whitney_galaxy_2021,margalef-bentabol_galaxy_2024}. 
In the first two $z-$bins, the change in $f_{merger}$ is very small ($\lesssim0.02$) after adding point sources. 
In $z-$bin 3, $f_{merger}$ is more or less the same as for the images without injected PSFs at $f_{PSF}=50\%$. 
At $f_{PSF}=80\%$ (90\%), $f_{merger}$ increases by $\sim 0.05$ ($\sim 0.1$).
Therefore, we can conclude that the overall merger classification is unaffected by point source contamination at $z<0.52$ at levels up to 90\%, and in the last $z$-bin at $f_{PSF}\leq50\%$. 
However, we do observe a $\sim 5-10\%$ increase in $f_{merger}$ in the last $z-$bin at $f_{PSF}\geq80\%$.

\begin{figure}
    \centering
    \includegraphics[width=.41\textwidth]{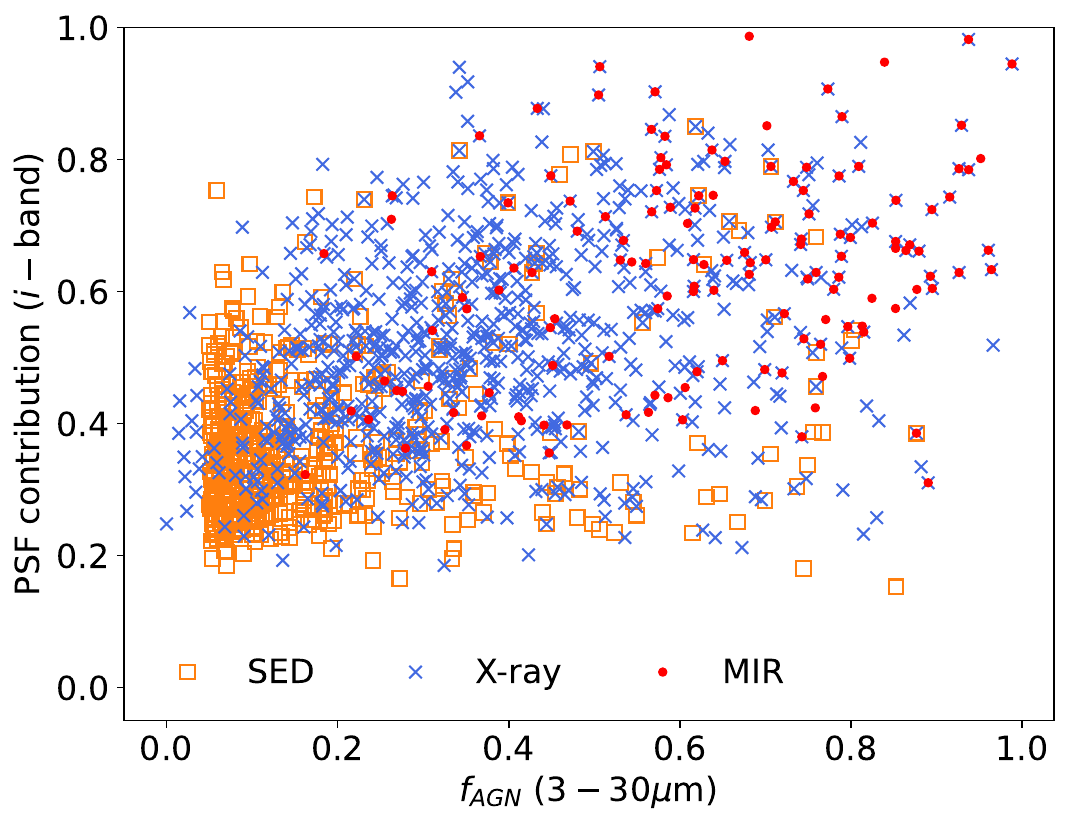}
    \includegraphics[width=.41\textwidth]{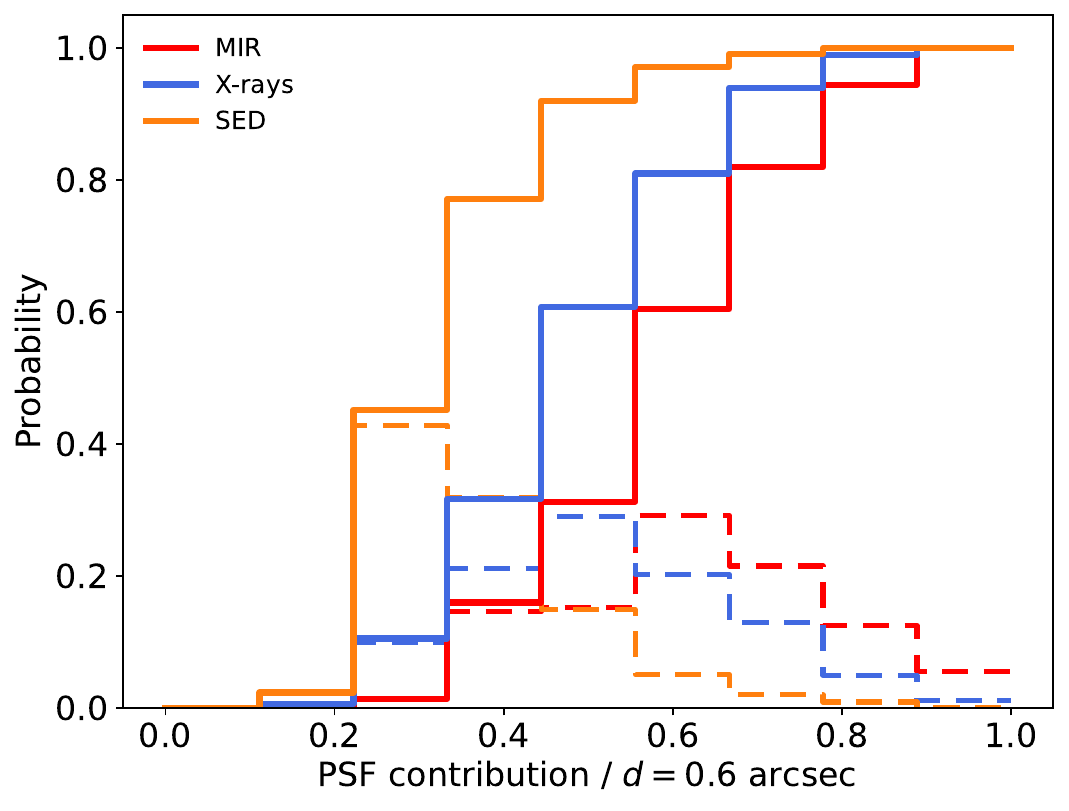}
    \caption{Point source contribution measured in  AGN host galaxies in $z$-bin 3. 
    \emph{Top}: PSF contribution measured in the $i$-band images vs $f_{AGN}$ obtained via SED fitting. 
    \emph{Bottom}: Cumulative (solid lines) and differential (dashed lines) distributions of the PSF contribution for the MIR, X-rays, and SED AGNs. 
    }
    \label{fig:PSF_contr}
\end{figure}

As our merger classifiers are mostly unaffected by point source contamination at $z<0.52$ based on the test on the simulations, we now focus on $z$-bin 3 to investigate how potential mis-classifications may affect the $f_{merger}-f_{AGN}$ relation in Fig.~\ref{fig:fmerg_bhar}.
First, we need to estimate $f_{PSF}$  in the AGN host galaxies to have an overview of the extent of PSF contamination in our AGN samples. We did this as follows: 
\emph{i)} For each image, we scaled the PSF to match the brightest central pixel and then subtracted it from the image;
\emph{ii)} We performed aperture photometry, using the same aperture as before, on the residual images and estimated the fractional point source contamination as $1-(F_{residual}/F_{tot})$.

In the top panel of Fig.~\ref{fig:PSF_contr}, we plot the measured $f_{PSF}$ vs $f_{AGN}$, for the three AGN types in $z$-bin 3. Given the larger number of the SED AGNs compared to the other two AGN types, we randomly selected 500 SED AGNs. 
While there does exist a general correlation, the scatter is very significant. 
In the bottom panel of Fig.~\ref{fig:PSF_contr}, we plot the differential and cumulative $f_{PSF}$ distributions, for the three AGN types in $z$-bin 3. 
It is clear that for all three AGN types, most host galaxies ($\sim90\%$) have a measured PSF contribution $f_{PSF}<0.80$.
From the previous experiment of injecting PSFs into the simulation test sets,  we measured a change in $f_{merger}\sim0.05$ at $f_{PSF}=80\%$ in $z-$bin 3. Therefore, even if we take the extreme scenario for the level of PSF contamination in our AGN samples,  this small change of $\sim0.05$ is insignificant compared to the much steeper increase in $f_{merger}$ ($\sim40\%$) in the regime of very dominant AGNs in Fig.~\ref{fig:fmerg_bhar}. 

To assess the significance of the steep $f_{merger}$ increase shown in Fig.~\ref{fig:fmerg_bhar}, we calculate the difference between $f_{merger}$ at $f_{AGN}>0.8$, and $f_{merger}$ at $f_{AGN} \leq 0.8$. The enhancement is statistically significant at $1.7\sigma$, $3.3\sigma$, and $2.9\sigma$ levels for MIR, X-ray, and SED AGNs, respectively. When considering the possible bias of $\sim 0.05$ at $f_{AGN}>0.8$, the significance is lowered to $1.45\sigma$, $2.9\sigma$, and $2.7\sigma$, in the corresponding AGNs. These findings confirm that our main results are robust against the possible classification bias.

\section{Summary and conclusions}\label{sect:Conclusions}

In this paper, we carefully investigate the connection between galaxy interactions and AGN triggering at $0.1\leq z\lesssim 0.8$, corresponding to the second half of the cosmic history.
We used the excellent imaging quality offered by the HSC-SSP survey to identify mergers and non-mergers in the KiDS-N-W2 field. 
We exploited multi-wavelength data from the X-ray to the sub-mm, including photo-$z$ from the KiDS-VIKING survey, and performed SED fitting to derive galaxy properties, such as stellar mass, AGN fraction ($f_{AGN}$), AGN bolometric luminosity, and BHAR. 
We selected three types of AGNs via MIR colours, X-ray detection, and SED fitting, with different levels of dust obscuration and AGN power. 
To classify HSC images into mergers and non-mergers, we implemented CNNs trained on mock HSC observations from two cosmological simulations and tested them on a small set of visually classified real galaxies. 
We built mass- and redshift-matched control samples to understand the role of mergers in triggering AGNs, for the three AGN types. 
We also extended the analysis of the merger-AGN connection by examining for the first time the continuous $f_{AGN}$ parameter. 
Our main results are summarised below: 

\begin{enumerate}
    \renewcommand{\labelenumi}{\it \roman{enumi})}
    
    \item A large excess (a factor of $\sim$2-3) of the MIR AGNs in mergers compared to non-mergers, and a lower excess ($\sim$1.4) of the SED AGNs in mergers, in every $z$-bin. 
    For the X-ray AGNs, there is a weak excess of $\sim$1.3 in mergers at $z\lesssim0.5$ and a slightly larger excess of 1.8 at $0.5\lesssim z \lesssim 0.8$.
    This indicates that mergers could trigger all three AGN types, but are more connected with the MIR AGNs, which represent dust-obscured AGNs and contain more dominant and/or luminous AGNs compared to the other two types. 
    
    \item The merger fraction in the MIR AGN host galaxies is much higher than in the corresponding controls ($45-55\%$ vs $20-30\%$).
    The merger fraction in the MIR AGNs is also higher than in the X-ray and SED AGNs. 
    Both the MIR and SED AGNs are more likely (by a factor of $\sim 2$) to be hosted by mergers than their respective controls.
    On the other hand, the excess of mergers in the X-ray AGNs compared to the corresponding non-AGNs is a factor of $\sim 1.2$ at $z\lesssim0.5$ and $\sim1.8$ at $0.5\lesssim z \lesssim 0.8$.
    We cannot yet conclude on the extent of mergers in triggering AGNs, but mergers could be the dominant mechanism in triggering the MIR AGNs.
    
    \item The AGN fraction $f_{AGN}$ distributions of the merger population are statistically different from those of the non-merger controls, for all $z$-bins and AGN types, with a clear excess at high $f_{AGN}$ values in mergers.
    The strongest excess is found in the MIR AGNs at $z\gtrsim0.3$.
    The BHAR distributions show a similar trend. 
    The incidence rate of SED AGNs with $f_{AGN}\geq0.5$ or $\geq0.7$ is a factor of $\sim 2-5$ larger in mergers than in non-mergers, which is much higher than the entire SED AGN sample and similar to the MIR AGNs. 
    These results imply that mergers play an increasingly more important role in triggering more dominant/luminous AGNs.
    
    \item We derived, for the first time, a relation between merger fraction $f_{merger}$ and AGN fraction $f_{AGN}$, which reveals two distinct regimes. 
    At $f_{AGN}<0.8$, we observe an almost flat $f_{merger}$ with increasing $f_{AGN}$, centred at $f_{merger}\sim 0.3-0.4$ for the SED and X-ray AGNs and at $\sim 0.5$ for the MIR AGNs. 
    At $f_{AGN}\geq0.8$, there is a very steep rise for all AGN types, up to $f_{merger}\simeq80-100\%$. 
    This could indicate that mergers are the dominant or even the sole fuelling mechanism for the most dominant/powerful AGNs and help explain why some studies found no dependence in $f_{merger}$ on AGN luminosity.

\end{enumerate}

In conclusion, our results show clear evidence that mergers are strongly connected to the presence of dust-obscured AGNs, as well as to the most dominant and powerful AGNs. 
This could be because mergers can funnel gas to the central regions to enable rapid accretion onto the SMBH. 
This rapid accretion phase is generally considered to be dust-obscured (thus promoting the detection of the AGN in the MIR). 
The sheer dominance of mergers in the most luminous and powerful AGNs, regardless of AGN type, could be explained by the possibility that mergers are by far the most likely or even the only viable way to bring in the large amount of gas needed to assemble the SMBH very quickly. 
This also highlights the need to use large-area surveys \citep[e.g. with {\it Euclid},][]{laureijs_2011} to study the most extreme AGNs. 
For the less dominant and powerful AGNs, secular processes could be more important than mergers, particularly for the less dust-obscured AGNs.
Finally, we have publicly released the multi-wavelength galaxy catalogue constructed in this work, including our detailed SED fitting results and merger classifications.

\section*{Data Availability}

The final sample of galaxies analysed in this work, including the merger classifications, is publicly released in two versions at \url{https://antolamarca.com/#data}. The catalogues are also available in electronic form at the CDS via anonymous ftp to cdsarc.u-strasbg.fr (130.79.128.5) or via \url{http://cdsweb.u-strasbg.fr/cgi-bin/qcat?J/A+A/}.

\begin{acknowledgements}
This publication is part of the project ‘Clash of the titans:
deciphering the enigmatic role of cosmic collisions’ (with project number VI.Vidi.193.113 of the research programme Vidi which is (partly) financed by the Dutch Research Council (NWO).
The Hyper Suprime-Cam (HSC) collaboration includes the astronomical communities of Japan and Taiwan, and Princeton University. The HSC instrumentation and software were developed by the National Astronomical Observatory of Japan (NAOJ), the Kavli Institute for the Physics and Mathematics of the Universe (Kavli IPMU), the University of Tokyo, the High Energy Accelerator Research Organization (KEK), the Academia Sinica Institute for Astronomy and Astrophysics in Taiwan (ASIAA), and Princeton University. Funding was contributed by the FIRST program from the Japanese Cabinet Office, the Ministry of Education, Culture, Sports, Science and Technology (MEXT), the Japan Society for the Promotion of Science (JSPS), Japan Science and Technology Agency (JST), the Toray Science Foundation, NAOJ, Kavli IPMU, KEK, ASIAA, and Princeton University. 
This paper is based on data collected at the Subaru Telescope and retrieved from the HSC data archive system, which is operated by the Subaru Telescope and Astronomy Data Center (ADC) at NAOJ. Data analysis was in part carried out with the cooperation of Center for Computational Astrophysics (CfCA), NAOJ. We are honored and grateful for the opportunity of observing the Universe from Maunakea, which has the cultural, historical and natural significance in Hawaii. 

Based on data products from observations made with ESO Telescopes at the La Silla Paranal Observatory under programme IDs 177.A-3016, 177.A-3017 and 177.A-3018, and on data products produced by Target/OmegaCEN, INAF-OACN, INAF-OAPD and the KiDS production team, on behalf of the KiDS consortium. OmegaCEN and the KiDS production team acknowledge support by NOVA and NWO-M grants. Members of INAF-OAPD and INAF-OACN also acknowledge the support from the Department of Physics \& Astronomy of the University of Padova, and of the Department of Physics of Univ. Federico II (Naples).

This publication makes use of data products from the Wide-field Infrared Survey Explorer, which is a joint project of the University of California, Los Angeles, and the Jet Propulsion Laboratory/California Institute of Technology, funded by the National Aeronautics and Space Administration.

This work is based on data from eROSITA, the soft X-ray instrument aboard SRG, a joint Russian-German science mission supported by the Russian Space Agency (Roskosmos), in the interests of the Russian Academy of Sciences represented by its Space Research Institute (IKI), and the Deutsches Zentrum für Luft- und Raumfahrt (DLR). 

Herschel is an ESA space observatory with science instruments provided by European-led Principal Investigator consortia and with important participation from NASA.

This work has made use of the Horizon cluster on which the Horizon-AGN simulation was post-processed, hosted by the Institut d’Astrophysique de Paris. We warmly thank S. Rouberol for running it smoothly.

We thank the Center for Information Technology of the University of Groningen for their support and for providing access to the Hábrók high performance computing cluster.

The SED fitting and modelling on such a large sample was possible thanks to the high-memory resources of the Dutch National Supercomputer (Snellius).
We thank SURF (www.surf.nl) for the support in using the National Supercomputer Snellius.

\end{acknowledgements}

\bibliography{aanda}


\begin{appendix}

\section{Examples of the best-fit SEDs from CIGALE}\label{app:CIGALE_examples}

\begin{figure*}[h]
    \centering
    \includegraphics[width=.42\textwidth]{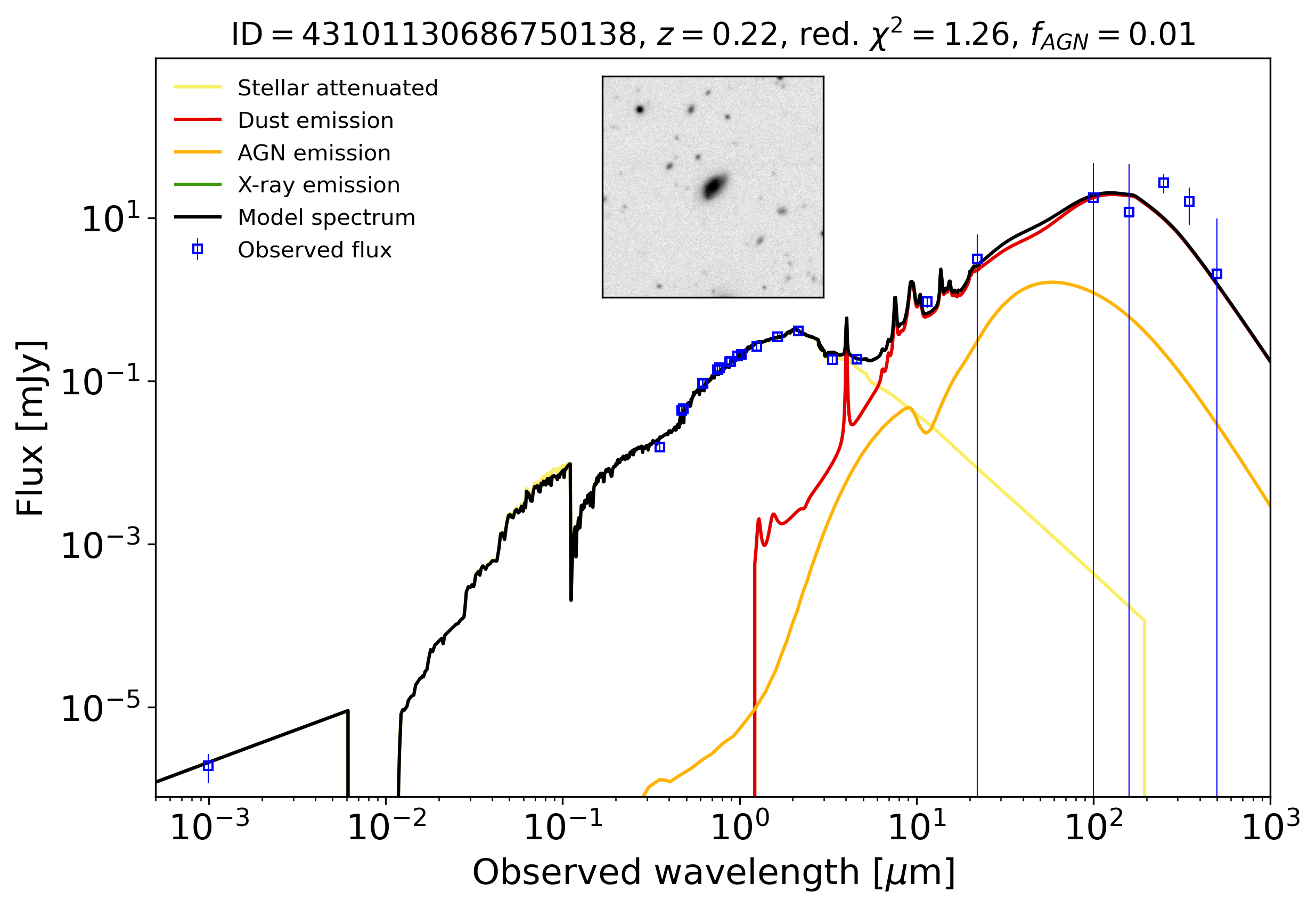}
    \includegraphics[width=.42\textwidth]{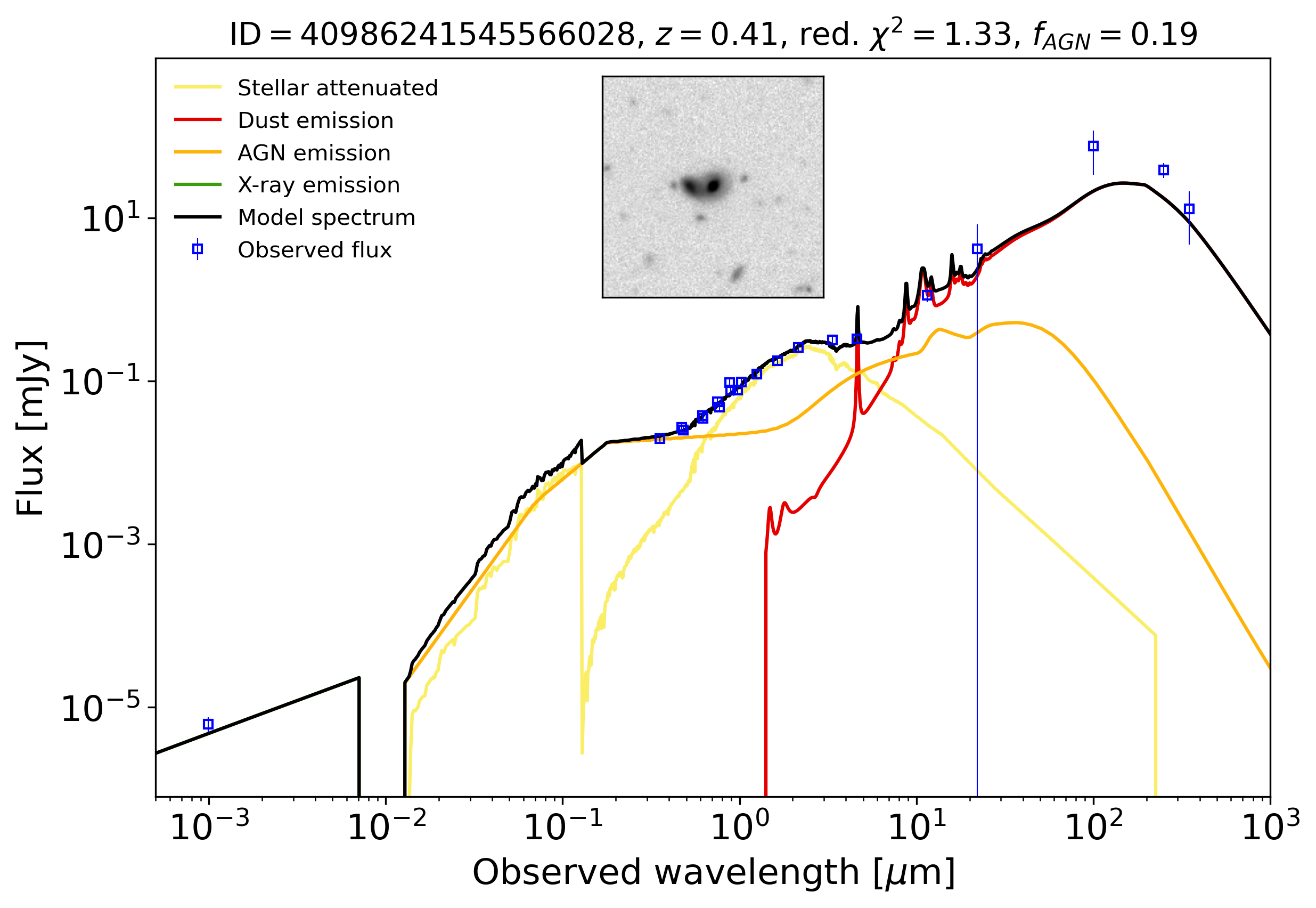}
    \includegraphics[width=.42\textwidth]{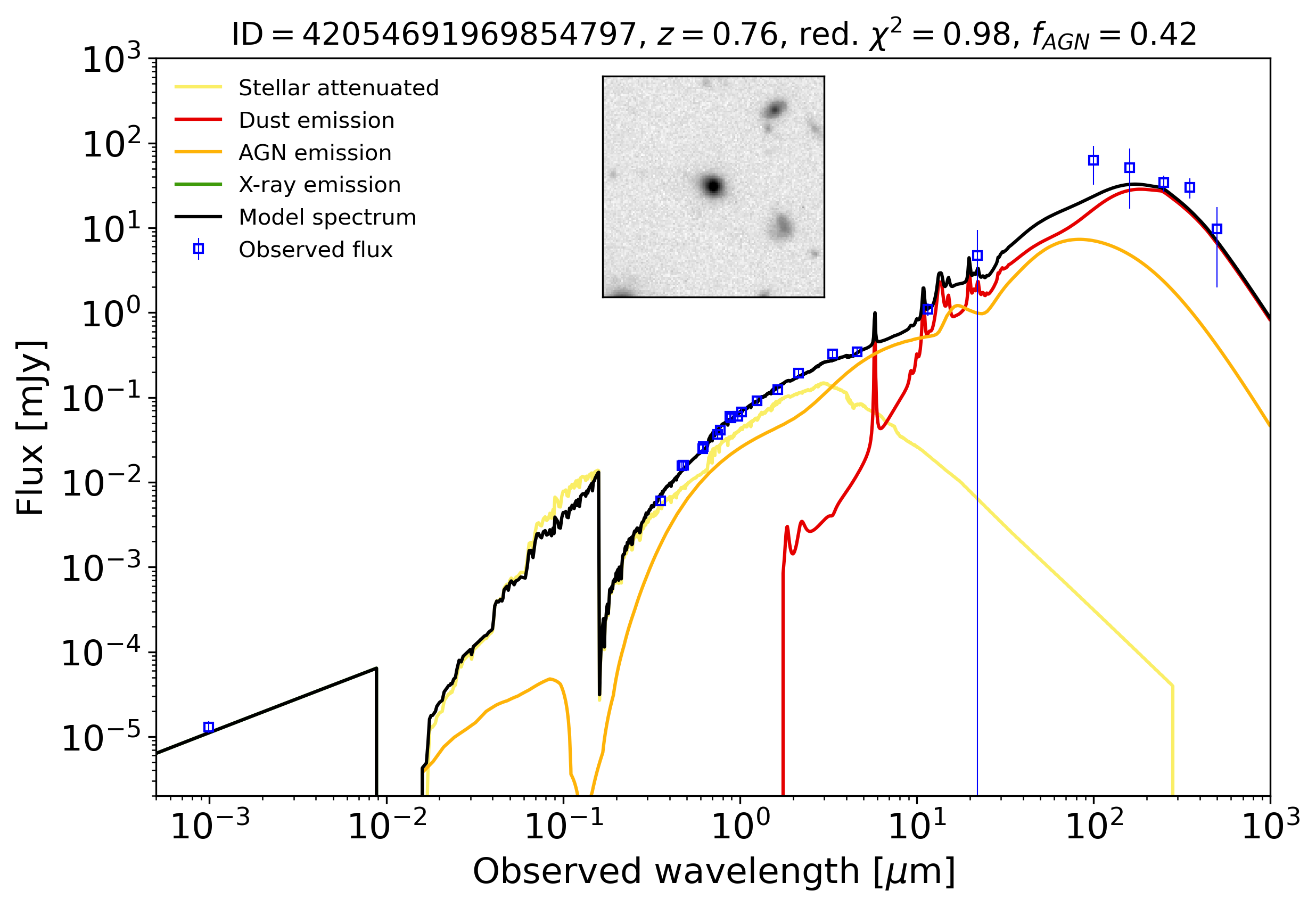}
    \includegraphics[width=.42\textwidth]{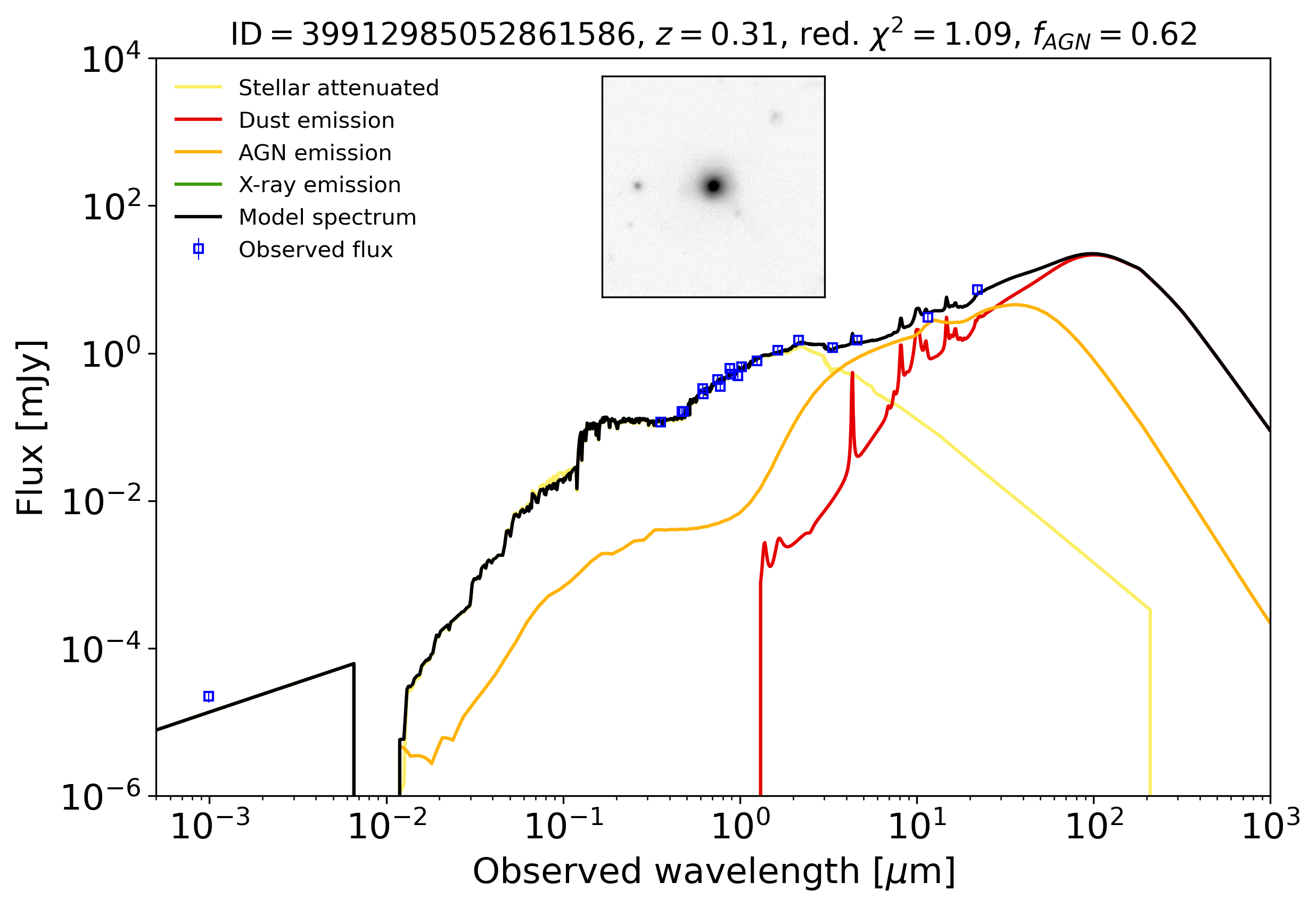}
    \includegraphics[width=.42\textwidth]{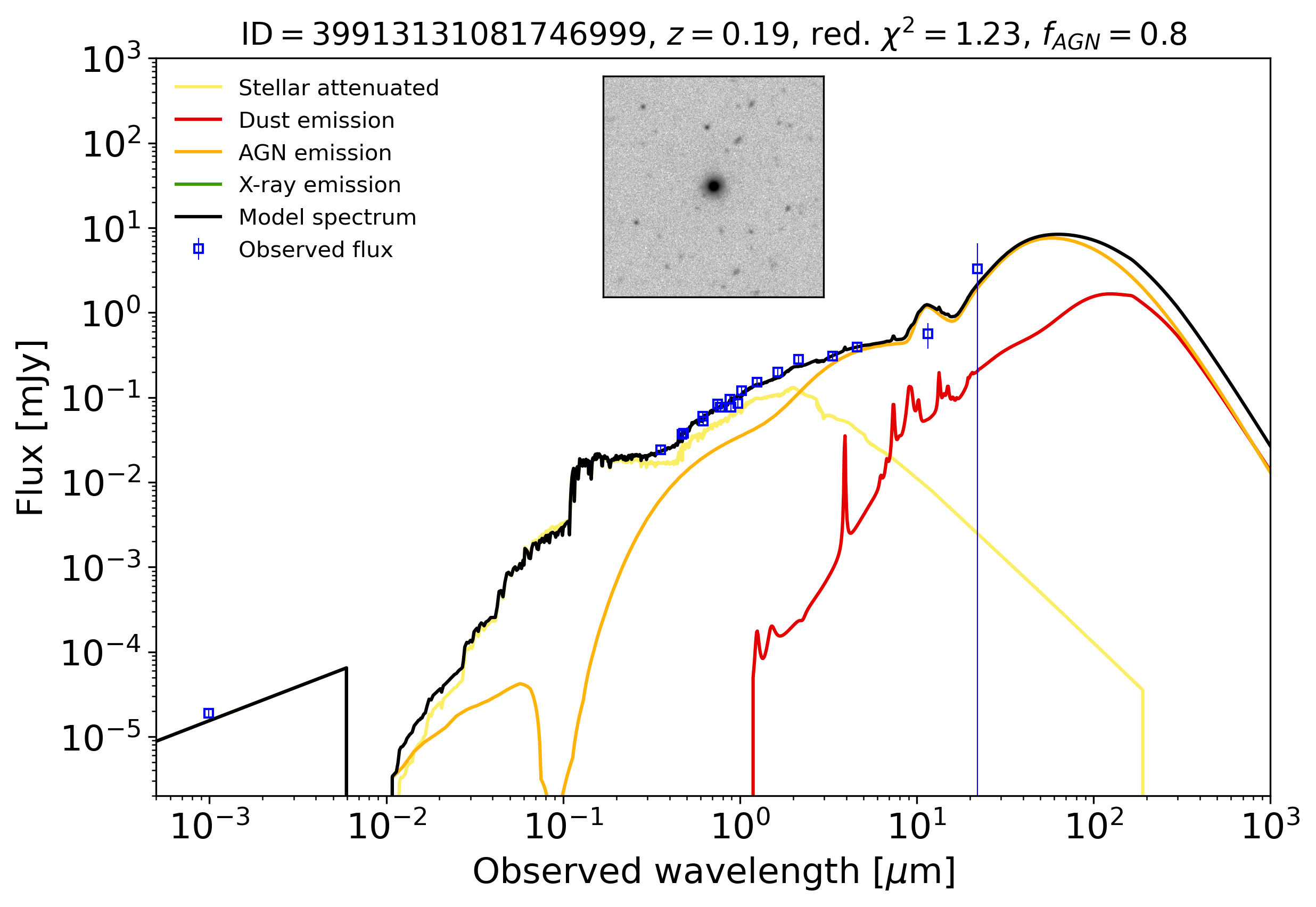}
    \includegraphics[width=.42\textwidth]{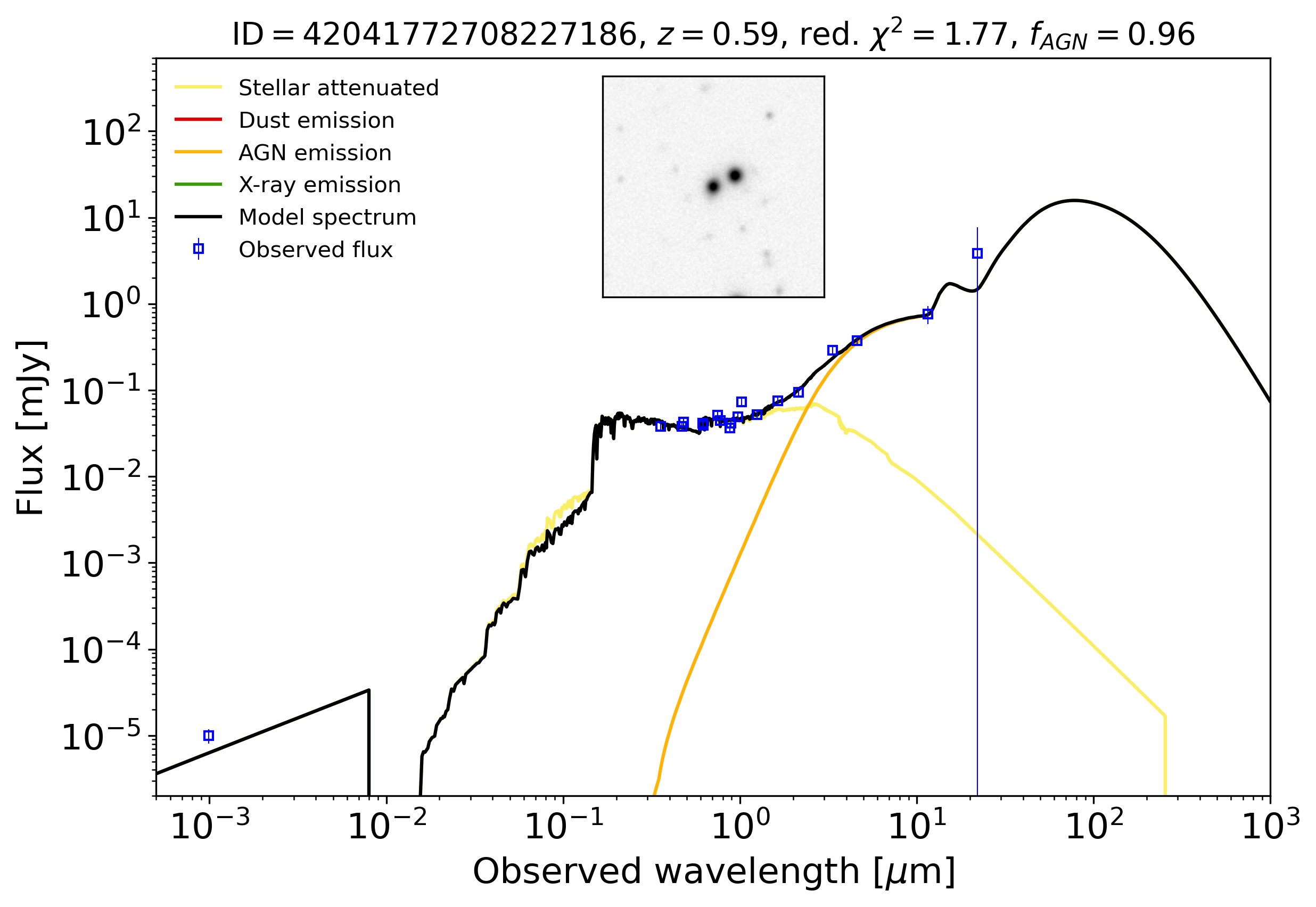}
    \caption{Example CIGALE best SED fits for galaxies with different AGN fractions (with increasing $f_{AGN}$ from left to right, and from top to bottom). Within each panel, the title lists the galaxy ID, redshift, reduced chi-squared, and $f_{AGN}$. The legend indicates the various SED components. The HSC $i$-band image is also shown, on an arcsinh inverted grey scale. 
    }
    \label{fig:SEDs}
\end{figure*}

To illustrate what the SEDs look like for galaxies with different AGN fractions, we show some examples of the best-fit SEDs from CIGALE for galaxies with increasing $f_{AGN}$ in Fig. \ref{fig:SEDs}. 

\section{Example of CNN activation maps}\label{sect:act_maps}

\begin{figure*}
    \centering
    \includegraphics[width=.13\textwidth]{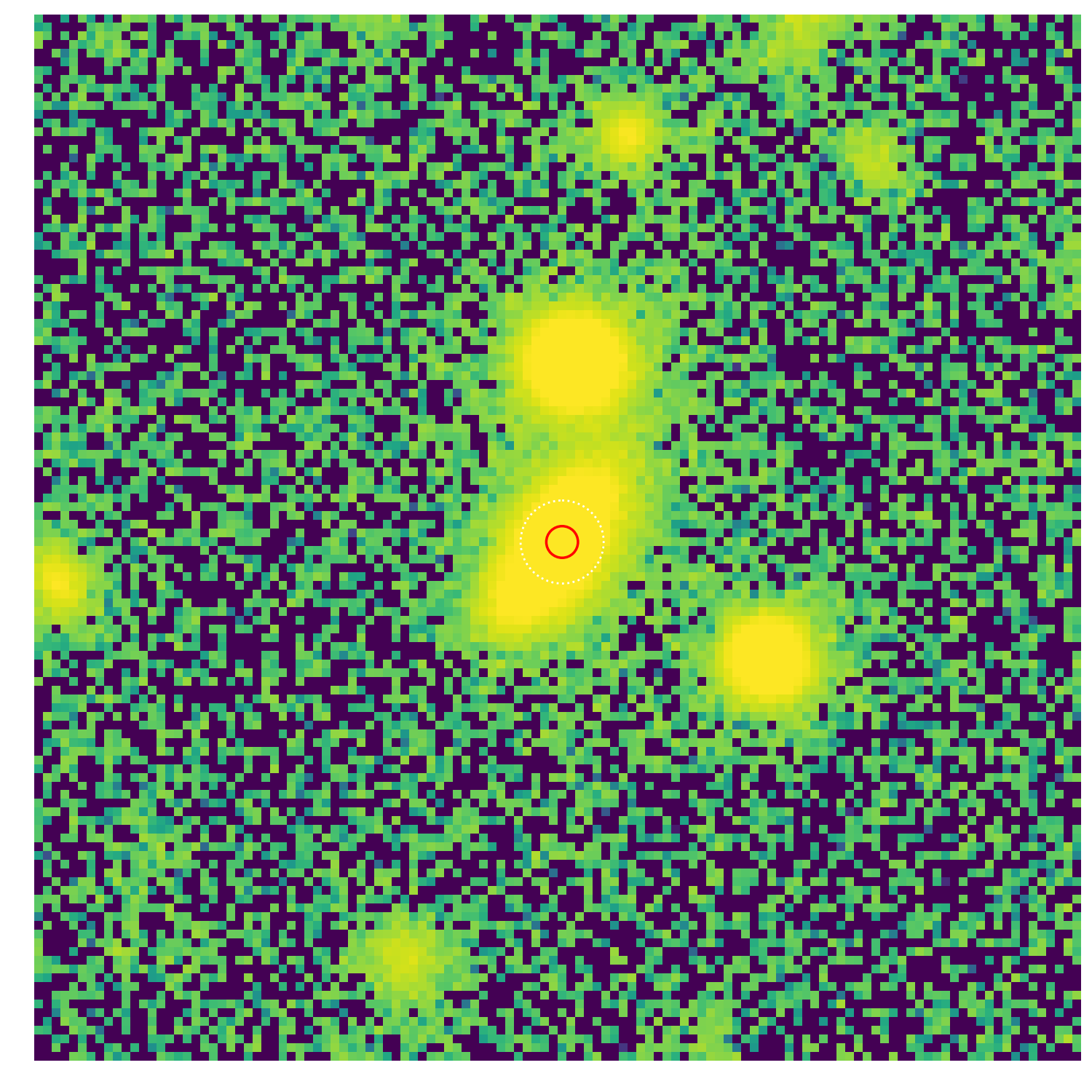}
    \includegraphics[width=.13\textwidth]{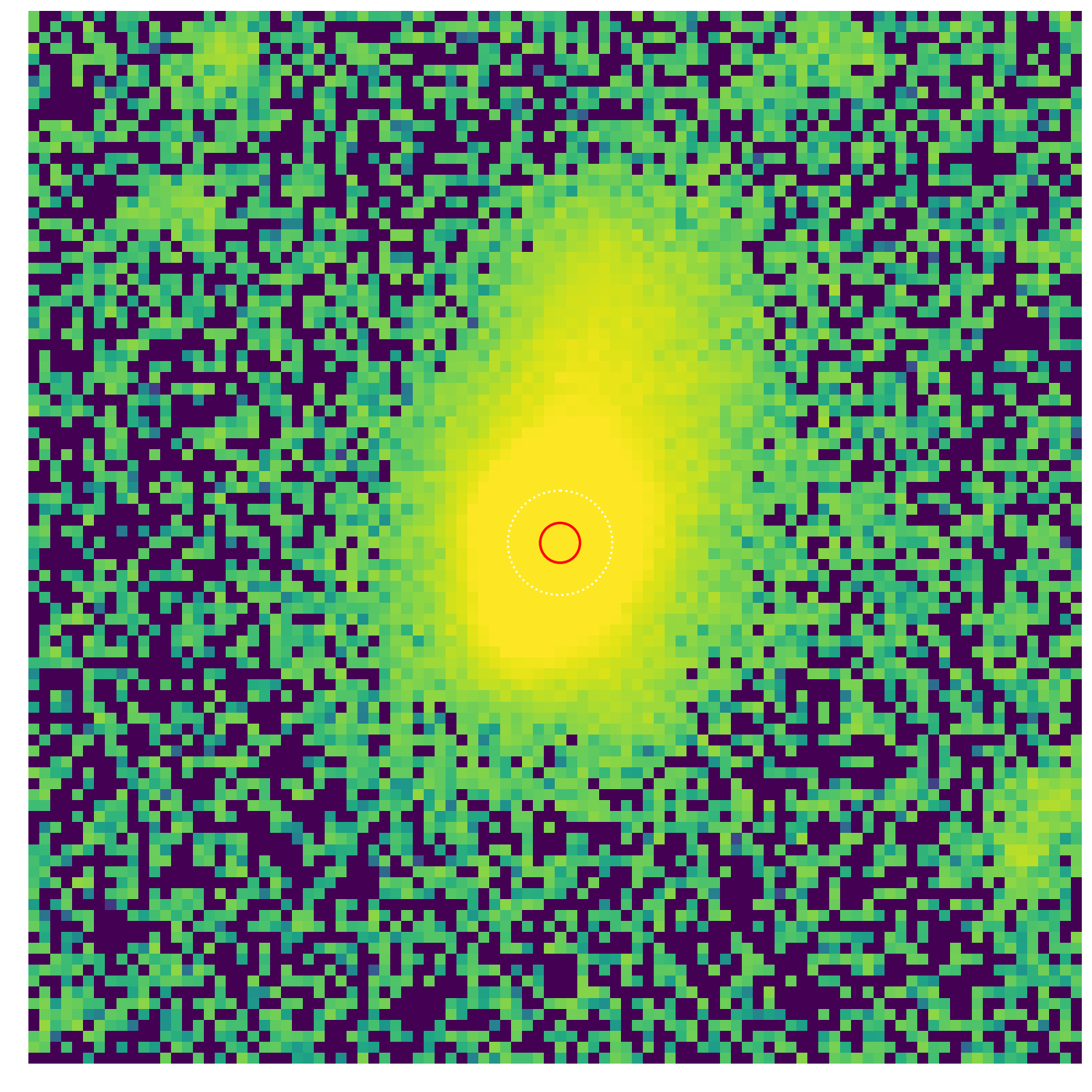}
    \includegraphics[width=.13\textwidth]{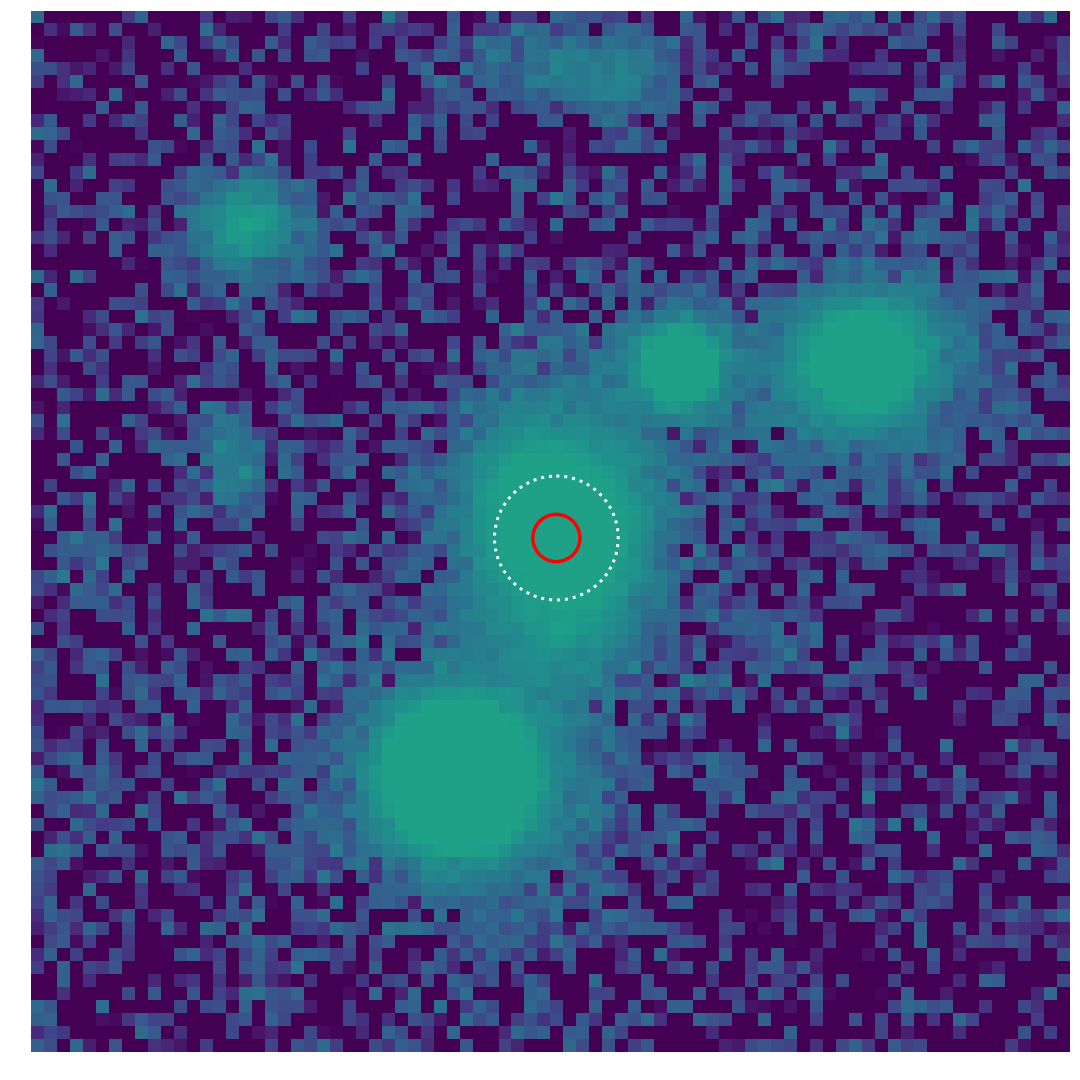}
    \hspace{1cm}
    \includegraphics[width=.13\textwidth]{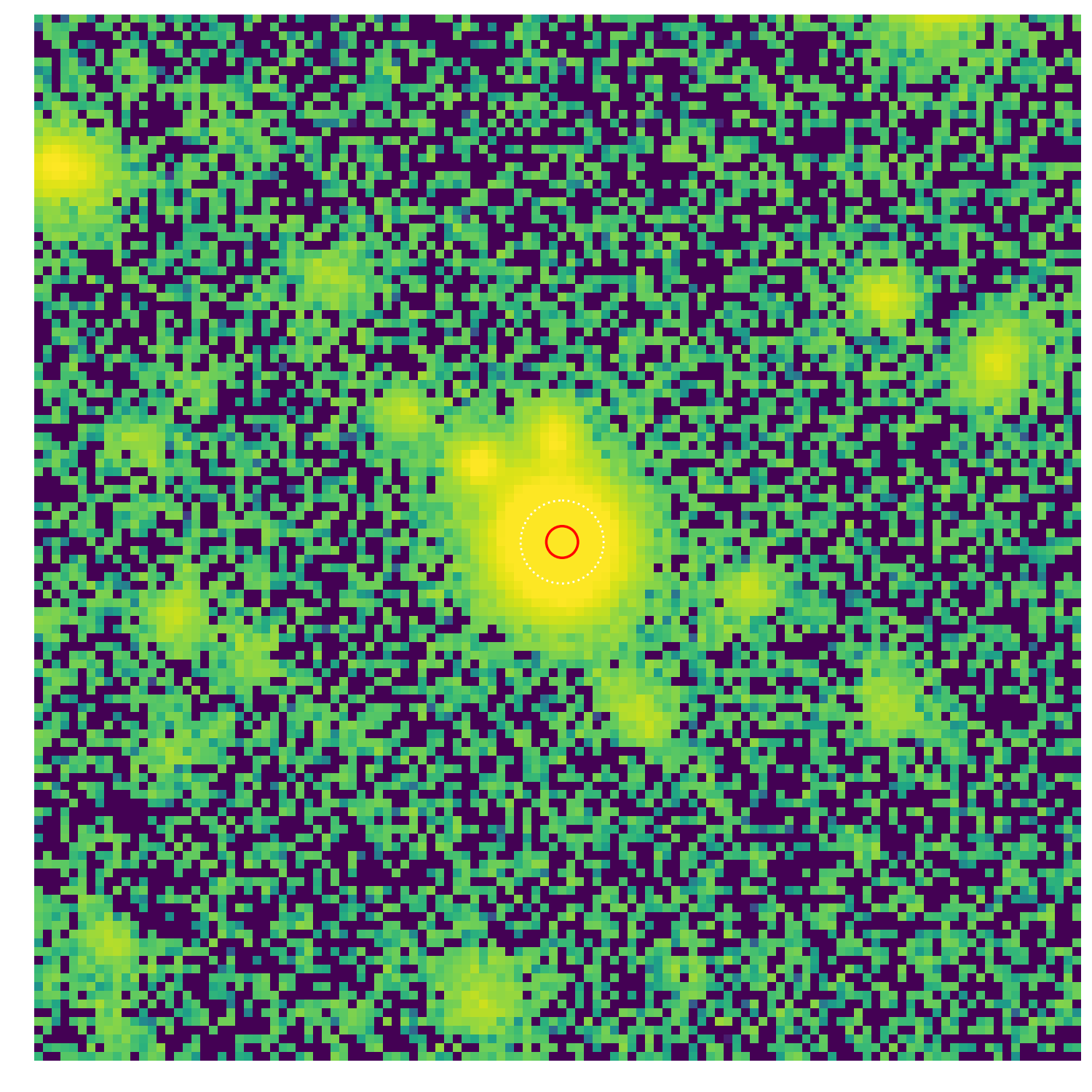}
    \includegraphics[width=.13\textwidth]{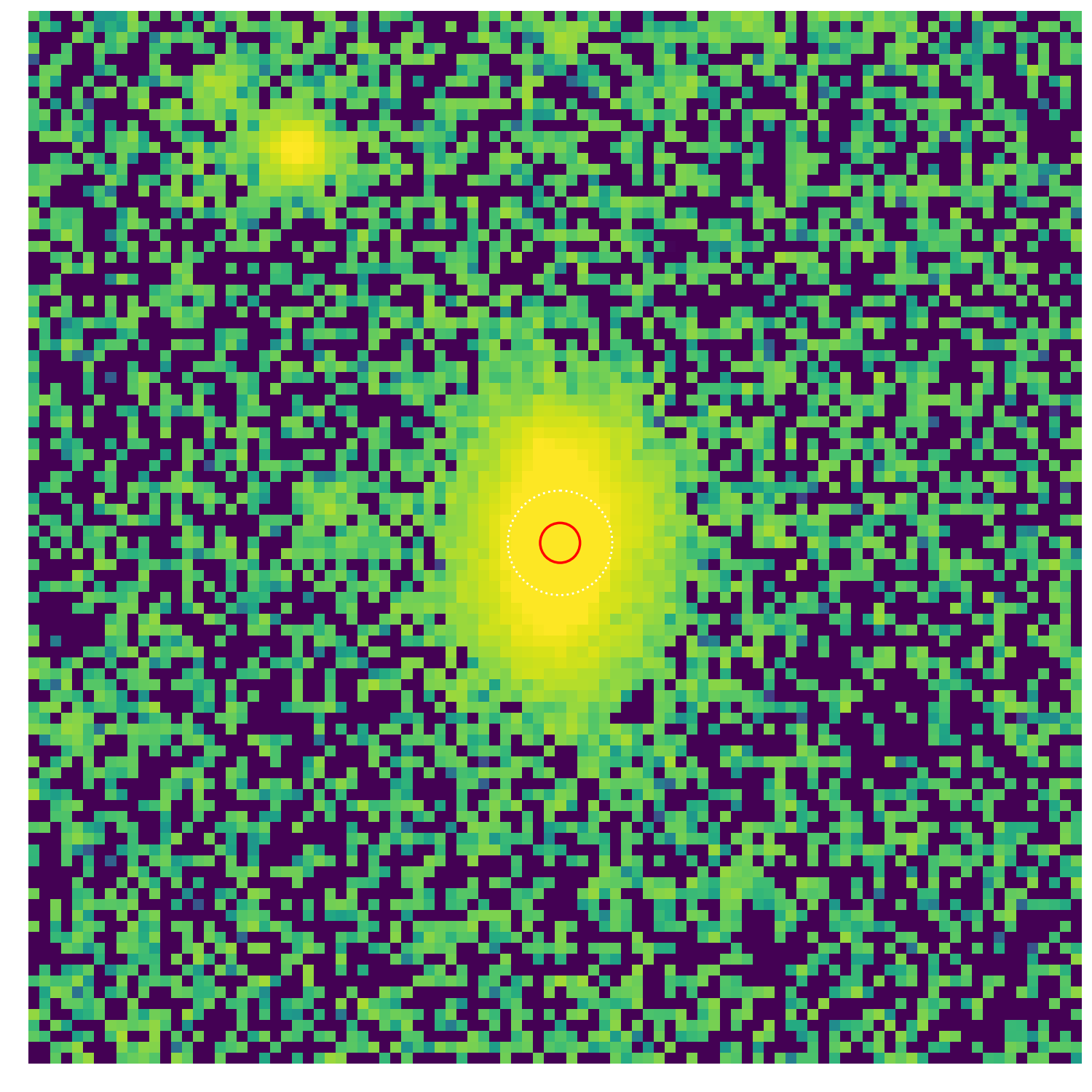}
    \includegraphics[width=.13\textwidth]{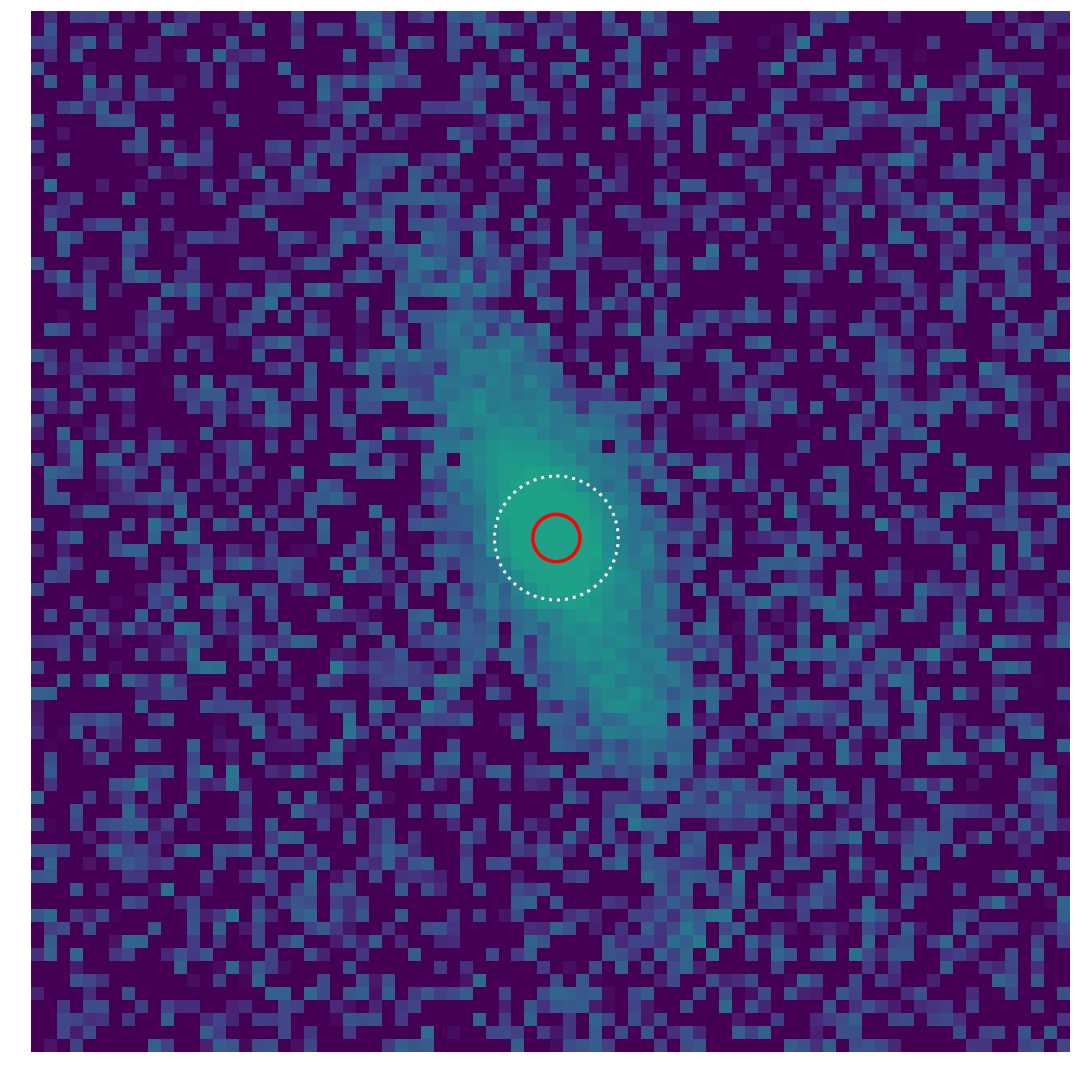}
    
    \includegraphics[width=.13\textwidth]{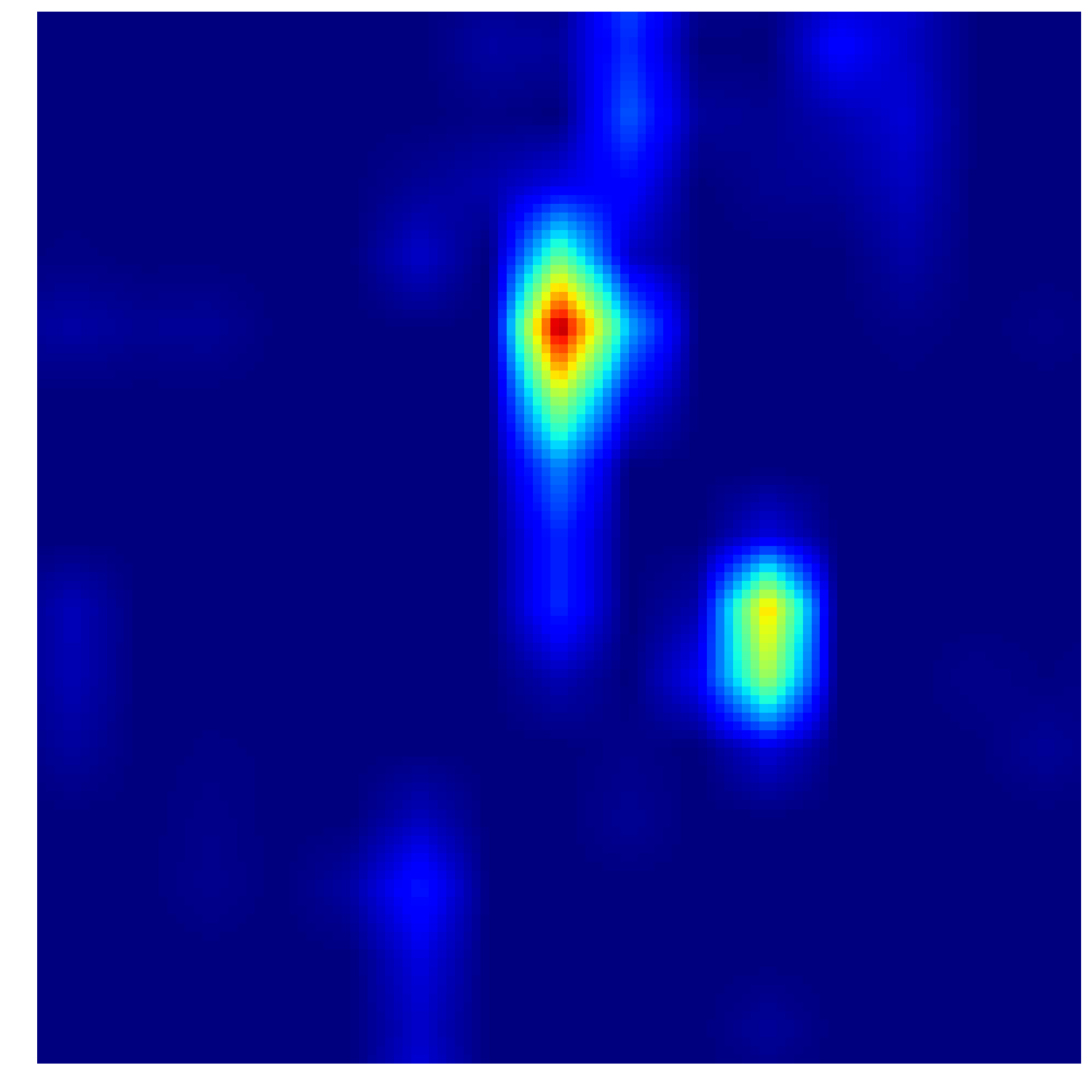}
    \includegraphics[width=.13\textwidth]{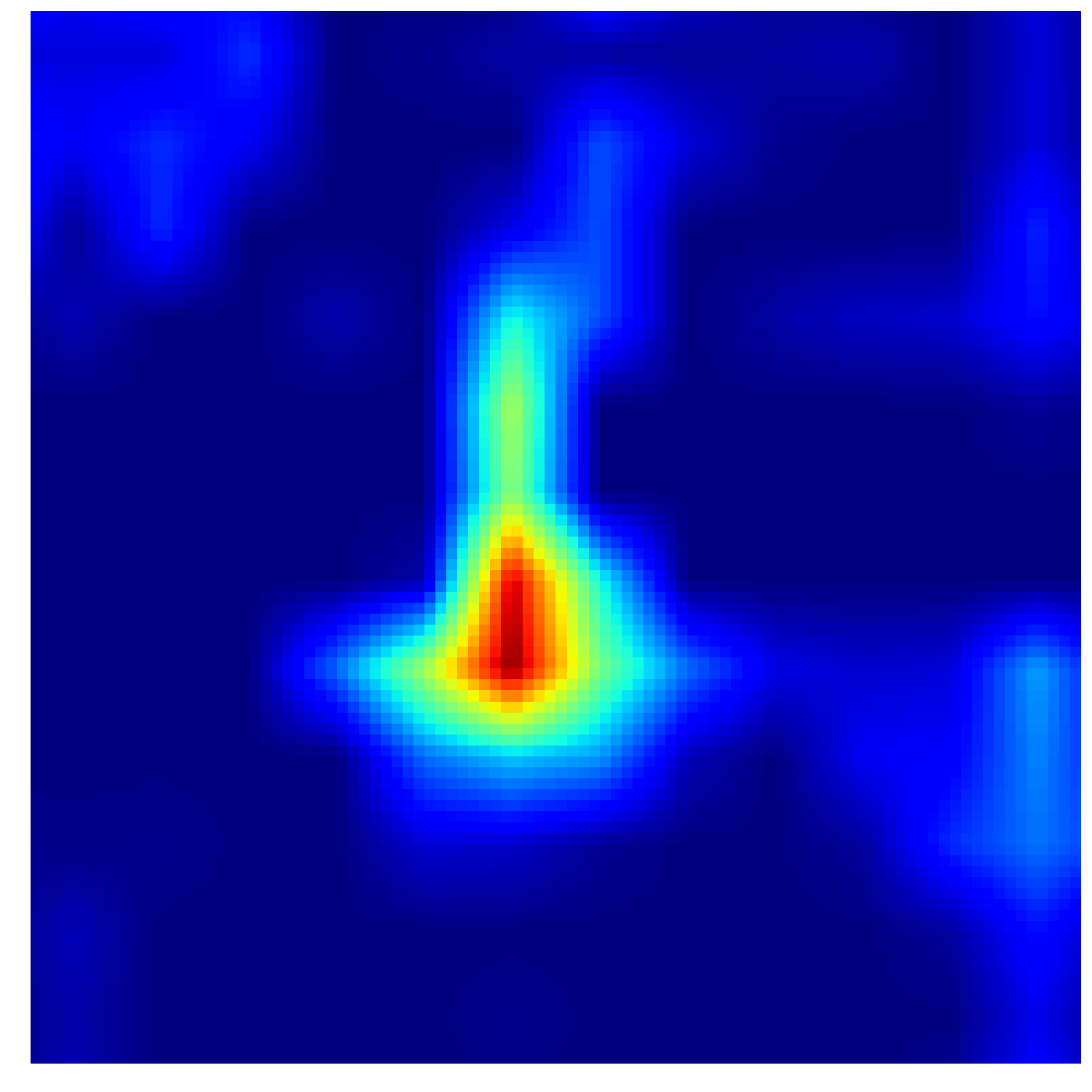}
    \includegraphics[width=.13\textwidth]{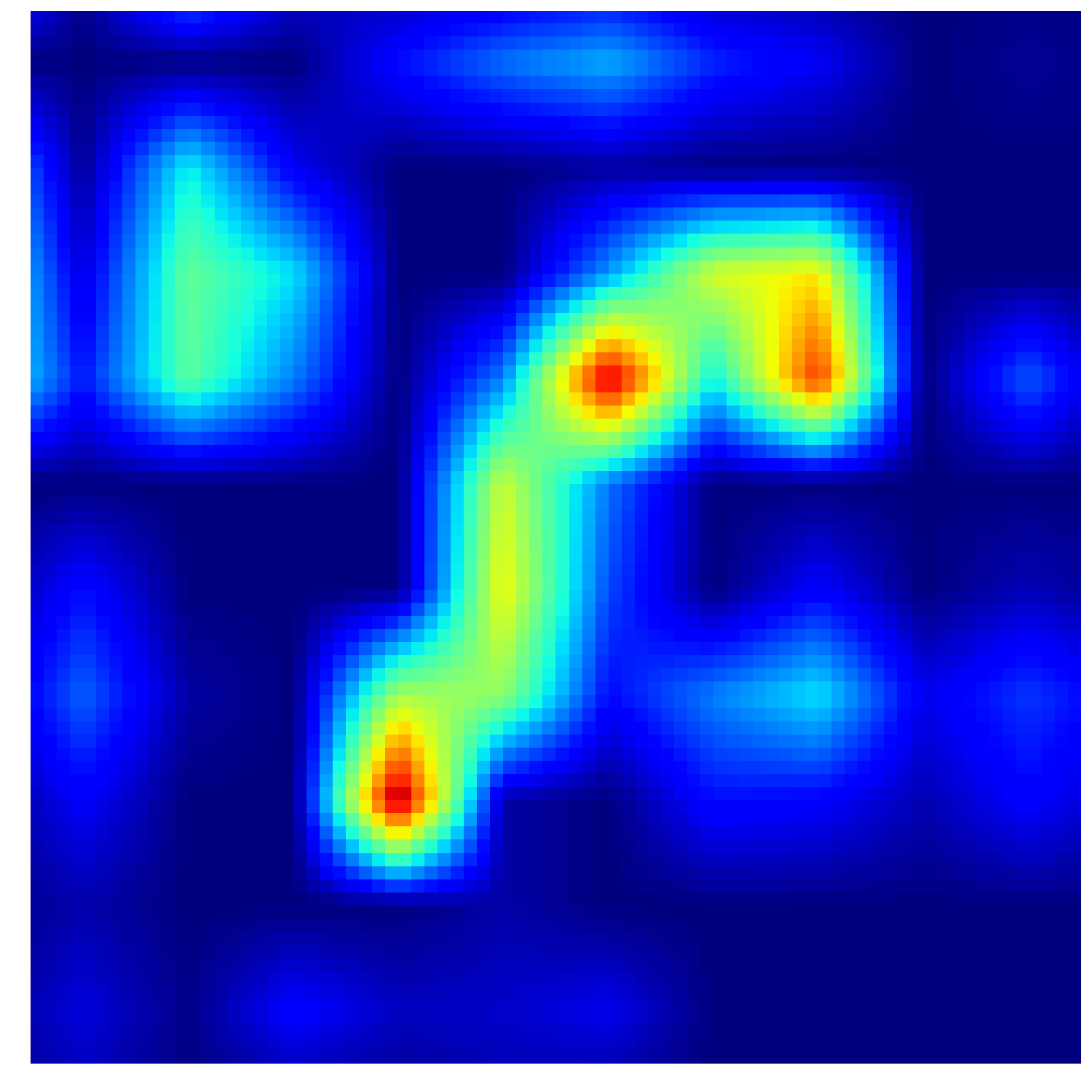}
    \hspace{1cm}
    \includegraphics[width=.13\textwidth]{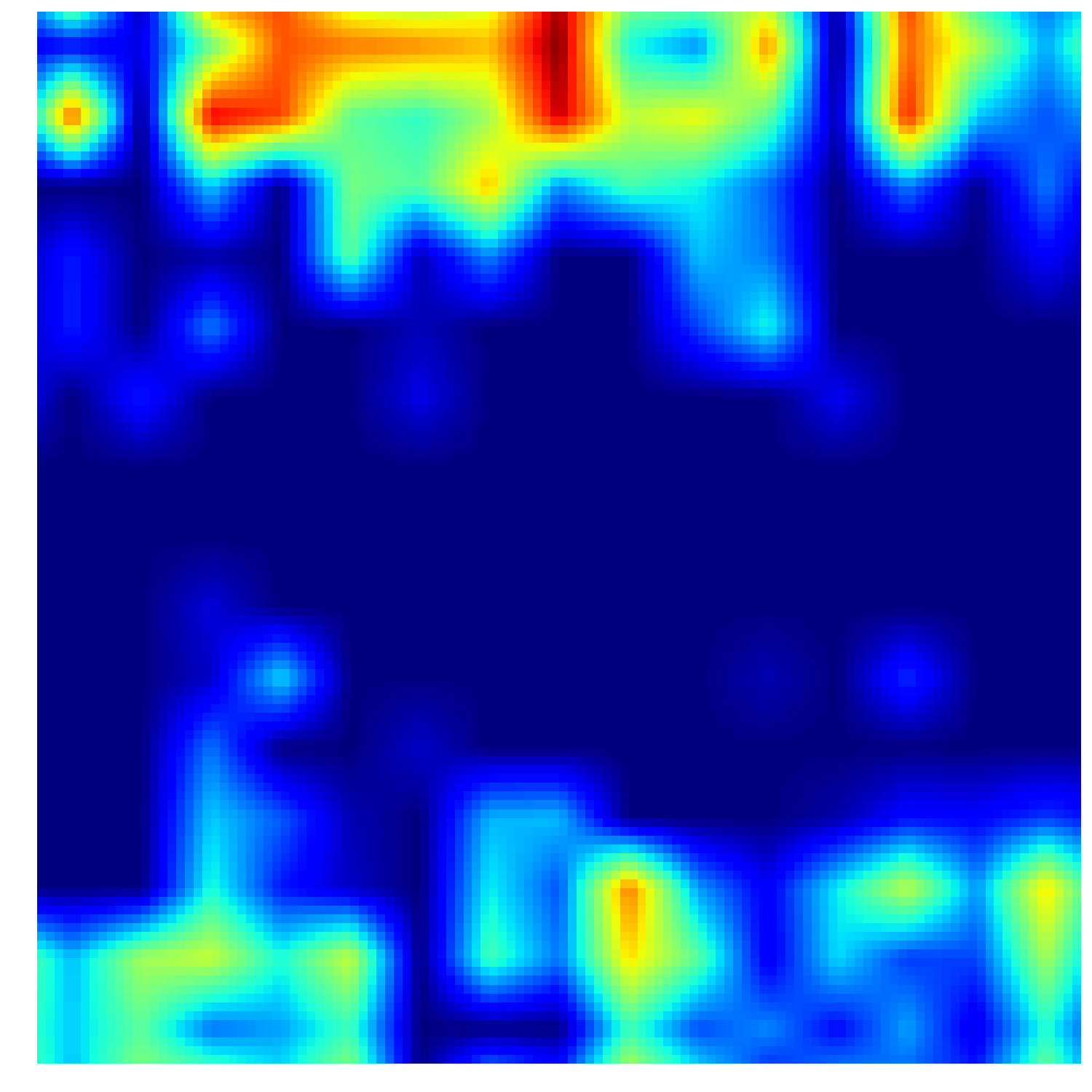}
    \includegraphics[width=.13\textwidth]{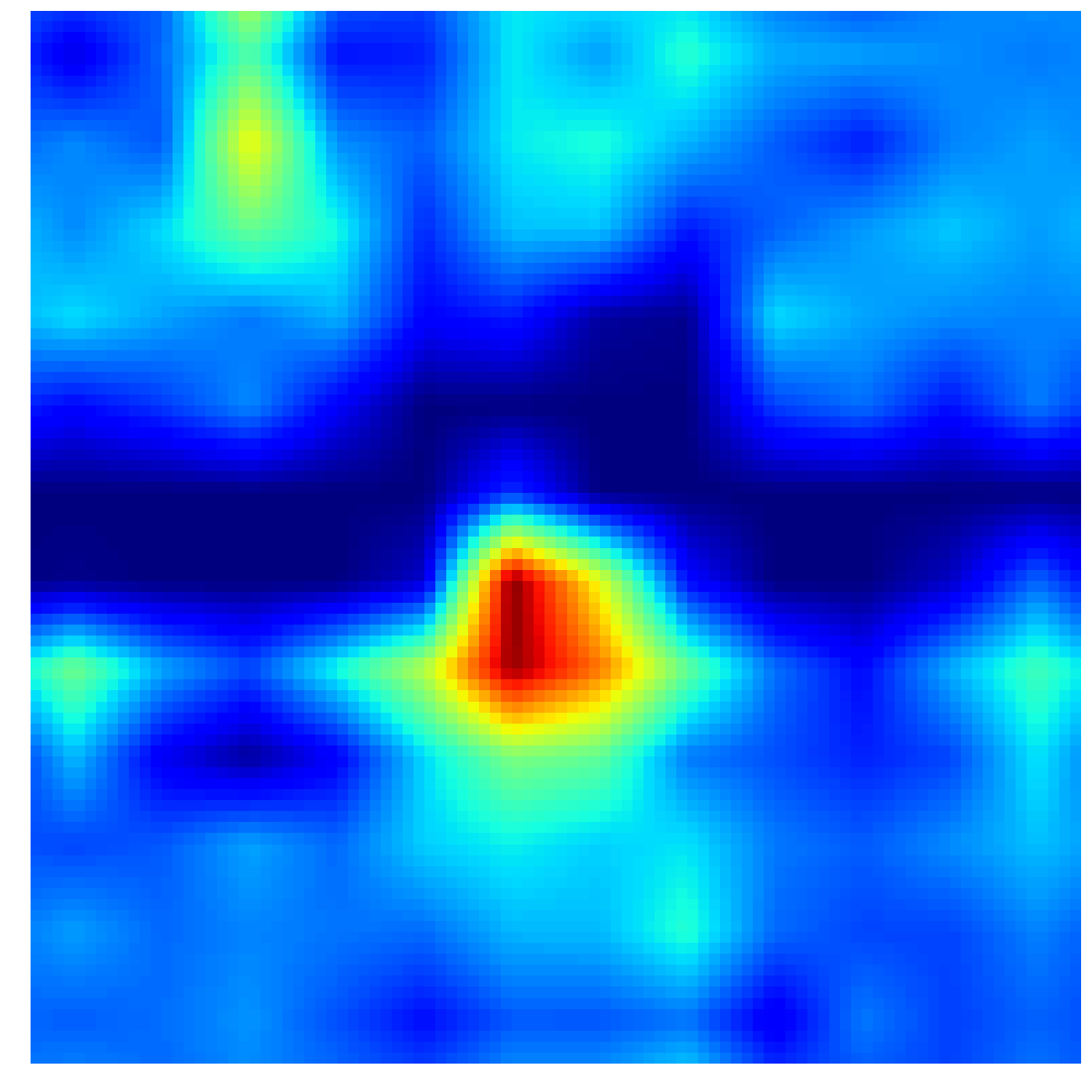}
    \includegraphics[width=.13\textwidth]{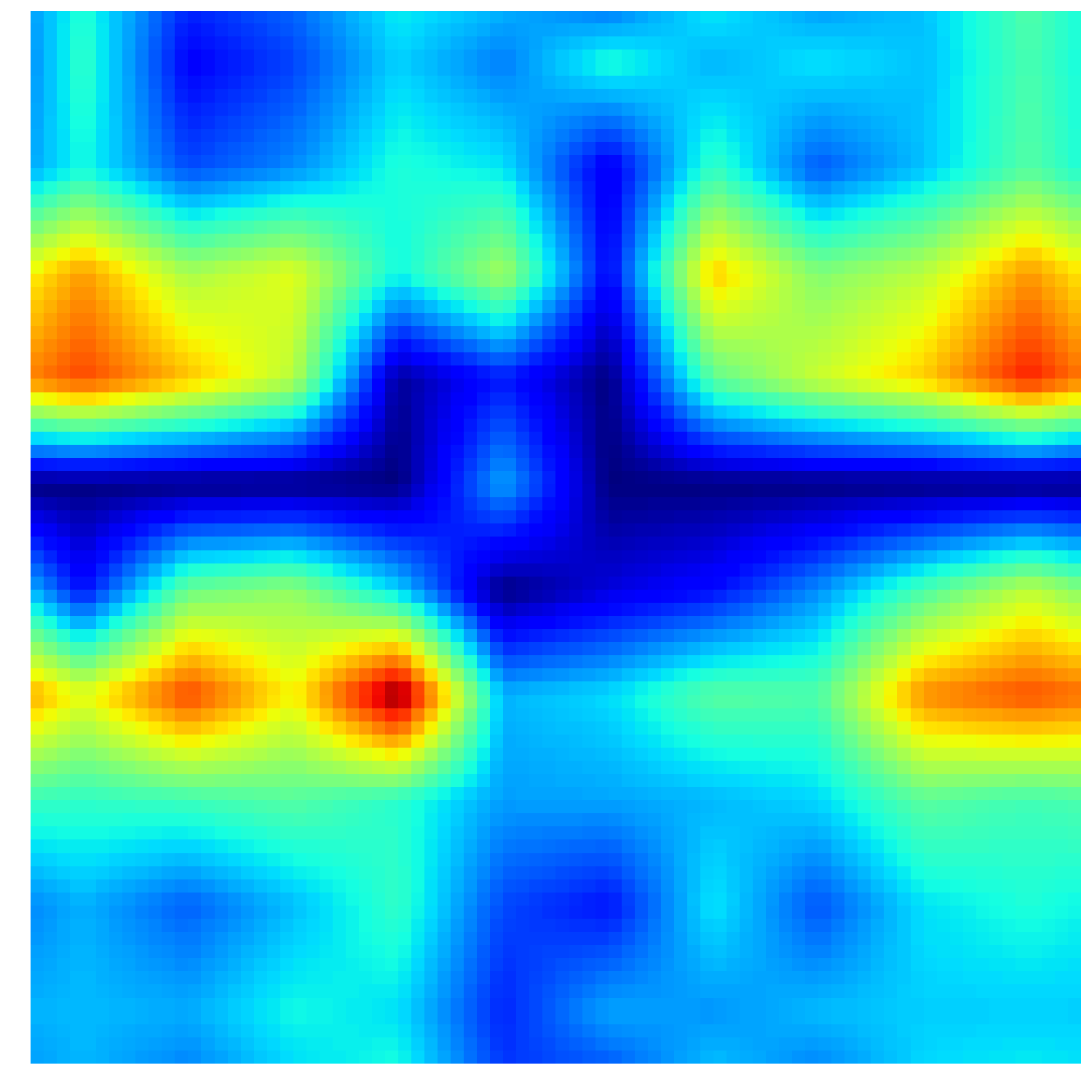}
    \includegraphics[width=.43\textwidth]{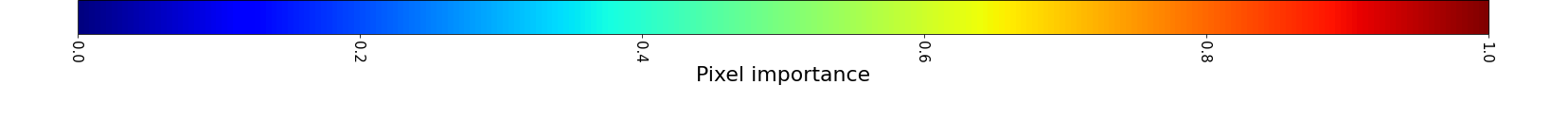}
    \hspace{0.5cm}
    \includegraphics[width=.43\textwidth]{Plots/colorbar.png}
    
    \caption{
    Examples of HSC $i-$band images and corresponding activation maps of the TNG-CNN. 
    The top row shows examples of mergers (images 1 to 3) and non-mergers (images 4 to 6) ordered in increasing $z$-bin. 
    The bottom row shows the activation maps. The redder the pixel colour the more the CNN pays attention to that region.
    The HSC images are cut and resized as detailed in Sect.~\ref{sect:img_prep}.
    }
    \label{fig:act_maps}
\end{figure*}

Activation maps help visualise which regions of the input data are important for a particular feature \citep{selvaraju_grad-cam_2016,zhou2016learning}.
Higher values in the activation map indicate areas where the feature is present and give a larger contribution to the model to make a decision.
Activation maps serve as a way to picture how features are detected and represented in different layers of the network, contributing to the interpretability and effectiveness of deep learning models.
We show example activation maps, generated through the Gradient-weighted Class Activation Mapping algorithm \citep[Grad-CAM;][]{selvaraju_grad-cam_2016}, for mergers and non-mergers for the TNG-CNN in Fig.~\ref{fig:act_maps}. 
Alongside the maps, we also provide the HSC $i-$band images given as input to the CNN to help interpret them. 
These maps confirm our expectations. 
In the case of mergers, it is clear that the model is focussing on possible close companions and asymmetric features in the central galaxy. 
For non-mergers, the model gives less weight to the central galaxy and gives a more uniform weight to the surrounding area. 

\section{Examples of galaxies in the visually inspected test set}\label{sect:example_img}

\begin{figure*}
  \centering
  \includegraphics[width=1.\textwidth]{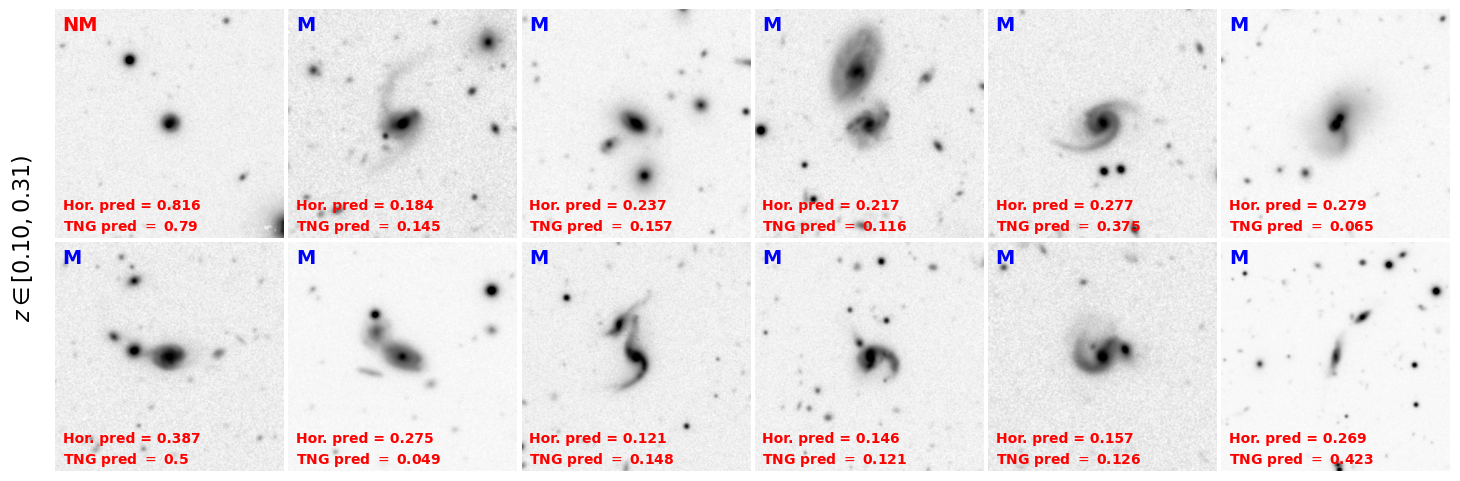}
  \includegraphics[width=1.\textwidth]{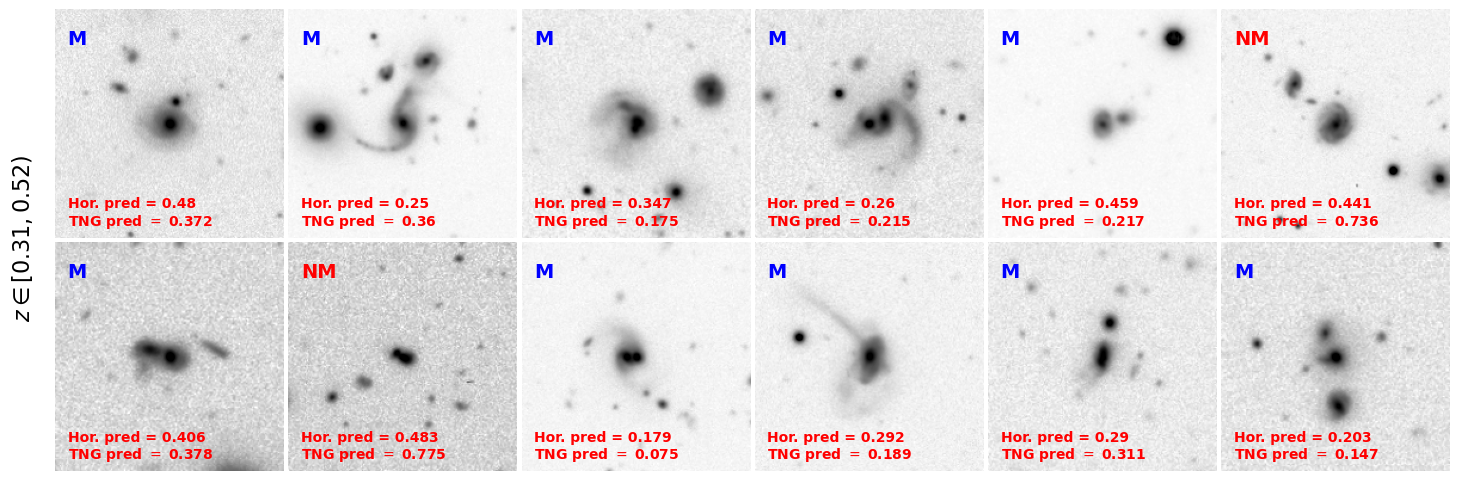}
  \includegraphics[width=1.\textwidth]{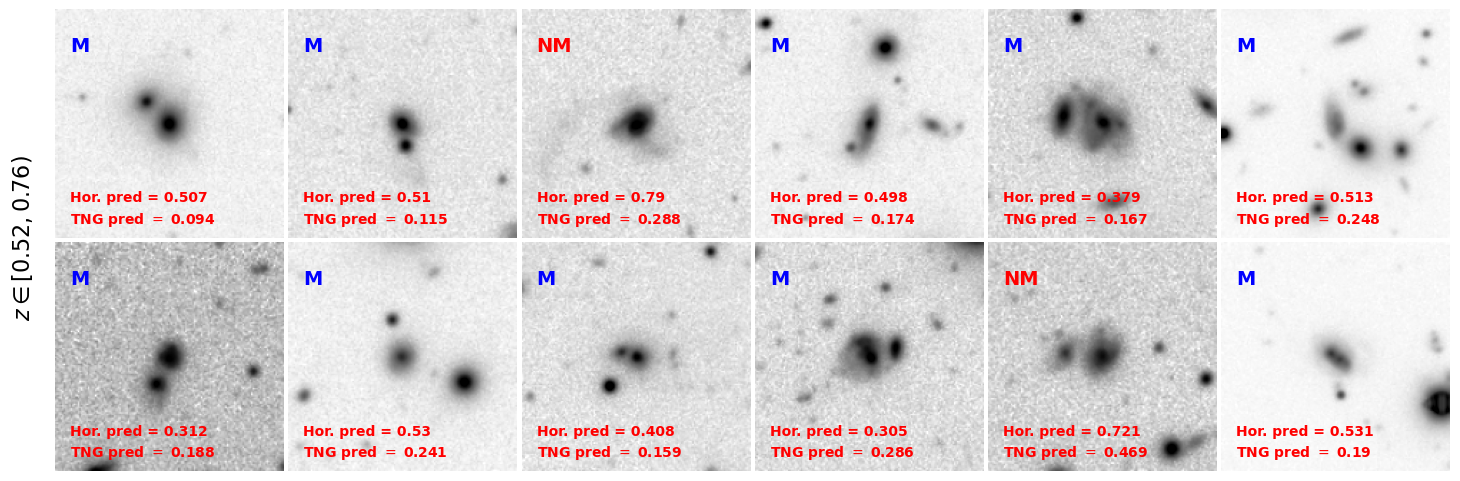}
  \caption{Example HSC $i$-band cutouts of the visually classified mergers ($z$-bin 1: the first two rows; $z$-bin 2: the third and fourth row; $z$-bin 3: the last two rows). At the bottom of each image, we report the predictions from the TNG-CNN and Horizon-CNN. At the top of each image, we list the predicted class according to our final classification (Comb-CNN with two thresholds). The blue 'M' indicates mergers, while the red 'NM' indicates non-mergers. The images have an approximate physical size of 160 kpc, resized to $160\times160$ pixels and displayed with an arcsinh greyscale.}
  \label{fig:merger_images}
\end{figure*}

\begin{figure*}
  \centering
  \includegraphics[width=1.\textwidth]{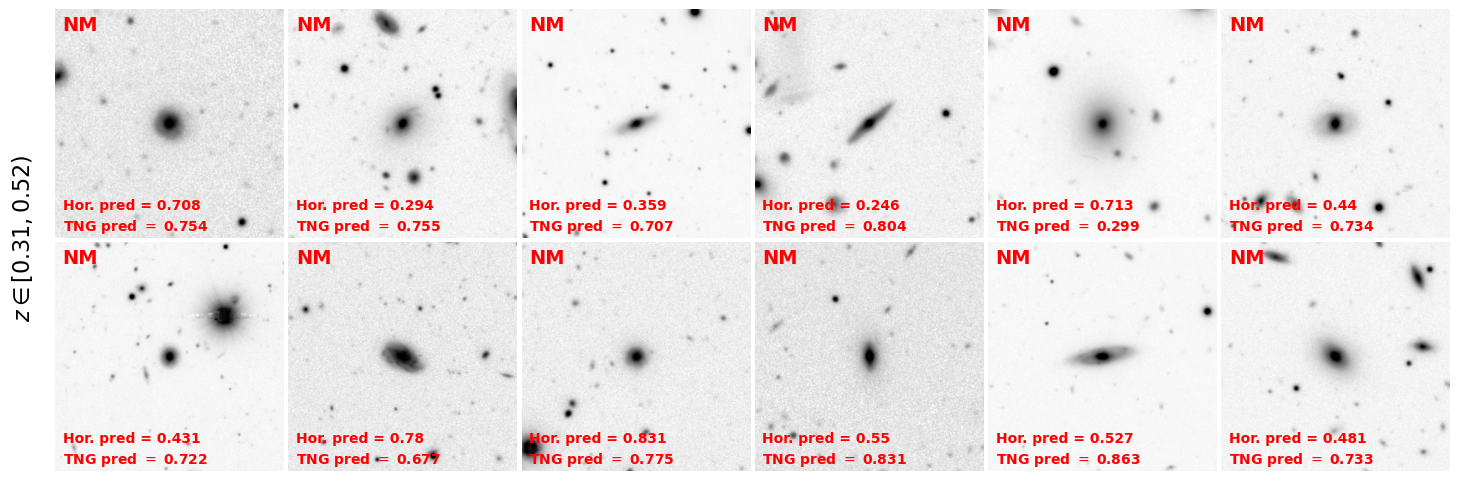}
  \includegraphics[width=1.\textwidth]{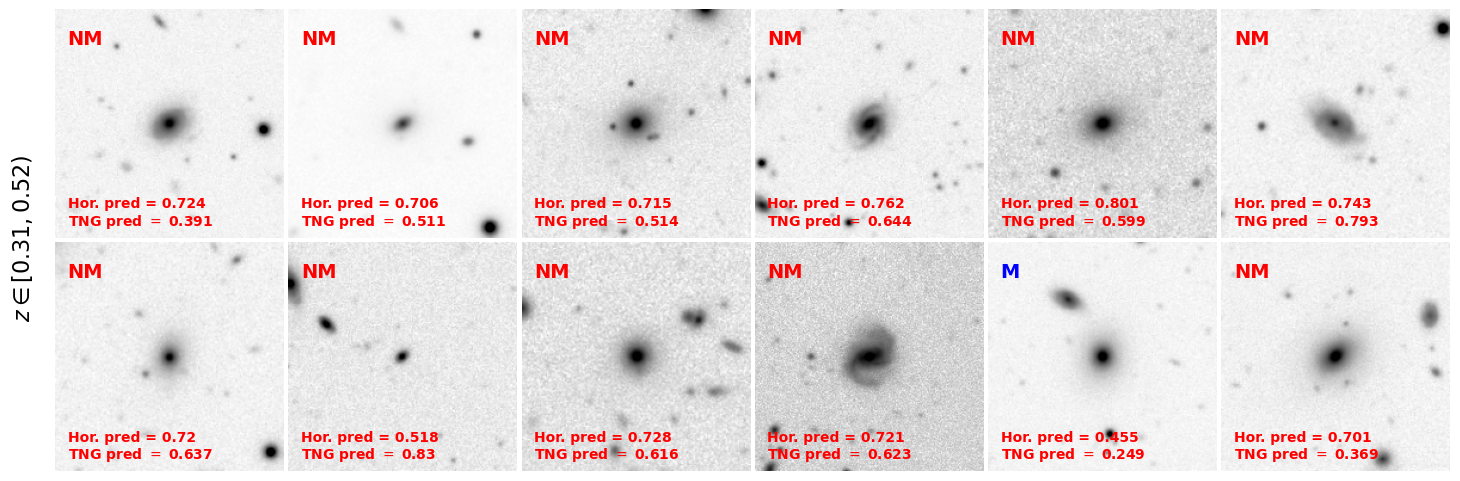}
  \includegraphics[width=1.\textwidth]{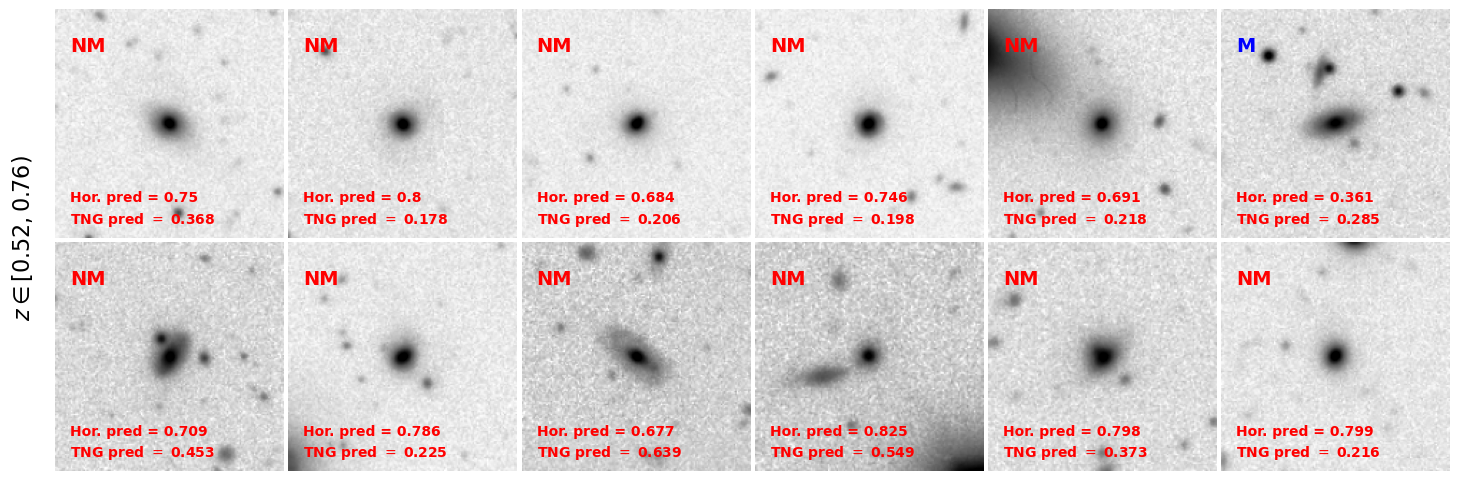}
  \caption{Similar to Fig. \ref{fig:merger_images}, but for visually classified non-mergers.}
  \label{fig:non-merger_images}
\end{figure*}

In this section, we show some examples of galaxies in the visually classified test set and how our classification scheme labels them. 
In Fig. \ref{fig:merger_images} we show 36 galaxies randomly selected from the visually inspected merger sample, 12 for each redshift bin. 
Generally speaking, there is a good agreement between the combined classifier and the visual labels. 
The model is able to correctly identify the most obvious mergers with prominent disturbances and tidal features. 
The classifier has some problems in identifying pre-mergers that do not show obvious merging signs. 
In Fig. \ref{fig:non-merger_images} we display 36 galaxies randomly chosen from the visually classified non-mergers. 
For the non-mergers, the CNNs classify correctly relatively isolated galaxies, while making some misclassifications when a nearby object is present. 

\section{Optically selected quasars}\label{sect:quasars}

Unobscured luminous AGNs can considerably contaminate the optical emission of the host galaxy.
To understand how our classification algorithm performs in the presence of a bright central point source, we analyse separately optically selected luminous quasars in our sample, as they represent the brightest optical AGNs (thus more likely to affect our merger classification). 
We selected quasars from the Sixteenth Data Release of the SDSS Quasar catalogue \citep{lyke_sloan_2020}, using the \textsc{is\_qso\_final} flag.
We matched the SDSS coordinates with the HSC coordinates of our sample using a $1\arcsec$ radius. 
In total, we found 112 quasars, of which 41 were classified as mergers, 26 as non-mergers, and 45 as unclassified by our algorithm.
Among the 112 optically-selected quasars, there were 111 SED AGNs, 58 MIR AGNs, and 91 X-ray AGNs. 
When restricting to the classified objects, we found 67, 36, and 54 SED, MIR, and X-ray AGNs, respectively.
These numbers also reflect the good quality of our SED fitting.

\begin{figure*}
    \centering
    \includegraphics[width=\textwidth]{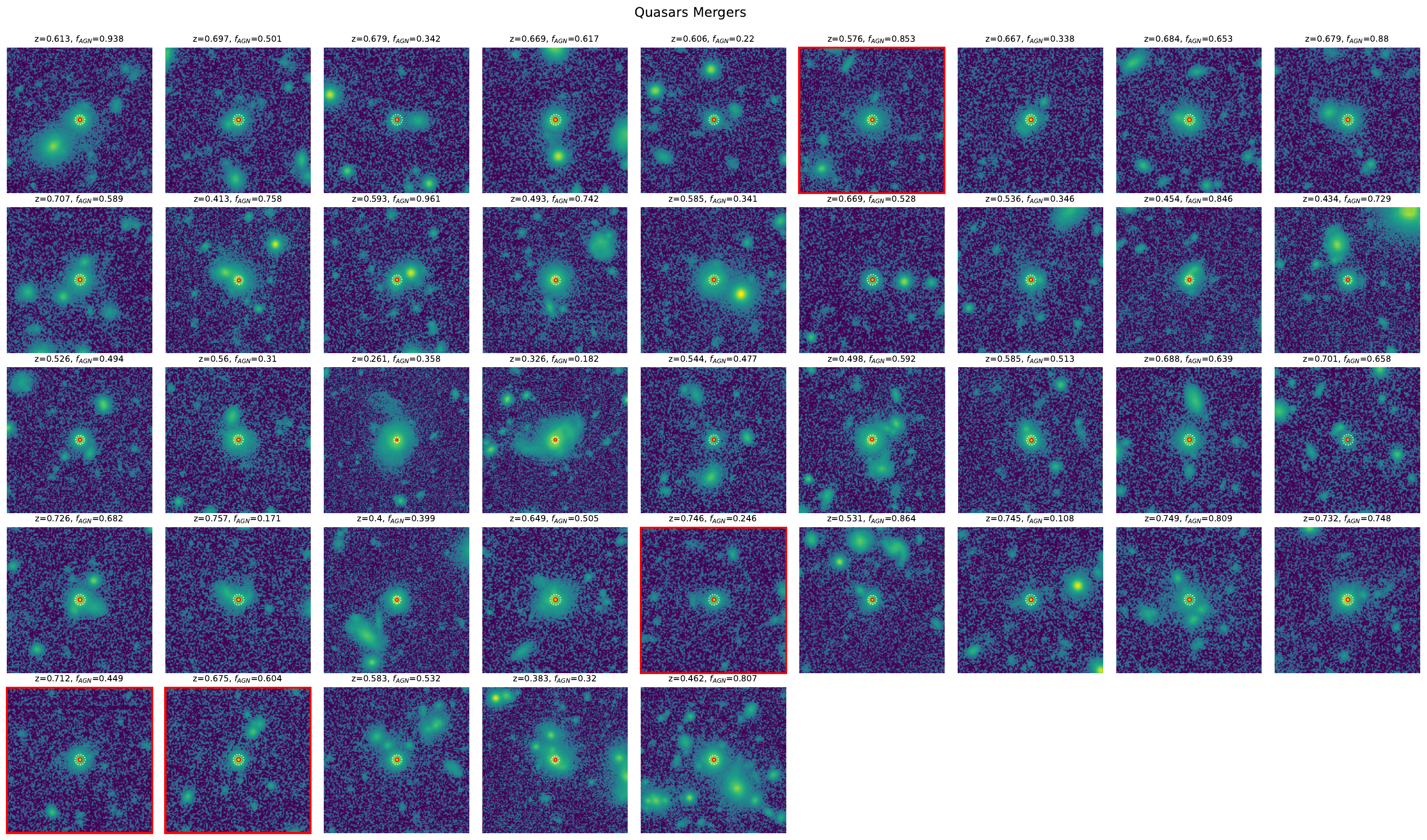}
    \includegraphics[width=\textwidth]{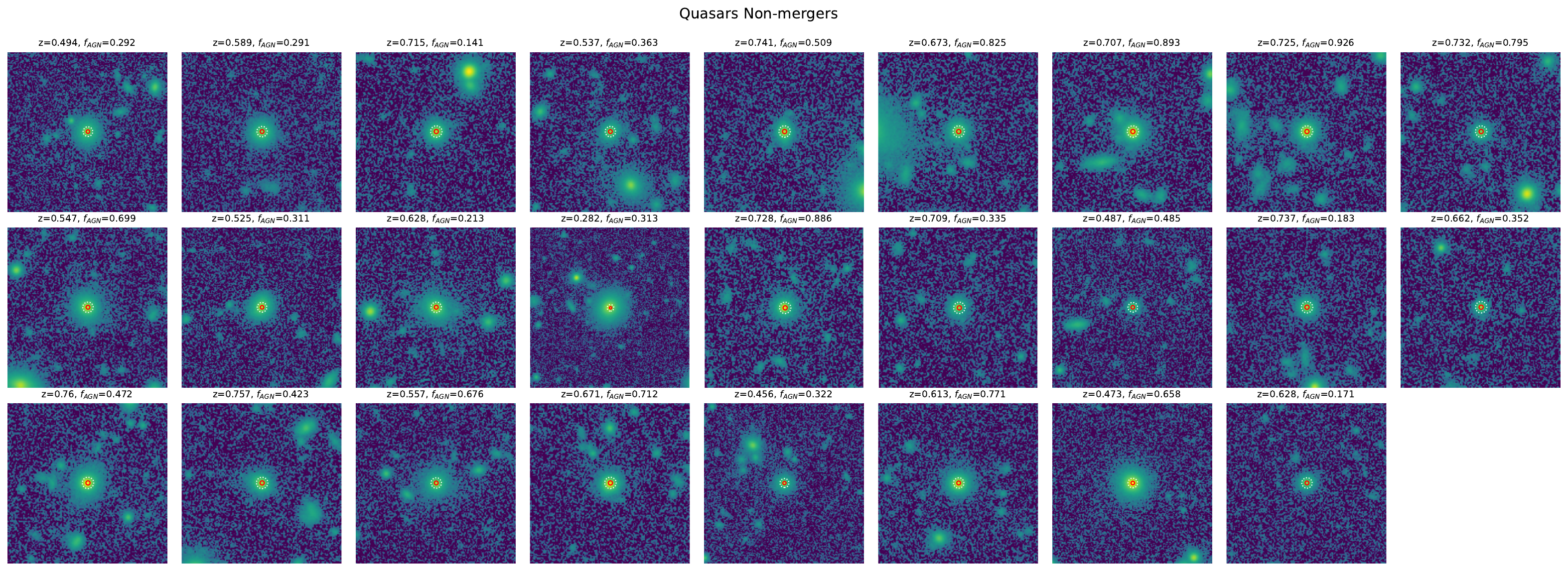}
    \caption{Optically selected quasars classified as mergers (\emph{Top}) and as non-mergers (\emph{Bottom}) by our algorithm. At the top of each cutout, we list the galaxy redshift and the AGN fraction measured by CIGALE. We show the HSC PSF as a red circle. The dotted white circle corresponds to the SDSS PSF (with width set to $1.6\arcsec$). Additionally, the galaxy mergers classified as non-mergers by visual inspection are highlighted with red edges.
    The cutouts are HSC $i-$band images with a physical size of 160 kpc and a logarithmic scaling.
    }
    \label{fig:QSOs}
\end{figure*}

In Fig.~\ref{fig:QSOs} we show HSC $i-$band cutouts of quasars classified as mergers and non-mergers. 
Based on a visual inspection of the quasars classified as mergers, it is clear that optical quasars do not particularly impact the host galaxy morphology. 
As \citet{ishino_subaru_2020} and \citet{tang_morphological_2023} showed, the angular resolution, superior seeing, and depth of HSC make it suitable for studying type-1 quasar host galaxies at $z<1$. 
The bright nucleus radiation is usually limited to a small number of pixels at the galaxy centre and the low noise level allows the detection of faint merging features in galaxy outskirts. 

\begin{figure*}[ht]
  \centering
  \includegraphics[width=0.49\textwidth]{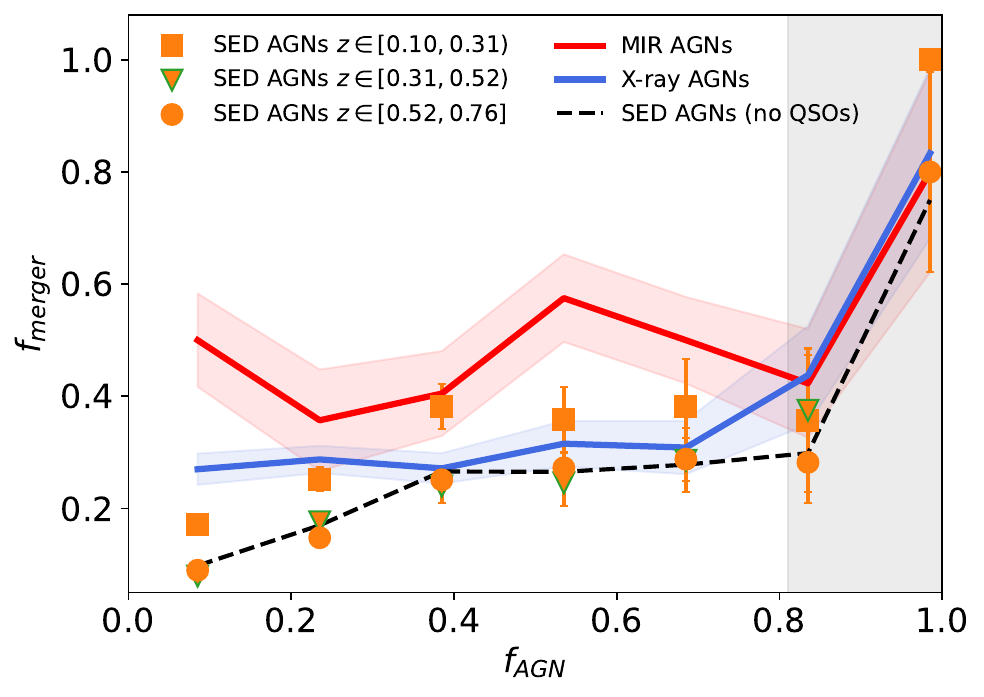}
  \includegraphics[width=.49\textwidth]{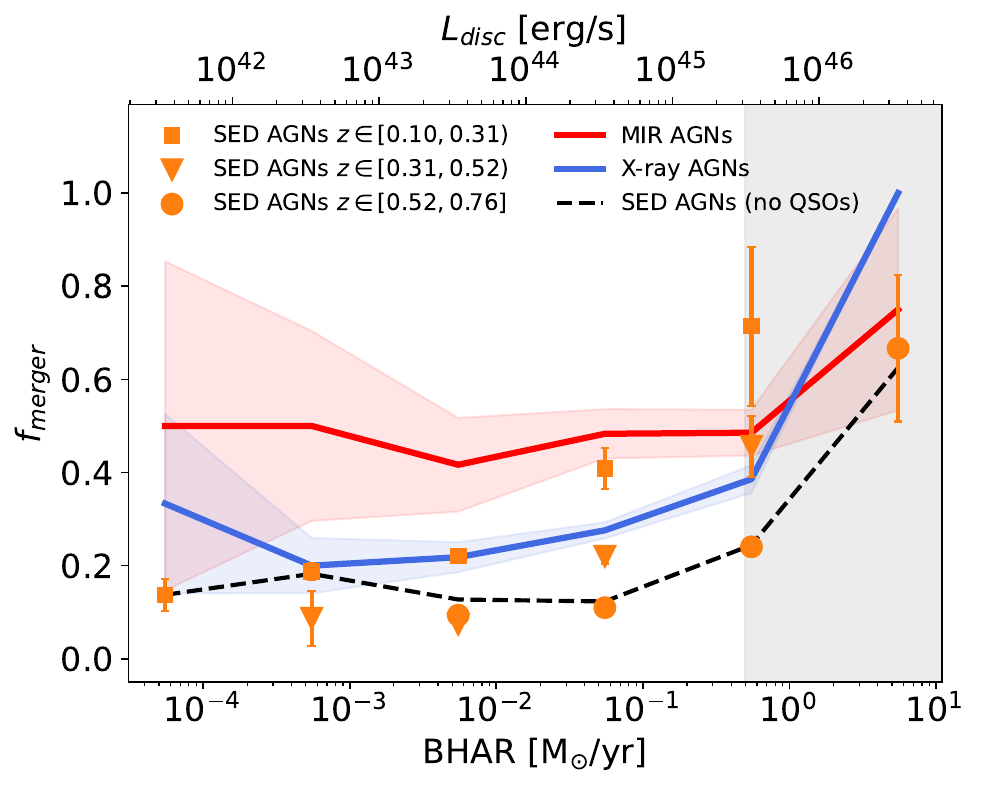}
  \caption{
  Same as Fig.~\ref{fig:fmerg_bhar} but for the sample of galaxies where quasars have been visually inspected. 
  The dashed black lines indicate the trends obtained after excluding all 67 classified optical quasars. 
  \textit{Left:} Merger fraction $f_{merger}$ as a function of $f_{AGN}$. 
  \textit{Right:} $f_{merger}$ as a function of BHAR or the equivalent AGN bolometric luminosity $L_{disc}$.
  These trends are similar to the ones presented for the main sample in Fig.~\ref{fig:fmerg_bhar}.  
  The overall trends hold even if we exclude all quasars from our sample. 
  }
    \label{fig:fmerg_bhar_QSO}
\end{figure*}

To ensure the robustness of our analysis, we also visually inspected the quasars classified as mergers and rejected those without any clear merging signs. 
These objects are highlighted with red edges in Fig.~\ref{fig:QSOs}.
Then, we performed the same analysis as for the entire sample in Sect.~\ref{sect:Results}.
We plot the merger fraction versus the AGN fraction for this sample in Fig.~\ref{fig:fmerg_bhar_QSO}. 
The relation reveals the same trend as presented in Fig.~\ref{fig:fmerg_bhar}.
For all three AGN types, there is a mildly increasing $f_{merger}$ with increasing $f_{AGN}$ at $f_{AGN}<80\%$. 
At $f_{AGN}\geq 80\%$, $f_{merger}$ quickly rises to $\sim 80-100\%$.
The same relation holds when considering the BHAR or AGN bolometric luminosity.
To further test our results, we also analysed our sample by excluding all 67 classified quasars (the dashed black line in Fig.~\ref{fig:fmerg_bhar_QSO}). 
Once again, the results qualitatively do not change. 
However, the sample size of the AGN-dominated galaxies is significantly reduced in this case. 

\section{Examples of galaxies with injected PSF}\label{appendix:psf_images}

\begin{figure*}[ht]
  \centering
  \includegraphics[width=.47\textwidth]{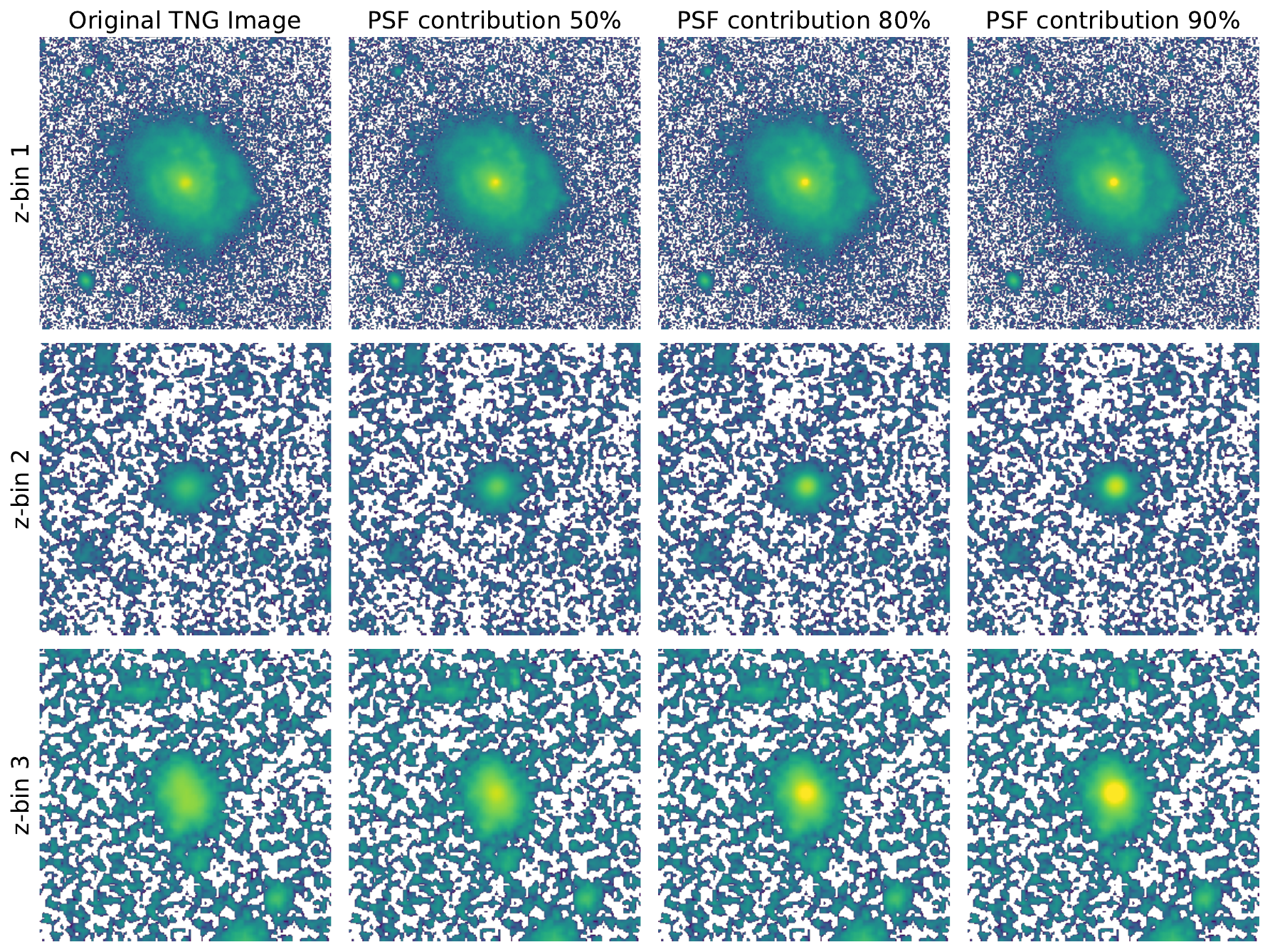}
  \includegraphics[width=.47\textwidth]{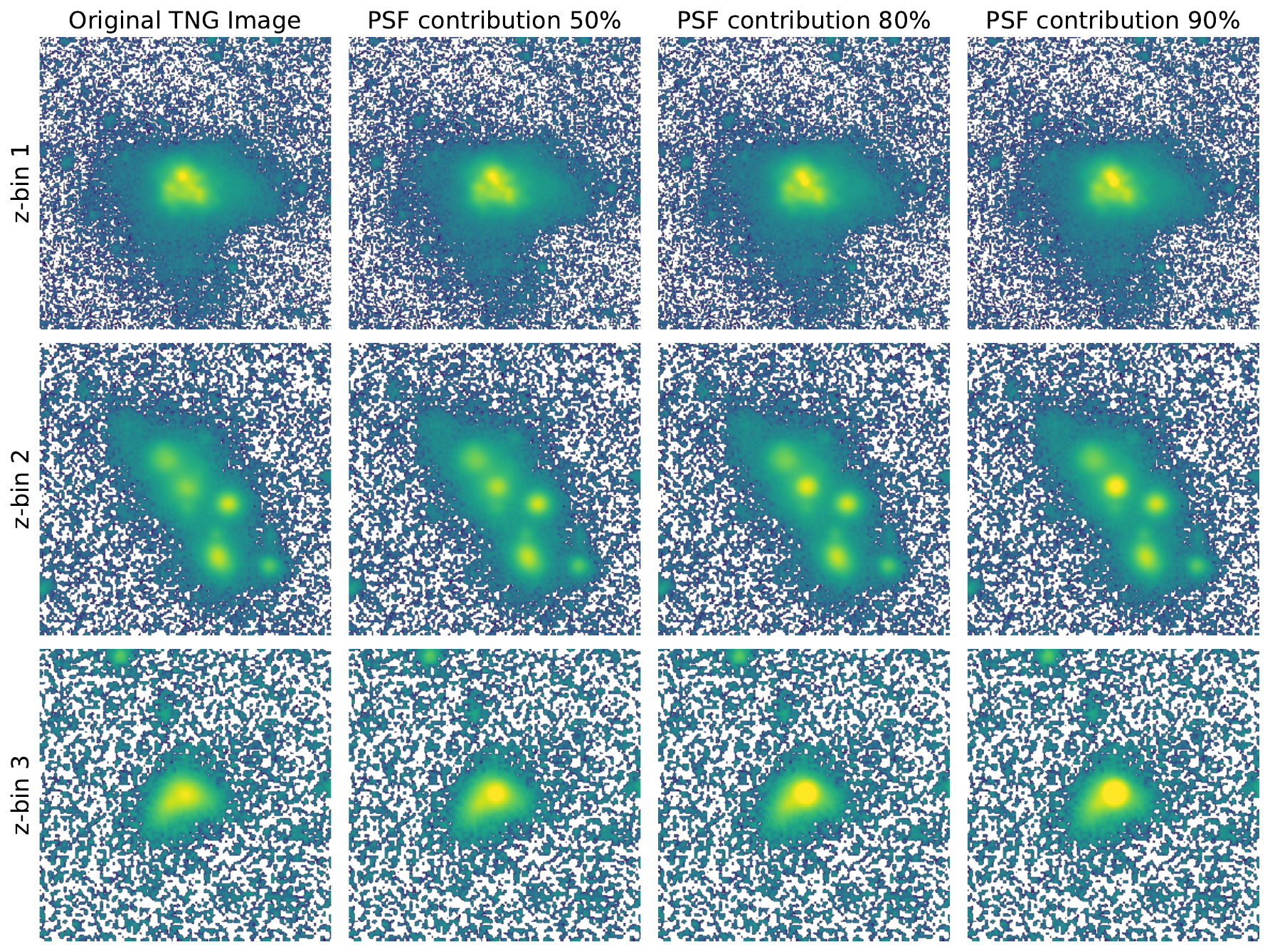}
  \includegraphics[width=.47\textwidth]{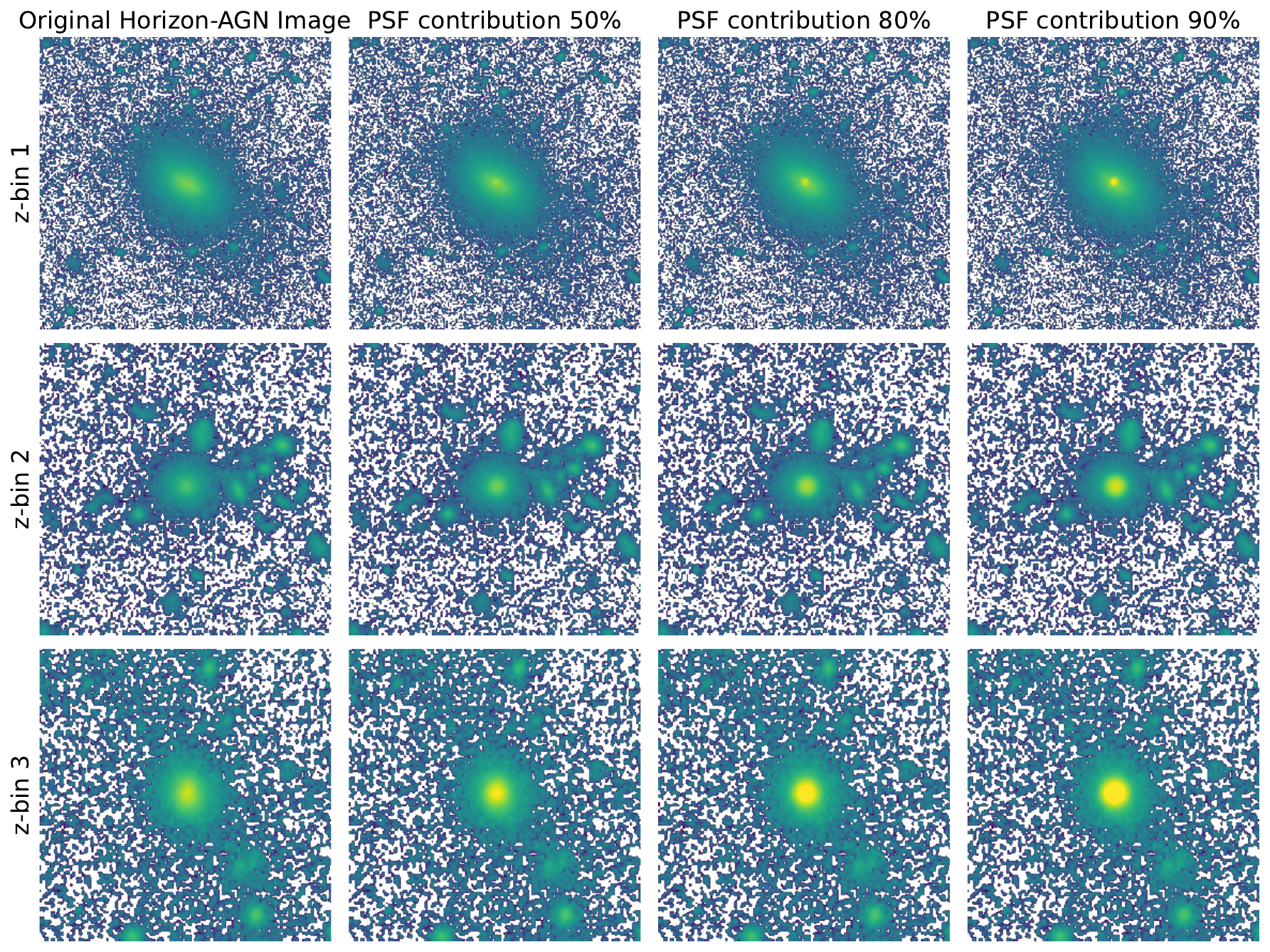}
  \includegraphics[width=.47\textwidth]{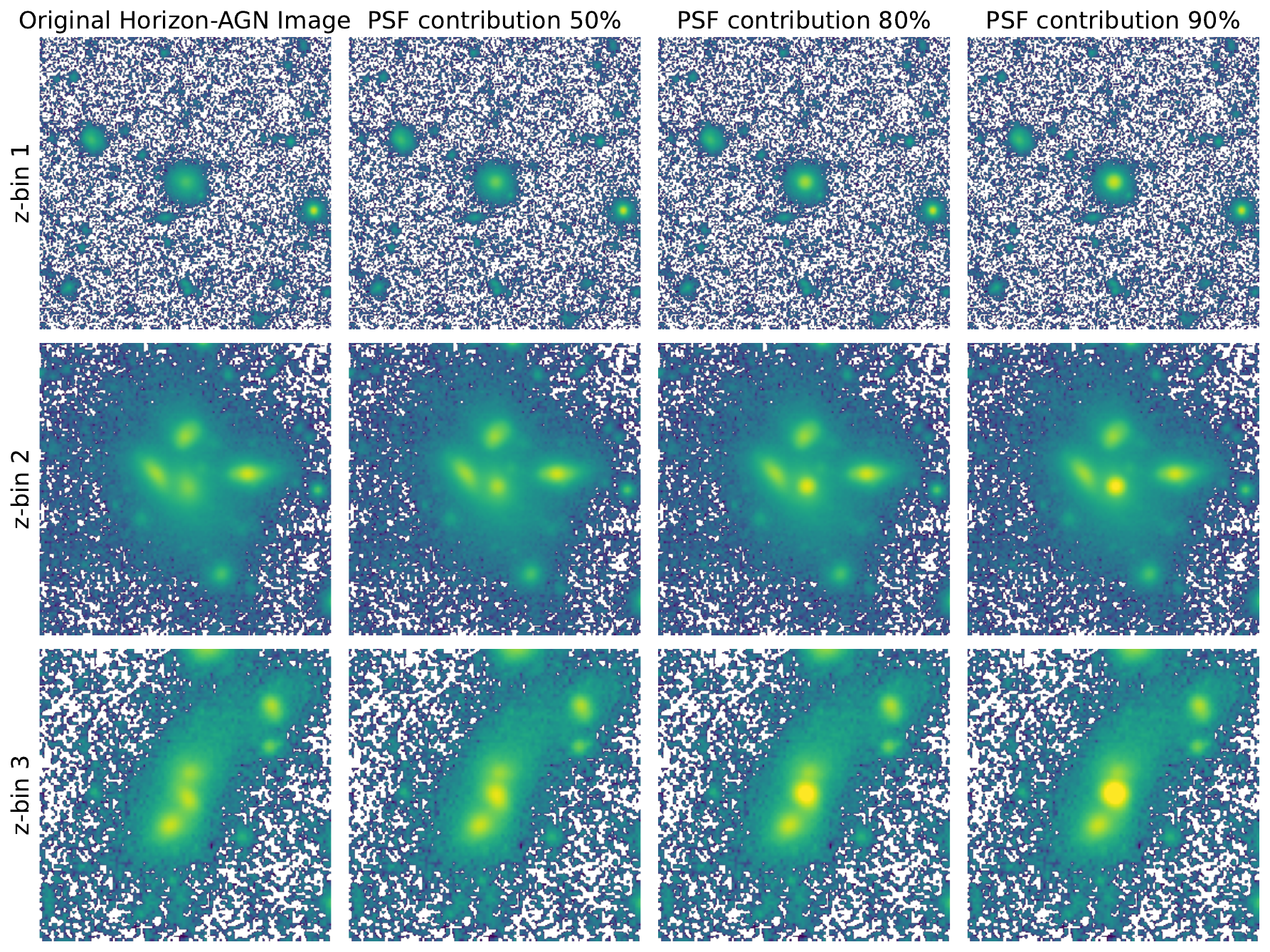}
  \caption{
  Examples of randomly chosen TNG (upper three rows) and Horizon-AGN (lower three rows) galaxies with injected point sources corresponding to different PSF contributions.
  The four left columns are examples of non-mergers, while the right four columns show examples of mergers. 
  The cutouts have a logarithmic scaling and have a physical size of $96\times 96$ kpc (to better illustrate the PSF contribution).
  }
    \label{fig:psf_examples}
\end{figure*}

In Fig. \ref{fig:psf_examples}, we show examples of the Illustris TNG and Horizon-AGN galaxies with different levels of injected PSF. 
For each simulation, we show examples of mergers and non-mergers for every redshift bin. 
The original images without PSF are also shown as a comparison. 

\end{appendix}


\end{document}